%% file: 00-main.tex
\documentclass[10pt]{article}
\usepackage[a4paper,margin=1in]{geometry}

\usepackage{mathptmx}

\usepackage{fancyhdr} 
\usepackage{titlesec} 
\usepackage{lipsum}

\usepackage{chngcntr}
\usepackage{etoolbox}

\usepackage{tikz}
\usetikzlibrary{arrows.meta}

\pretocmd{\chapter}{%
\renewcommand\thefigure{\thechapter.\arabic{figure}}%
\renewcommand\thetable{\thechapter.\arabic{table}}%
\renewcommand\theequation{\thechapter.\arabic{equation}}%

\usepackage{amsmath}

}{}{}

\usepackage{geometry}

\usepackage[utf8]{inputenc}

\usepackage{bm}
\usepackage{amsmath}
\usepackage{amssymb}
\usepackage{graphicx}

\usepackage[T1]{fontenc}
\usepackage{caption}
\usepackage[english]{babel}
\usepackage{afterpage}
\usepackage{abstract}
\usepackage[obeyDraft, bordercolor=red, backgroundcolor=white]{todonotes}
\usepackage{amsmath, amssymb, amsfonts, amsthm}               
\usepackage{graphicx}
\usepackage{subcaption}
\usepackage[toc,page]{appendix}
\usepackage[hidelinks]{hyperref}
\usepackage{bm}
\usepackage{colortbl}
\usepackage{longtable}
\usepackage{rotating}
\usepackage{lscape}
\usepackage{setspace}
\usepackage{flafter}
\usepackage{morefloats}
\usepackage{array}
\usepackage{tabularx,booktabs}
\usepackage{xr}
\usepackage{mathtools}

\usepackage[
  backend=biber,
  style=numeric,
  sorting=none,
  citestyle=numeric-comp,
  autocite=superscript,
  natbib=true,
  maxbibnames=99,
  doi=true,
  url=true,
  eprint=false
]{biblatex}

\usepackage{placeins}
\usepackage{enumitem}
\usepackage[normalem]{ulem}
\usepackage{floatrow}
\usepackage{dcolumn}
\usepackage{booktabs}
\usepackage{array}
\usepackage{booktabs}

\newcounter{referredassumption}

\usepackage[bottom]{footmisc}

\usepackage{caption}
\usepackage{lineno}
\makeatletter
\newcommand{\manuallabel}[2]{\def\@currentlabel{#2}\label{#1}}
\makeatother

\usepackage{tikz}
\usetikzlibrary{arrows.meta}

\begin{document}

\singlespacing

\title{\vspace*{4cm}Living forwards or understanding backwards? A comparison of Inverse Probability of Treatment Weighting and G-estimation methods for targeting hypothetical full adherence estimands in longitudinal cohort studies}

\author{Xiaoran Liang$^{\ast a}$,
Deniz Türkmen$^{a}$, Jane A H Masoli$^{a, b}$, Luke C Pilling$^{a}$, and Jack Bowden$^{a, c}$\\
%EndAName
$^{a}${\small Department of Clinical and Biomedical Sciences, University of Exeter, Exeter, UK}\\
$^{b}${\small Royal Devon University Healthcare NHS Foundation Trust, Barrack Road, Exeter, UK}\\
$^{c}${\small Novo Nordisk Research Centre (NNRCO), Oxford, U.K}\\[1em]
{\small Xiaoran Liang (correspondence): x.liang2@exeter.ac.uk}}

\date{\vspace{-0.9em}6 March, 2026\endgraf\bigskip\bigskip}

\maketitle
\thispagestyle{empty}
\pagebreak

\begin{abstract}\vspace{-1em}
%\doublespacing
\noindent
Medication adherence is essential to ensure treatment effectiveness, but too often in routine care non-adherence compromises the desired outcome.
We explore longitudinal causal modelling using observational data to estimate the time-varying effects of continuous drug adherence measures on health outcomes over a sustained period. 
The goal of such analyses is to quantify the potential impact of interventions to improve adherence on long-term health. 
We consider two established longitudinal causal approaches designed to handle time-varying confounding under the ``no unmeasured confounding'' (NUC) assumption: G-estimation and inverse probability of treatment weighting (IPTW). 
In randomized trials, NUC-based methods have been applied to address non-adherence as an intercurrent event, and instrumental variable (IV) extensions of G-estimation have also been introduced for settings where the NUC assumption may fail. 
We adapt these methods to observational data settings and illustrate their use for assessing how adherence over time impacts health outcomes.
We align the causal parameters across methods and show they can target the same causal estimand: the average effect among treated individuals of full adherence versus zero adherence. We set out the identification conditions for IPTW and G-estimation under NUC, and for an IV-based extension that has specific utility when the NUC assumption is implausible. 
We assess the statistical properties, strengths and weaknesses of each approach through Monte Carlo simulations designed to reflect longitudinal studies with a continuous exposure.
We demonstrate these methods by quantifying the effect of full statin adherence on LDL cholesterol control in 13,000 UK Biobank participants with linked primary care data.

\medskip 

\noindent%
\end{abstract}

\pagenumbering{gobble}
\pagebreak
\pagenumbering{arabic} 
\setcounter{page}{1}%

\manuallabel{app:distance}{A}
\manuallabel{app:Q}{B}
\manuallabel{app:model_setup}{C}
\manuallabel{app:model_pleiotropy}{C.1}
\manuallabel{app:direct_causality}{C.2}
\manuallabel{app:simulations}{D}
\manuallabel{app:simudesign}{D.1}
\manuallabel{app:simumethod}{D.2}
\manuallabel{app:simuresults}{D.3}
\manuallabel{tab:app_simulation3}{S1}
\manuallabel{tab:app_simulation2_correlation}{S2}
\manuallabel{tab:app_simulation3_correlation}{S3}
\manuallabel{tab:app_simulation1}{S4}
\manuallabel{app:extend_simu}{E}
\manuallabel{app:inference}{E.1}
\manuallabel{app:alt_pleiotropy}{E.2}
\manuallabel{app:weak_iv}{E.3}
\manuallabel{app:application}{F}
\manuallabel{app:application_estimation}{F.1}
\manuallabel{tab:app_estimation}{S14}
\manuallabel{tab:app_estimation_mclust}{S15}
\manuallabel{tab:app_estimation_navmix}{S16}
\manuallabel{app:pathway}{F.2}
\manuallabel{tab:app_cluster1_mr}{S17}
\manuallabel{tab:app_cluster4_mr}{S18}

%\onehalfspacing
\begin{refsection}
\input{01-Introduction}

\input{02-Methods}
\input{03-Simulations}
\input{04-Application}
\input{05-Discussion}

\singlespacing
\printbibliography
\end{refsection}

\doublespacing

\clearpage

\begin{refsection}
\input{06-Appendix}

\newpage 
\singlespacing
\clearpage
\printbibliography

\end{refsection}

\end{document}

%% file: 01-Introduction.tex
\section{Introduction}\label{sec:intro}
Drug adherence refers to the degree to which patients comply with prescribed therapeutic regimens when taking medications \citep{sabate2003adherence,andrade2006methods, cordioli2023socio}. Whilst a high level of adherence is essential for ensuring the expected efficacy of a given treatment, in routine care settings, low adherence is commonplace and is a major obstacle that compromises this desired process \citep{andrade2006methods,sperber2017upper, bosworth2011medication, lam2015medication}. For instance, real-world studies report that adherence to statin therapy can drop below 50\% within the first year of treatment, which is substantially lower than observed in the controlled trials that led to their original approval \citep{maningat2013we, rodriguez2019association}. With routinely collected patient data available in electronic health records (EHRs), we cannot directly measure how much drug the patients actually take. However, we can estimate adherence using patients’ prescription records and the time they are treated using the {\it Medication Possession Ratio} (MPR), which is defined as sum of the days’ supply of medication divided by the number of days in the observation period, expressed as a percentage \citep{andrade2006methods, hess2006measurement}. If a clinically significant benefit would be achieved by increasing adherence, the focus can then shift to deciding how to realize such an intervention in a cost-effective manner – hence the need for causal methodology.\\
\\
EHRs provide the longitudinal data necessary for such analyses. By tracking both prescription history and health outcomes over time, they create the opportunity to evaluate how sustained adherence impacts long-term health.
However, this longitudinal structure introduces a key methodological challenge when estimating the joint effect of a sequence of adherence interventions: feedback between the time-varying exposure and time-varying confounders.
For example, a patient’s evolving health status often influences both their subsequent adherence and their future health outcomes, and thus acts as a confounder. Crucially, this same health status is often itself affected by prior adherence.
In this setting, standard regression methods face a dilemma: adjusting for this intermediate health status blocks part of the effect of past adherence mediated through health improvement, whereas failing to adjust for it leaves the effect of current adherence confounded \citep{robins1992g,robins2000marginal, vansteelandt2014structural, blackwell2018make}. 
To address the time-varying confounding problem, longitudinal causal methods designed to handle such feedback loops are required. \\
\\
These methods have been widely applied in clinical trials to address non-adherence, classified by the ICH E9(R1) addendum \citep{ICH2020} as an `intercurrent event', i.e., an event occurring post-randomization that affects the straightforward interpretation of the treatment effect. The addendum formalizes this perspective by introducing the Estimand Framework, which emphasizes the importance of clearly defining the treatment effect of interest by explicitly acknowledging and accounting for intercurrent events within the analysis.
Estimating the treatment effect which would have been realized if (possibly contrary to fact) all patients fully adhered is a clear example of a Hypothetical Strategy when mapped to the E9 addendum. 
Longitudinal causal modelling approaches have been proposed to estimate hypothetical estimands in randomized trials under the assumption that there are no unmeasured confounders of the intercurrent event-outcome relationship. Examples include inverse probability of treatment weighting (IPTW) and the parametric g-formula  \citep{olarte2023hypothetical} and more recently, structural nested mean models fitted by G-estimation (e.g., \citet{lasch2025comparison, olarte2025estimating}). 
Whilst the ``no unmeasured confounding'' (NUC) assumption is unquestionably valid at the point of randomization in a properly conducted large trial, it does not ensure that the outcomes of patients who adhere fully and those who do not are exchangeable as a trial progresses, especially as the length of follow-up increases. For this reason, instrumental variable (IV) methods have also been developed to target hypothetical estimands within the Estimand Framework: initially for single-time outcomes \citep{bowden2021connecting} and more recently for longitudinal outcomes \citep{Bowden2025}. Trial-based IV approaches exploit random assignment as an instrument because it is generally a strong predictor of adherence and, by definition, is independent of other variables that could confound adherence and the outcome, thereby supporting identification of the target estimand even when the NUC assumption is violated.\\
\\
While the problem of non-adherence is most formally addressed in trials, it is equally pervasive in routine care \citep{andrade2006methods,sperber2017upper, bosworth2011medication}. We therefore adapt the trial-based framework to address non-adherence in the observational setting, using statin adherence as a clinically prominent example from our recent work in \citet{turkmen2025understanding}. In that study, we conducted a detailed epidemiological and genetic investigation of approximately 69{,}000 UK Biobank participants with linked primary care data who were prescribed statins and had pre- and post-treatment low-density lipoprotein cholesterol (LDL-c) measurements, alongside an MPR-based measure of adherence. For a subset of roughly 13{,}000 individuals with three post-treatment LDL-c measurements over a maximum of 10 years follow-up after statin initiation, our central questions were:\\
\\
{`{\it Over routine follow-up, how much lower would LDL-c be on average if these 13,000 statin patients had adhered fully at every interval (MPR=100\%) compared with if they did not adhere at all (MPR=0\%)?'}}\\
\\
Furthermore, by estimating time-varying effects, the longitudinal framework allows us to decompose this cumulative effect and investigate:\\
\\
{`{\textit{How does statin adherence at different follow-up intervals contribute to the final LDL-c outcome?'}}}\\
\\
We previously applied IPTW of Marginal Structural Models (MSM) \citep{robins2000marginal} to address these questions, given its widespread use in applied longitudinal studies \citep{clare2019causal}. Specifically, we estimated weights for continuous adherence using the Covariate Balancing Propensity Score (CBPS) method \citep{fong2018covariate}, adjusting for baseline covariates as well as prior LDL-c, prior MPR and follow-up time as time-varying confounders.
Building on this work and our observational IPTW application, here we additionally consider G-estimation of Structural Nested Mean Models (SNMM) \citep{robins1994correcting} under the NUC assumption. 
This is motivated by the fact that adherence is measured on a continuous scale: while IPTW is widely used and useful, it can be sensitive to practical positivity violations and weight instability in continuous-exposure settings, whereas G-estimation can offer a natural alternative with practical advantages \citep{vansteelandt2014structural, vansteelandt2016revisiting, goetgeluk2008estimation, westreich2010invited}.
A key challenge in the observational setting is the absence of randomization, meaning adherence behaviours are likely influenced by both observed and unobserved confounders. We therefore also consider an IV extension of G-estimation for settings where the NUC assumption may fail \citep{Bowden2025}.
Unlike in randomized trials where treatment assignment provides a natural instrument for adherence, IV analyses in observational data require careful justification of candidate instruments and may be limited by instrument strength.\\
\\
We place IPTW, G-estimation under NUC, and IV-based G-estimation within a unified longitudinal framework and define the causal estimand of interest. We introduce the underlying structural models: MSM for IPTW and SNMM for G-estimation, demonstrating that both can be specified to target the same estimand (Sections 2.1-2.2).
We then detail the identification assumptions and general estimation strategies for each approach (Sections 2.3–2.4). 
We also show how G-estimation can be implemented within a generalized method of moments (GMM) framework, allowing efficient estimation and standard sandwich-based inference.
Kierkegaard's adage that `life must be lived forwards but understood backwards' is particularly apt for our exposition: IPTW sequentially re-weights observations forwards in time from the study beginning to its end, whereas G-estimation subtracts adherence effects backwards in time from the study end to its beginning, but (unlike the adage) both approaches can furnish valid causal estimates.
In Section 2.5, we provide specific implementation details and inferential procedures for these methods. 
In Section 3, we assess the finite-sample performance of these methods via extensive Monte Carlo simulations, specifically evaluating their estimation accuracy, statistical efficiency, and robustness to strong feedback between exposure and covariates, alongside the utility of diagnostic statistics for examining the estimated weights.
In Section 4, we illustrate the approaches by estimating the effect of statin adherence on LDL-c control using the same UK Biobank patient subset of 13,000 participants, using their genotypic data in an effort to construct an instrument for statin adherence. \textsf{R} code to implement our approaches is provided and in Section 5 we discuss their potential to be applied to future observational cohort studies on the horizon.

%% file: 02-Methods.tex
\section{Methods}\label{sec:method}
\subsection{Longitudinal Modelling Framework and the Causal Estimand of Interest}
\begin{figure}[htbp]
    \centering

    % Subfigure (a)
    %\begin{subfigure}{\textwidth}
        \centering
         \resizebox{10cm}{!}{
         \begin{tikzpicture}
    \begin{scope}[every node/.style={rectangle,thick,draw}]
        \node (U1) at (0,4) {$\bm{L}_1, U_1$};
        \node (U2) at (3,4) {$\bm{L}_2, U_2$};
        \node (UK) at (6,4) {$\bm{L}_3, U_3$};
        \node (A1) at (-1,2) {$A_1$};
        \node (A2) at (2,2) {$A_2$};
        \node (AK) at (5,2) {$A_3$};
        \node (Y1) at (-1,0) {$Y_1$};
        \node (Y2) at (2,0) {$Y_2$};
        \node (YK) at (5,0) {$Y_3$};
    \end{scope}

    \begin{scope}[>={Stealth[black]}, every node/.style={fill=white,circle}]
        \draw[->,blue,thick,-{Stealth}] (Y2)  --   (YK);
        \draw[->,black,thick,-{Stealth}] (Y2)  --  (AK);
        \draw[->,black,thick,-{Stealth}] (Y1)  --  (AK);
        
        \draw[->,black,thick,-{Stealth}] (U1) -- (A1); 
        \draw[->,black,thick,-{Stealth}] (U1) -- (Y1); 
        \draw[->,black,thick,-{Stealth}] (U2) -- (A2); 
        \draw[->,red,thick,-{Stealth}] (U2) -- (Y2); 
        \draw[->,black,thick,-{Stealth}] (UK) -- (AK); 
        \draw[->,blue,thick,-{Stealth}] (UK) -- (YK); 

        \draw[->,red,thick,-{Stealth}] (A1) -- (Y1);
        \draw[->,red,thick,-{Stealth}] (A1) -- (Y2);
        \draw[->,red,thick,-{Stealth}] (A1) -- (YK);
        \draw[->,black,thick,-{Stealth}] (A1) -- (A2); 
        \draw[->,black,thick,-{Stealth}] (Y1) -- (A2); 
    
        \draw[->,red,thick,-{Stealth}] (Y1) -- (Y2);

        \draw[->,blue,thick,-{Stealth}] (A2) -- (Y2);
        \draw[->,blue,thick,-{Stealth}] (A2) -- (YK);
        \draw[->,black,thick,-{Stealth}] (A2) -- (AK); 

        \draw[->,green,thick,-{Stealth}] (AK) -- (YK); 

        \draw[->,red,thick,-{Stealth}] (A1) -- (U2); 
        \draw[->,black,thick,-{Stealth}] (Y1) -- (U2); 

        \draw[->,blue,thick,-{Stealth}] (A2) -- (UK); 
        \draw[->,black,thick,-{Stealth}] (A1) to[out=20,in=160] (AK);
    \end{scope}
\end{tikzpicture}
         }
    \caption{Illustration of the assumed longitudinal causal structure with three time points.
    Arrows are colour-coded to reflect the components of the total causal effect of adherence at each time point: red for $A_1$, blue for $A_2$ and green for $A_3$. For visual clarity, we omit some arrows (such as the direct effect of $A_1$ on $\bm{L}_{3}$), but the framework allows for general relationships where variables may depend on the entire past history.}
    \label{fig:causal-DAG}
\end{figure}
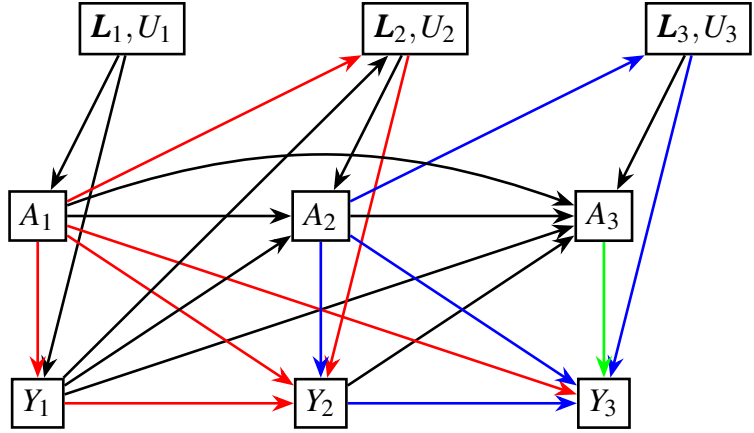

Suppose we have a sample of $n$ 
on-medication patients followed longitudinally across $K$ time points. For each individual $i$ and time point $k$ ($k \in \{1, ..., K\}$), we assume complete information on the following variables is available in theory:
\begin{itemize}
    \item Drug adherence status $A_{i, k}$.
    \item A continuous clinical outcome measure $Y_{i, k}$.
    \item An $l \times 1$ vector $\bm{L}_{i, k}$ which includes all $l$ observed covariates (both baseline and time-varying) at time point $k$ that confound the $A-Y$ relationship.
\end{itemize}
Additionally, to formalize the potential for confounding, we posit the existence of:
\begin{itemize}
    \item An `unobserved' confounder $U_{i, k}$ between $A_{i, k}$ and $Y_{i, k}$.
\end{itemize}
The longitudinal data structure is illustrated in Figure~\ref{fig:causal-DAG} with three time points. 
In this setting, adherence at a given time may affect later outcomes both directly and indirectly through mediators (such as through its influence on intermediate outcomes). When we refer to the ``effect of adherence at time $j$ on outcome at time $k$'' ($j \leq k$), we mean the total effect that operates along all pathways except those that pass through future adherence status.
The arrows in Figure~\ref{fig:causal-DAG} use different colours to highlight the components of the total causal effect of adherence at each time point on the longitudinal outcomes: red for $A_1$, blue for $A_2$ and green for $A_3$.
\\
\\
Let the vector $\bar{\bm{A}}_{i,k}$ denote the adherence history of individual $i$ up to time point $k$, so that $\bar{\bm{A}}_{i,k} = \left(A_{i,1}, ..., A_{i,k}\right)$. Define $Y_{i,k}(\bar{\bm{a}}_k)$ as the potential outcome at time point $k$ given that the adherence history took, possibly contrary to fact, the value $\bar{\bm{a}}_k$. We are interested in quantifying the difference between mean potential outcomes under different intervention histories in order to inform future policy decisions and, more broadly, to compare different statistical approaches for estimating them from observational data. 
In principle, we can consider contrasts between any two adherence histories $\bar {\bm{a}}_k$ and $\bar {\bm{a}}_k^{\prime}$.
In this paper our primary focus is the comparison between sustained full adherence and sustained non-adherence, represented by the constant histories $\bar {\bm{a}}_k = \mathbf{1}_k$ (full adherence at every time up to $k$) and $\bar {\bm{a}}_k^{\prime} = \mathbf{0}_k$ (no adherence at any time up to $k$), respectively.
Our primary estimand of interest is therefore the difference in mean potential outcomes among the medication initiators:
\begin{equation}\label{eq:estimand}
    \mathbb{E}[Y_{i,k}(\bm{1}_k) - Y_{i,k}(\bm{0}_k) \mid S = 1] \text{ for $k=1,...,K$}.
\end{equation}
In Equation (1), $S = 1$ denotes that we are conditioning on being treated with the chosen medication (e.g., statin), and the target estimand is defined among those prescribed the medication. For brevity, we implicitly assume and therefore omit this conditioning from now on in subsequent expressions. 
For the remainder of the methodology sections, we anchor our theoretical development to an arbitrary target time point $k$. 
To simplify notation, we drop the subscript $k$ and denote this target outcome simply as $Y$ (i.e., $Y \equiv Y_{i,k}$), with the understanding that the framework applies identically to any chosen time horizon $k \in \{1, \dots, K\}$. 
Throughout, we also suppress the individual patient subscript $i$ for notational convenience.
%We also drop the subscript $k$ for the final chosen outcome and denote simply as $Y$. 

\subsection{Parameterizing the Causal Estimand via MSM and SNMM}\label{sec:MSM-and-SNMM}
To estimate the causal contrast defined in Equation \eqref{eq:estimand}, we link potential outcomes to observed data using two modelling frameworks commonly used for longitudinal causal inference: SNMM and MSM \citep{robins1994correcting, robins2000marginal, HernanRobins2020}.
We begin with the SNMM which models the causal effect of adherence at each time point through the blip function. 
We define patients' observed history $\bm{H}_j$ as the collection of all information available up to time $j$:
\begin{equation*}
\bm{H}_j=(\bar {\bm{A}}_{j-1}, \bar {\bm{Y}}_{j-1}, \bar {\bm{L}}_j)
\end{equation*}
for $j = 1, \dots, k$. Here, $\bar {\bm{L}}_j$ denotes the observed covariate history up to time $j$, and $\bar{\bm{Y}}_{j-1}$ denotes the observed intermediate outcome history up to time $j-1$.
The blip at time $j$ is defined as the contrast in conditional mean potential outcomes when adherence at time $j$ is set to $a_j$ versus set to $0$, given the observed history $\bm{H}_j$, holding the past adherence fixed and setting the future adherence as $0$ \citep{robins1994correcting, vansteelandt2014structural}:
\begin{equation}\label{eq:SNMM}
    \mathbb{E}[Y(\bar {\bm{a}}_{j-1}, a_j,\bm{0}_{k-j})-Y(\bar {\bm{a}}_{j-1}, 0, \bm{0}_{k-j})\mid \bm{H}_j, A_j=a_j] = \psi_j(a_j,\bm{H}_j).
\end{equation}
It is a conditional causal effect that compares switching on/off adherence at time $j$ among individuals with the same observed history $\bm{H}_j$.
Here we assume no effect modification by exposure, outcome and covariate histories, and let the blip function depend only on the current adherence level $a_j$. Additionally, we assume ``no current treatment interaction'' (NCTI), which states that the causal effect is invariant to the actual adherence status $A_j$. Following the notation in \citet{tompsett2025instrumental}, we formalize the NCTI assumption as:
\begin{align*}
& \mathbb{E}[Y(\bar {\bm{a}}_{j-1}, a_j,\bm{0}_{k-j})-Y(\bar {\bm{a}}_{j-1}, 0, \bm{0}_{k-j})\mid \bm{H}_j,A_j=a^{\prime}]\\
&=
\mathbb{E}[Y(\bar {\bm{a}}_{j-1}, a_j,\bm{0}_{k-j})-Y(\bar {\bm{a}}_{j-1}, 0, \bm{0}_{k-j})\mid \bm{H}_j,A_j=a^{\prime\prime}] \quad \forall a^{\prime},a^{\prime\prime}.
\end{align*}
Collectively, under these structural assumptions, the blip function simplifies to the following form:
\begin{equation}\label{eq:blip}
    \psi_j(a_j,\bm{H}_j)=\beta_j a_j \quad \text{   for all $\bm{H}_j$}.
\end{equation}
In Appendix A.1, we show that under the linear blip specification in Equation (\ref{eq:blip}), the estimand in Equation (\ref{eq:estimand}) can be written as a function of the SNMM parameters $\bm{\beta}_k = (\beta_1,\ldots,\beta_k)$ such that:
\begin{equation}\label{eq:general-snmm}
    \mathbb{E}[Y(\bar {\bm{a}}_k) - Y(\bm{0}_k)]
= \sum_{j=1}^k \beta_j a_j.
\end{equation}
Each $\beta_j$ represents the incremental causal effect of adherence at time $j$, and the total effect of sustained full adherence (setting all $a_j = 1$) versus sustained non-adherence is obtained by summing all the $\beta_j$.\\
\\
Next we turn to the MSM \citep{robins2000marginal}, which specifies the marginal mean of the potential outcome $Y(\bar {\bm{a}}_k)$ as a function of the adherence history $\bar {\bm{a}}_k$ with the parameter vector $\bm{\eta}_k = \left(\eta_0, \eta_1, ..., \eta_k\right)$. Here we focus on a linear additive MSM, which is a natural choice given that the outcome $Y$ is continuous. We write:
\begin{equation}\label{eq:MSM}
    \mathbb{E}[Y(\bar {\bm{a}}_k)] = \eta_0 + \sum_{j = 1}^k \eta_j a_j \text{.}
\end{equation}
Under this specification, $\eta_0$ represents the mean potential outcome under sustained non-adherence (i.e., $\bar {\bm{a}}_k = \mathbf{0}_k$), and each coefficient $\eta_j$ can be interpreted as the marginal contribution to the mean outcome of adhering at time $j$ rather than not adhering, holding adherence at all other times fixed \citep{HernanRobins2020}.
For the causal estimand in (\ref{eq:estimand}), it follows directly from (\ref{eq:MSM}) that
\begin{align*}
     \mathbb{E}[Y(\bm{1}_k)] - \mathbb{E}[Y(\bm{0}_k)] = \sum_{j = 1}^k \eta_j \text{.}
\end{align*}
Thus, under their respective specifications, both SNMM and MSM allow us to estimate the target causal estimand in Equation (\ref{eq:estimand}) by summing their time-specific coefficients, ensuring that both frameworks align with the same estimand.
Furthermore, in Appendix A.1, we show that under these model specifications, the SNMM parameters $\beta_j$ and the MSM parameters $\eta_j$ coincide (i.e., $\beta_j = \eta_j$ for all $j$). This equivalence implies that both models target the same time-varying causal effects, allowing us to evaluate their performance against a common ground truth in our simulation study in Section~\ref{sec:main_simu}.
In practice, the MSM parameter $\bm{\eta}_k$ and the SNMM blip parameters $\bm{\beta}_k$ can be estimated from the observed data using IPTW and G-estimation, respectively \citep{robins1994correcting, robins1997causal, robins2000marginal}, under suitable identification assumptions, which are described in the next subsection.

\subsection{IPTW and G-estimation under NUC}
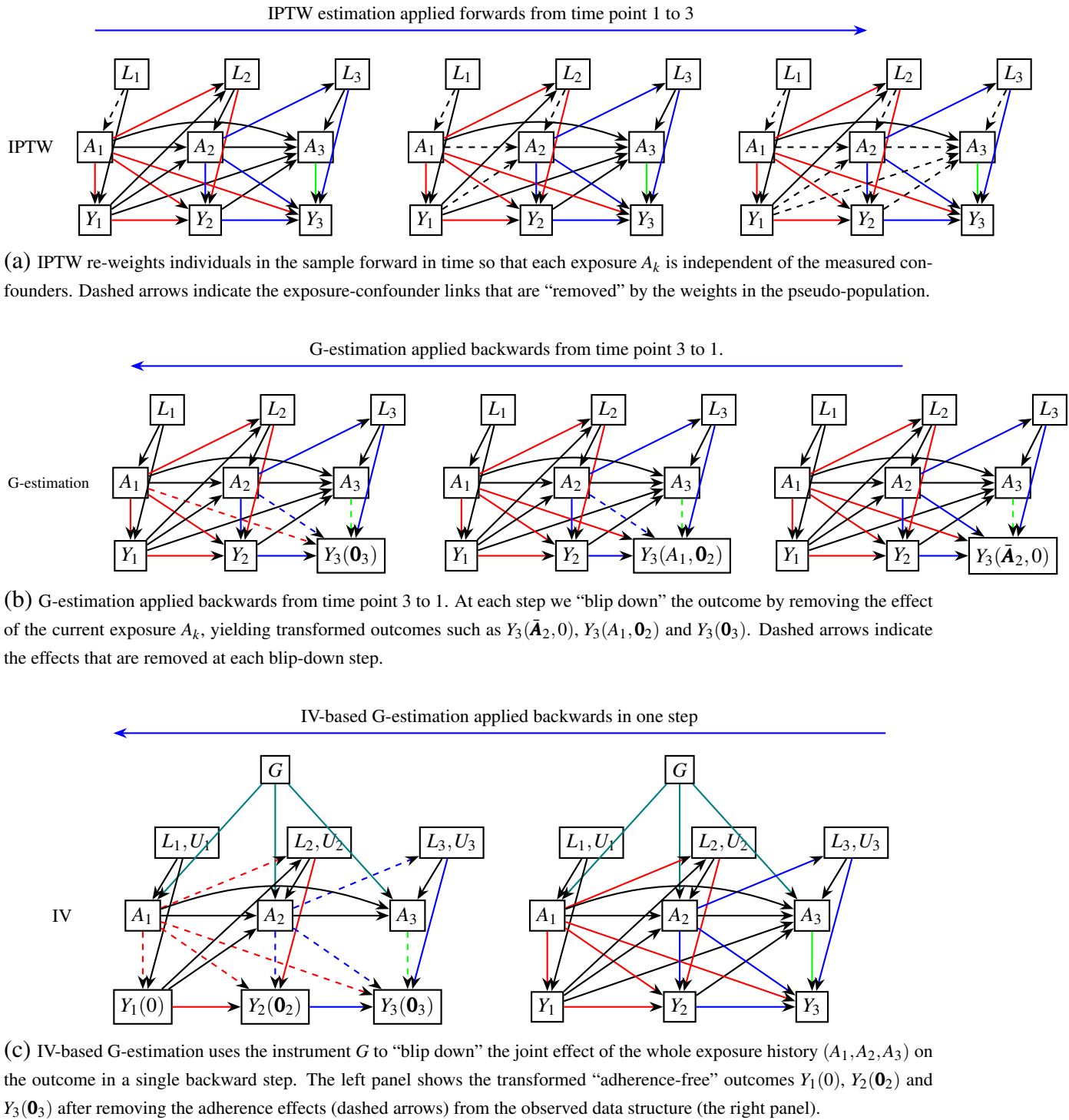
\begin{figure}[!htbp]
\centering

% =========================================================
% (a) IPTW subfigure
% =========================================================
\begin{subfigure}{\textwidth}
\centering
\begin{tikzpicture}[>=Stealth,scale=0.63, every node/.style={transform shape}]

% ======================
% IPTW row label
% ======================
\node[anchor=east] at (-2,2) {\Large IPTW};

% ======================
% IPTW Panel 1
% ======================
\begin{scope}
  \begin{scope}[every node/.style={rectangle,thick,draw}]
    \node (L1) at (0,4) {$L_1$};
    \node (L2) at (3,4) {$L_2$};
    \node (L3) at (6,4) {$L_3$};
    \node (A1) at (-1,2) {$A_1$};
    \node (A2) at (2,2) {$A_2$};
    \node (A3) at (5,2) {$A_3$};
    \node (Y1) at (-1,0) {$Y_1$};
    \node (Y2) at (2,0) {$Y_2$};
    \node (Y3) at (5,0) {$Y_3$};
  \end{scope}
  \begin{scope}[>={Stealth[black]}, every node/.style={fill=white,circle}]
    \draw[->,blue,thick]  (Y2) -- (Y3);
    \draw[->,black,thick] (Y2) -- (A3);
    \draw[->,black,thick] (Y1) -- (A3);

    % L's (L1->A1 removed)
    \draw[->,black,thick] (L1) -- (Y1);
    \draw[->,black,thick] (L2) -- (A2);
    \draw[->,red,thick]   (L2) -- (Y2);
    \draw[->,black,thick] (L3) -- (A3);
    \draw[->,blue,thick]  (L3) -- (Y3);
    \draw[->,black,thick,dashed] (L1) -- (A1);

    % A's and Y's
    \draw[->,red,thick]   (A1) -- (Y1);
    \draw[->,red,thick]   (A1) -- (Y2);
    \draw[->,red,thick]   (A1) -- (Y3);
    \draw[->,black,thick] (A1) -- (A2);
    \draw[->,black,thick] (Y1) -- (A2);
    \draw[->,red,thick]   (Y1) -- (Y2);

    \draw[->,blue,thick]  (A2) -- (Y2);
    \draw[->,blue,thick]  (A2) -- (Y3);
    \draw[->,black,thick] (A2) -- (A3);

    \draw[->,green,thick] (A3) -- (Y3);

    \draw[->,red,thick]   (A1) -- (L2);
    \draw[->,black,thick] (Y1) -- (L2);
    \draw[->,blue,thick]  (A2) -- (L3);
     \draw[->,black,thick,-{Stealth}] (A1) to[out=20,in=160] (A3);
  \end{scope}
\end{scope}

% ======================
% IPTW Panel 2
% ======================
\begin{scope}[xshift=9cm]
  \begin{scope}[every node/.style={rectangle,thick,draw}]
    \node (L1) at (0,4) {$L_1$};
    \node (L2) at (3,4) {$L_2$};
    \node (L3) at (6,4) {$L_3$};
    \node (A1) at (-1,2) {$A_1$};
    \node (A2) at (2,2) {$A_2$};
    \node (A3) at (5,2) {$A_3$};
    \node (Y1) at (-1,0) {$Y_1$};
    \node (Y2) at (2,0) {$Y_2$};
    \node (Y3) at (5,0) {$Y_3$};
  \end{scope}
  \begin{scope}[>={Stealth[black]}, every node/.style={fill=white,circle}]
    \draw[->,blue,thick]  (Y2) -- (Y3);
    \draw[->,black,thick] (Y2) -- (A3);
    \draw[->,black,thick] (Y1) -- (A3);

    % L1->A1 removed, L2->A2 removed
    \draw[->,black,thick] (L1) -- (Y1);
    \draw[->,red,thick]   (L2) -- (Y2);
    \draw[->,black,thick] (L3) -- (A3);
    \draw[->,blue,thick]  (L3) -- (Y3);
    \draw[->,black,thick,dashed] (L1) -- (A1);
    \draw[->,black,thick,dashed] (L2) -- (A2);

    \draw[->,red,thick]   (A1) -- (Y1);
    \draw[->,red,thick]   (A1) -- (Y2);
    \draw[->,red,thick]   (A1) -- (Y3);
    % A1->A2, Y1->A2 removed
    \draw[->,red,thick]   (Y1) -- (Y2);
    \draw[->,black,thick,dashed] (A1) -- (A2);
    \draw[->,black,thick,dashed] (Y1) -- (A2);

    \draw[->,blue,thick]  (A2) -- (Y2);
    \draw[->,blue,thick]  (A2) -- (Y3);
    \draw[->,black,thick] (A2) -- (A3);

    \draw[->,green,thick] (A3) -- (Y3);

    \draw[->,red,thick]   (A1) -- (L2);
    \draw[->,black,thick] (Y1) -- (L2);
    \draw[->,blue,thick]  (A2) -- (L3);
    \draw[->,black,thick,-{Stealth}] (A1) to[out=20,in=160] (A3);
  \end{scope}
\end{scope}

% ======================
% IPTW Panel 3
% ======================
\begin{scope}[xshift=18cm]
  \begin{scope}[every node/.style={rectangle,thick,draw}]
    \node (L1) at (0,4) {$L_1$};
    \node (L2) at (3,4) {$L_2$};
    \node (L3) at (6,4) {$L_3$};
    \node (A1) at (-1,2) {$A_1$};
    \node (A2) at (2,2) {$A_2$};
    \node (A3) at (5,2) {$A_3$};
    \node (Y1) at (-1,0) {$Y_1$};
    \node (Y2) at (2,0) {$Y_2$};
    \node (Y3) at (5,0) {$Y_3$};
  \end{scope}
  \begin{scope}[>={Stealth[black]}, every node/.style={fill=white,circle}]
    \draw[->,blue,thick]  (Y2) -- (Y3);
    % Y2->A3, Y1->A3 removed

    % L1->A1, L2->A2, L3->A3 removed
    \draw[->,black,thick] (L1) -- (Y1);
    \draw[->,red,thick]   (L2) -- (Y2);
    \draw[->,blue,thick]  (L3) -- (Y3);

    \draw[->,red,thick]   (A1) -- (Y1);
    \draw[->,red,thick]   (A1) -- (Y2);
    \draw[->,red,thick]   (A1) -- (Y3);
    % A1->A2, Y1->A2 removed
    \draw[->,red,thick]   (Y1) -- (Y2);

    \draw[->,blue,thick]  (A2) -- (Y2);
    \draw[->,blue,thick]  (A2) -- (Y3);
    % A2->A3 removed

    \draw[->,green,thick] (A3) -- (Y3);

    \draw[->,red,thick]   (A1) -- (L2);
    \draw[->,black,thick] (Y1) -- (L2);
    \draw[->,blue,thick]  (A2) -- (L3);

    \draw[->,black,thick,dashed] (L1) -- (A1);
    \draw[->,black,thick,dashed] (L2) -- (A2);
    \draw[->,black,thick,dashed] (A1) -- (A2);
    \draw[->,black,thick,dashed] (Y1) -- (A2);
    \draw[->,black,thick,dashed] (L3) -- (A3);
    \draw[->,black,thick,dashed] (A2) -- (A3);
    \draw[->,black,thick,dashed] (Y2) -- (A3);
    \draw[->,black,thick,dashed] (Y1) -- (A3);
     \draw[->,black,thick,-{Stealth}] (A1) to[out=20,in=160] (A3);
  \end{scope}
\end{scope}

% Big arrow above IPTW row
\draw[->,thick,blue]
  (-1,5.2) -- (20,5.2)
  node[midway,above=2pt,black, font=\Large]
  {IPTW estimation applied forwards from time point $1$ to $3$};

\end{tikzpicture}
\caption{{\small IPTW re-weights individuals in the sample forward in time so that each exposure $A_k$ is independent of the measured confounders. Dashed arrows indicate the exposure-confounder links that are ``removed'' by the weights in the pseudo-population.}}
\label{fig:iptw-row}
\end{subfigure}

\vspace{1.5em}

% =========================================================
% (b) SNMM subfigure
% =========================================================
\begin{subfigure}{\textwidth}
\centering
\begin{tikzpicture}[>=Stealth,scale=0.63, every node/.style={transform shape}]

\node[anchor=east] at (-2,2) {\large G-estimation};

% ----------------------
% SNMM Panel 1 (left)
% ----------------------
\begin{scope}
  \begin{scope}[every node/.style={rectangle,thick,draw}]
    \node (L1) at (0,4) {$L_1$};
    \node (L2) at (3,4) {$L_2$};
    \node (L3) at (6,4) {$L_3$};
    \node (A1) at (-1,2) {$A_1$};
    \node (A2) at (2,2) {$A_2$};
    \node (A3) at (5,2) {$A_3$};
    \node (Y1) at (-1,0) {$Y_1$};
    \node (Y2) at (2,0) {$Y_2$};
    \node (Y3) at (5,0) {$Y_3(\bm{0}_3)$};
  \end{scope}

  \begin{scope}[>={Stealth[black]}, every node/.style={fill=white,circle}]
    \draw[->,blue,thick]  (Y2) -- (Y3);
    \draw[->,black,thick] (Y2) -- (A3);
    \draw[->,black,thick] (Y1) -- (A3);

    \draw[->,black,thick] (L1) -- (A1);
    \draw[->,black,thick] (L1) -- (Y1);
    \draw[->,black,thick] (L2) -- (A2);
    \draw[->,red,thick]   (L2) -- (Y2);
    \draw[->,black,thick] (L3) -- (A3);
    \draw[->,blue,thick]  (L3) -- (Y3);

    \draw[->,red,thick]   (A1) -- (Y1);
    \draw[->,red,thick]   (A1) -- (Y2);
    % blipped A1 effect on Y3
    \draw[->,red,thick,dashed]   (A1) -- (Y3);

    \draw[->,black,thick] (A1) -- (A2);
    \draw[->,black,thick] (Y1) -- (A2);
    \draw[->,red,thick]   (Y1) -- (Y2);

    \draw[->,blue,thick]        (A2) -- (Y2);
    % blipped A2 effect on Y3
    \draw[->,blue,thick,dashed] (A2) -- (Y3);
    \draw[->,black,thick]       (A2) -- (A3);

    % blipped A3 effect on Y3
    \draw[->,green,thick,dashed](A3) -- (Y3);

    \draw[->,red,thick]   (A1) -- (L2);
    \draw[->,black,thick] (Y1) -- (L2);
    \draw[->,blue,thick]  (A2) -- (L3);
     \draw[->,black,thick,-{Stealth}] (A1) to[out=20,in=160] (A3);
  \end{scope}
\end{scope}

% ----------------------
% SNMM Panel 2 (middle)
% ----------------------
\begin{scope}[xshift=9cm]
  \begin{scope}[every node/.style={rectangle,thick,draw}]
    \node (L1) at (0,4) {$L_1$};
    \node (L2) at (3,4) {$L_2$};
    \node (L3) at (6,4) {$L_3$};
    \node (A1) at (-1,2) {$A_1$};
    \node (A2) at (2,2) {$A_2$};
    \node (A3) at (5,2) {$A_3$};
    \node (Y1) at (-1,0) {$Y_1$};
    \node (Y2) at (2,0) {$Y_2$};
    \node (Y3) at (5,0) {$Y_3(A_1, \bm{0}_2)$};
  \end{scope}

  \begin{scope}[>={Stealth[black]}, every node/.style={fill=white,circle}]
    \draw[->,blue,thick]  (Y2) -- (Y3);
    \draw[->,black,thick] (Y2) -- (A3);
    \draw[->,black,thick] (Y1) -- (A3);

    \draw[->,black,thick] (L1) -- (A1);
    \draw[->,black,thick] (L1) -- (Y1);
    \draw[->,black,thick] (L2) -- (A2);
    \draw[->,red,thick]   (L2) -- (Y2);
    \draw[->,black,thick] (L3) -- (A3);
    \draw[->,blue,thick]  (L3) -- (Y3);

    \draw[->,red,thick]   (A1) -- (Y1);
    \draw[->,red,thick]   (A1) -- (Y2);
    % A1->Y3 not yet blipped
    \draw[->,red,thick]   (A1) -- (Y3);

    \draw[->,black,thick] (A1) -- (A2);
    \draw[->,black,thick] (Y1) -- (A2);
    \draw[->,red,thick]   (Y1) -- (Y2);

    \draw[->,blue,thick]        (A2) -- (Y2);
    % blipped A2 effect on Y3
    \draw[->,blue,thick,dashed] (A2) -- (Y3);
    \draw[->,black,thick]       (A2) -- (A3);

    % blipped A3 effect on Y3
    \draw[->,green,thick,dashed](A3) -- (Y3);

    \draw[->,red,thick]   (A1) -- (L2);
    \draw[->,black,thick] (Y1) -- (L2);
    \draw[->,blue,thick]  (A2) -- (L3);
     \draw[->,black,thick,-{Stealth}] (A1) to[out=20,in=160] (A3);
  \end{scope}
\end{scope}

% ----------------------
% SNMM Panel 3 (right)
% ----------------------
\begin{scope}[xshift=18cm]
  \begin{scope}[every node/.style={rectangle,thick,draw}]
    \node (L1) at (0,4) {$L_1$};
    \node (L2) at (3,4) {$L_2$};
    \node (L3) at (6,4) {$L_3$};
    \node (A1) at (-1,2) {$A_1$};
    \node (A2) at (2,2) {$A_2$};
    \node (A3) at (5,2) {$A_3$};
    \node (Y1) at (-1,0) {$Y_1$};
    \node (Y2) at (2,0) {$Y_2$};
    \node (Y3) at (5,0) {$Y_3(\bar{\bm{A}}_2, 0)$};
  \end{scope}

  \begin{scope}[>={Stealth[black]}, every node/.style={fill=white,circle}]
    \draw[->,blue,thick]  (Y2) -- (Y3);
    \draw[->,black,thick] (Y2) -- (A3);
    \draw[->,black,thick] (Y1) -- (A3);

    \draw[->,black,thick] (L1) -- (A1);
    \draw[->,black,thick] (L1) -- (Y1);
    \draw[->,black,thick] (L2) -- (A2);
    \draw[->,red,thick]   (L2) -- (Y2);
    \draw[->,black,thick] (L3) -- (A3);
    \draw[->,blue,thick]  (L3) -- (Y3);

    \draw[->,red,thick]   (A1) -- (Y1);
    \draw[->,red,thick]   (A1) -- (Y2);
    \draw[->,red,thick]   (A1) -- (Y3);

    \draw[->,black,thick] (A1) -- (A2);
    \draw[->,black,thick] (Y1) -- (A2);
    \draw[->,red,thick]   (Y1) -- (Y2);

    \draw[->,blue,thick]  (A2) -- (Y2);
    \draw[->,blue,thick]  (A2) -- (Y3);
    \draw[->,black,thick] (A2) -- (A3);

    % blipped A3 effect on Y3
    \draw[->,green,thick,dashed](A3) -- (Y3);

    \draw[->,red,thick]   (A1) -- (L2);
    \draw[->,black,thick] (Y1) -- (L2);
    \draw[->,blue,thick]  (A2) -- (L3);
     \draw[->,black,thick,-{Stealth}] (A1) to[out=20,in=160] (A3);
  \end{scope}
\end{scope}

% Arrow above SNMM row
\draw[->,thick,blue]
  (20,5.2) -- (-1,5.2)
  node[midway,above=2pt,black, font=\Large]
  {G-estimation applied backwards from time point $3$ to $1$.};

\end{tikzpicture}
\caption{{\small G-estimation applied backwards from time point $3$ to $1$. At each step we ``blip down'' the outcome by removing the effect of the current exposure $A_k$, yielding transformed outcomes such as $Y_3(\bar{\bm{A}}_2, 0)$, $Y_3(A_1, \bm{0}_2)$ and $Y_3(\bm{0}_3)$. Dashed arrows indicate the effects that are removed at each blip-down step.}}
\label{fig:snmm-row}
\end{subfigure}

\vspace{1.5em}

% =========================================================
% (c) IV subfigure
% =========================================================
\begin{subfigure}{\textwidth}
\centering
\begin{tikzpicture}[>=Stealth,scale=0.63, every node/.style={transform shape}]

\node[anchor=east] at (-2,2) {\Large IV};

% IV panels (two-panel version, with G above A/Y)
\begin{scope}[xshift=1cm]

% Left IV panel
\begin{scope}
  \begin{scope}[every node/.style={rectangle,thick,draw}]
    % Instrument
    \node (G1)     at (2.4,6) {$G$};

    % L, A, Y layers (wider spacing)
    \node (L1iv1)  at (0,4)   {$L_1,U_1$};
    \node (L2iv1)  at (3.6,4) {$L_2,U_2$};
    \node (L3iv1)  at (7.2,4) {$L_3,U_3$};

    \node (A1iv1)  at (-1.2,2) {$A_1$};
    \node (A2iv1)  at (2.4,2)  {$A_2$};
    \node (A3iv1)  at (6.0,2)  {$A_3$};

    \node (Y1iv1)  at (-1.2,-0.5) {$Y_1(0)$};
    \node (Y2iv1)  at (2.4,-0.5)  {$Y_2(\bm{0}_2)$};
    \node (Y3iv1)  at (6.0,-0.5)  {$Y_3(\bm{0}_3)$};
  \end{scope}

  \begin{scope}[>={Stealth[black]}, every node/.style={fill=white,circle}]
    % G arrows (teal)
    %\draw[->,teal,thick] (G1) -- (L1iv1);
    %\draw[->,teal,thick] (G1) -- (L2iv1);
    %\draw[->,teal,thick] (G1) -- (L3iv1);
    \draw[->,teal,thick] (G1) -- (A1iv1);
    \draw[->,teal,thick] (G1) -- (A2iv1);
    \draw[->,teal,thick] (G1) -- (A3iv1);

    % Usual longitudinal structure
    \draw[->,blue,thick]  (Y2iv1) -- (Y3iv1);

    \draw[->,black,thick] (L1iv1) -- (A1iv1);
    \draw[->,black,thick] (L1iv1) -- (Y1iv1);
    \draw[->,black,thick] (L2iv1) -- (A2iv1);
    \draw[->,red,thick]   (L2iv1) -- (Y2iv1);
    \draw[->,black,thick] (L3iv1) -- (A3iv1);
    \draw[->,blue,thick]  (L3iv1) -- (Y3iv1);

    \draw[->,red,thick,dashed]   (A1iv1) -- (Y1iv1);
    \draw[->,red,thick,dashed]   (A1iv1) -- (Y2iv1);
    \draw[->,red,thick,dashed]   (A1iv1) -- (Y3iv1);

    \draw[->,black,thick] (A1iv1) -- (A2iv1);
    \draw[->,black,thick] (Y1iv1) -- (A2iv1);
    \draw[->,red,thick]   (Y1iv1) -- (Y2iv1);

    \draw[->,blue,thick,dashed]  (A2iv1) -- (Y2iv1);
    \draw[->,blue,thick,dashed]  (A2iv1) -- (Y3iv1);
    \draw[->,black,thick]        (A2iv1) -- (A3iv1);

    \draw[->,green,thick,dashed] (A3iv1) -- (Y3iv1);

    \draw[->,red,thick,dashed]   (A1iv1) -- (L2iv1);
    \draw[->,black,thick]        (Y1iv1) -- (L2iv1);
    \draw[->,blue,thick,dashed]  (A2iv1) -- (L3iv1);
    \draw[->,black,thick,-{Stealth}] (A1iv1) to[out=20,in=160] (A3iv1);
  \end{scope}
\end{scope}

% Right IV panel
\begin{scope}[xshift=11cm]
  \begin{scope}[every node/.style={rectangle,thick,draw}]
    \node (G2)     at (2.4,6) {$G$};

    \node (L1iv2)  at (0,4)   {$L_1,U_1$};
    \node (L2iv2)  at (3.6,4) {$L_2,U_2$};
    \node (L3iv2)  at (7.2,4) {$L_3,U_3$};

    \node (A1iv2)  at (-1.2,2) {$A_1$};
    \node (A2iv2)  at (2.4,2)  {$A_2$};
    \node (A3iv2)  at (6.0,2)  {$A_3$};

    \node (Y1iv2)  at (-1.2,-0.5) {$Y_1$};
    \node (Y2iv2)  at (2.4,-0.5)  {$Y_2$};
    \node (Y3iv2)  at (6.0,-0.5)  {$Y_3$};
  \end{scope}

  \begin{scope}[>={Stealth[black]}, every node/.style={fill=white,circle}]
    %\draw[->,teal,thick] (G2) -- (L1iv2);
    %\draw[->,teal,thick] (G2) -- (L2iv2);
    %\draw[->,teal,thick] (G2) -- (L3iv2);
    \draw[->,teal,thick] (G2) -- (A1iv2);
    \draw[->,teal,thick] (G2) -- (A2iv2);
    \draw[->,teal,thick] (G2) -- (A3iv2);

    \draw[->,blue,thick]  (Y2iv2) -- (Y3iv2);
    \draw[->,black,thick] (Y2iv2) -- (A3iv2);
    \draw[->,black,thick] (Y1iv2) -- (A3iv2);

    \draw[->,black,thick] (L1iv2) -- (A1iv2);
    \draw[->,black,thick] (L1iv2) -- (Y1iv2);
    \draw[->,black,thick] (L2iv2) -- (A2iv2);
    \draw[->,red,thick]   (L2iv2) -- (Y2iv2);
    \draw[->,black,thick] (L3iv2) -- (A3iv2);
    \draw[->,blue,thick]  (L3iv2) -- (Y3iv2);

    \draw[->,red,thick]   (A1iv2) -- (Y1iv2);
    \draw[->,red,thick]   (A1iv2) -- (Y2iv2);
    \draw[->,red,thick]   (A1iv2) -- (Y3iv2);

    \draw[->,black,thick] (A1iv2) -- (A2iv2);
    \draw[->,black,thick] (Y1iv2) -- (A2iv2);
    \draw[->,red,thick]   (Y1iv2) -- (Y2iv2);

    \draw[->,blue,thick]  (A2iv2) -- (Y2iv2);
    \draw[->,blue,thick]  (A2iv2) -- (Y3iv2);
    \draw[->,black,thick] (A2iv2) -- (A3iv2);

    \draw[->,green,thick] (A3iv2) -- (Y3iv2);

    \draw[->,red,thick]   (A1iv2) -- (L2iv2);
    \draw[->,black,thick] (Y1iv2) -- (L2iv2);
    \draw[->,blue,thick]  (A2iv2) -- (L3iv2);
    \draw[->,black,thick,-{Stealth}] (A1iv2) to[out=20,in=160] (A3iv2);
    
  \end{scope}
\end{scope}

\end{scope} % end overall IV scope

% Big arrow ABOVE the IV row
\draw[->,thick,blue]
  (20,7.0) -- (-1,7.0)
  node[midway,above=2pt,black, font=\Large]
  {IV-based G-estimation applied backwards in one step};

\end{tikzpicture}
\caption{{\small IV-based G-estimation uses the instrument $G$ to ``blip down'' the joint effect of the whole exposure history $\left(A_1, A_2, A_3\right)$ on the outcome in a single backward step. The left panel shows the transformed ``adherence-free'' outcomes $Y_1(0)$, $Y_2(\bm{0}_2)$ and $Y_3(\bm{0}_3)$ after removing the adherence effects (dashed arrows) from the observed data structure (the right panel).}}
\label{fig:iv-row}
\end{subfigure}

\caption{Conceptual illustration of (a) IPTW, (b) SNMM G-estimation, and (c) IV-based G-estimation for the effect of adherence history $(A_1,A_2,A_3)$ on the selected final outcome $Y_3$. For visual clarity, we omit some arrows (such as the direct effect of $A_1$ on $\bm{L}_{3}$), but the framework allows for general relationships where variables may depend on the entire past history.}
\label{fig:all-methods-subfig}
\end{figure}

Now that we have established that both the MSM and the SNMM provide parametric representations of the same causal estimand, we turn to its identification and estimation from the observed data under both models. 
We invoke three standard identification conditions: \textit{consistency}, \textit{positivity}, and \textit{conditional exchangeability} \citep{rosenbaum1983central,fitzmaurice2008longitudinal}.\\
\\
\textbf{Assumption 1: Sequential Exchangeability} \citep{robins1997causal,HernanRobins2020}. This assumption can be viewed as the longitudinal formulation of NUC \citep{fitzmaurice2008longitudinal,daniel2013methods,HernanRobins2020}. 
It states that at each time point $j$, once we condition on the observed history, no unmeasured confounders remain that influence both the adherence decision and the potential outcome:
\begin{equation}\label{eq:se}
    Y(\bar{\bm{a}}_k) \perp\!\!\!\perp A_j \mid \bm{H}_j = \left(\bm{\bar{A}}_{j-1}, \bm{\bar{Y}}_{j-1}, \bm{\bar{L}}_j\right)\quad \text{for all } \bar{\bm{a}}_k \text{ and } j \in \{1, \dots, k\} \text{.}
\end{equation}
Thus, within strata of $\bm{H}_{j}$, the adherence choice at time $j$ can be treated as if it were “as good as random” with respect to the potential outcomes.
While $\bm{H}_j$ formally includes the entire observed history, in practice, identification may only require conditioning on a sufficient subset of $\bm{H}_j$ that renders the adherence decision independent of the potential outcomes.\\
\\
\textbf{Assumption 2: Consistency}. This assumption links the potential outcomes to the observed data. It states that for any adherence history $\bar{\bm{a}}_k$, if a patient's observed adherence path matches $\bar{\bm{a}}_k$, then their observed outcome coincides with the corresponding potential outcome \citep{cole2009consistency, HernanRobins2020}:
\begin{equation}\label{eq:consistency}
    % Y_i = Y_i(\bm{\bar{a}}_k) \quad \text{if} \quad \bar {\bm{A}}_{i,k} = \bm{\bar{a}}_k.
    Y = Y(\bm{\bar{a}}_k) \quad \text{if} \quad \bar {\bm{A}}_{k} = \bm{\bar{a}}_k.
\end{equation}
\\
\textbf{Assumption 3: Positivity}. This assumption states that for each time $j$ and any history $\bm{H}_j$ that occurs in the population, the conditional density of adherence must be positive over the range of adherence levels relevant to the study \citep{westreich2010invited, cole2008constructing}. 
Formally, for any $\bm{h}_j = (\bar {\bm{a}}_{j-1}, \bar {\bm{y}}_{j-1}, \bar {\bm{\ell}}_j)$ with $f(\bar {\bm{A}}_{j-1}=\bar {\bm{a}}_{j-1}, \bar {\bm{Y}}_{j-1}=\bar {\bm{y}}_{j-1}, \bm{\bar{L}}_j = \bar {\bm{\ell}}_j) \neq 0$,
\begin{equation}\label{eq:positivity}
    % f\big(A_j = a_j \mid \bar {\bm{A}}_{j-1}=\bar {\bm{a}}_{j-1}, \bar {\bm{Y}}_{j-1}=\bar {\bm{y}}_{j-1}, \bm{\bar{L}}_j = \bar {\bm{\ell}}_j\big) > 0 
    f\big(A_j = a_j \mid \bm{H}_j = \bm{h}_j \big) > 0 
\end{equation}
for all $a_j$ in the support considered by the model. Intuitively, this assumption requires that within every observed history stratum we would actually see in the data that there is sufficient variation in adherence across the range of interest.
This is particularly important when estimating fully time-specific effects, where a distinct causal parameter is associated with each pair of $A_j$ and $Y_k$ ($j \leq k$).
Later we show under more parsimonious (unsaturated) model specifications, the positivity assumption can be less stringent in the sense that estimation may rely on sufficient adherence variation across the overall follow-up period, rather than requiring positivity to hold for every single time point.
In practice, practical violations or `near-violations' of the positivity assumption are particularly common with continuous exposures in finite samples \citep{schomaker2024causal, bao2025addressing, petersen2012diagnosing, westreich2010invited}. This issue may be exacerbated in the presence of strong exposure–covariate feedback \citep{robins2000marginal, vansteelandt2014structural}, where adherence levels become highly predictable based on patient history, thereby restricting the observed adherence distribution to narrow ranges. Such weak positivity can substantially destabilize estimation, particularly for weighting-based approaches such as IPTW, which may yield highly variable or extreme weights \citep{robins2000marginal, seaman2021using, petersen2012diagnosing, spreafico2025impact}. We illustrate this impact in our simulation study in Section~\ref{sec:main_simu}.\\
\\
Under the identification assumptions above, IPTW estimates $\bm{\eta}_k$ by modelling the conditional distribution of adherence at each time point to construct inverse probability weights. We then fit the MSM using weighted regression, effectively creating a pseudo-population where adherence is unconfounded \citep{robins2000marginal, HernanRobins2020}. 
To implement this, we model the conditional density of adherence $A_j$ given the observed history $\bm{H}_{j} = \left(\bm{\bar{A}}_{j-1}, \bm{\bar{Y}}_{j-1}, \bm{\bar{L}}_j\right)$. For a continuous exposure, the stabilized inverse probability weight $W_j$ for each individual at time $j$ is defined as:
\begin{equation}\label{eq:weight}
    % W_{j} = \prod_{j^{\prime} = 1}^j w_{j^{\prime}} = \prod_{j^{\prime} = 1}^j \frac{f(A_{j^{\prime}} \mid \bm{I}_{j^\prime})}{f(A_{j^{\prime}} \mid \bm{H}_{j^{\prime}})}
    W_{j} = \prod_{s = 1}^j w_{s} = \prod_{s = 1}^j \frac{f(A_{s} \mid \bm{I}_{s})}{f(A_{s} \mid \bm{H}_{s})}
\end{equation}
for $j \in \{1, \dots, k\}$, where $\bm{I}_{s}$ is a reduced conditioning set typically containing baseline covariates and/or past adherence history, or empty (in which case, $f(A_{s} \mid \bm{I}_{s})$ corresponds to the marginal density $f(A_{s})$). The denominator is the conditional density of adherence given the full observed history, while the numerator is included to stabilize the weights and reduce variance \citep{robins2000marginal, naimi2014constructing, fong2018covariate, cole2008constructing, daniel2013methods}.\\
\\
%In the post-weighting pseudo-population, $A_j$ is by construction independent of $\bm{H}_j$. 
Since the weights are constructed from the conditional adherence distribution given $\bm{H}_j$, they create a pseudo-population in which adherence is independent of the measured history \citep{robins2000marginal, HernanRobins2020}. Under sequential exchangeability (that $\bm{H}_j$ contains all confounders of adherence and the potential outcomes), adherence is unconfounded in this weighted pseudo-population (see Figure~\ref{fig:all-methods-subfig}, Top Panel (a) for illustration) and the MSM parameters can therefore be interpreted causally. 
We then estimate the MSM parameters $\bm{\eta}_k $ using weighted least squares regression:
    \begin{equation}\label{eq:IPTW_mini}
        \widehat{\bm{\eta}}_k = \arg\min_{\bm{\eta}_k} \sum_{i = 1}^n W_{i, k}\left(Y_{i, k} - \bm{\tilde{A}}^{\top}_{i, k}\bm{\eta}_k\right)^2  \quad \text{for  } \bm{\eta}_k = (\eta_0, \eta_1, ..., \eta_k)^{\top}
    \end{equation}
where $\bm{\tilde{A}}_{i, k} = (1, A_{i, 1} ..., A_{i, k})^{\top}$.\\
\\    
Conversely, G-estimation exploits the structure of the SNMM together with \textit{sequential exchangeability} to construct a “blipped-down” (or “adherence-free”) outcome, serving as a proxy for the relevant counterfactual outcome. The SNMM parameter $\bm{\beta}_k$ is then chosen so that, under the correctly specified model, the blipped-down outcome behaves like the relevant counterfactual in conditional expectation, and therefore should be mean independent of adherence conditional on the observed history \citep{robins1997causal, daniel2013methods}. 
For each time point $j \in \{1, \dots, k\}$, based on our specification of the blip function in Equation (\ref{eq:blip}), define the blipped-down outcome as
\begin{equation*}
    Y^{(j)}(\bm{\beta}_k) = Y - \sum_{s=j}^k \beta_s A_s,
\end{equation*}
which subtracts from the observed outcome $Y$ the cumulative effect of adherence from time $j$ onward implied by the SNMM with parameter $\bm{\beta}_k$. According to \citet{robins1994correcting} and \citet{robins1997causal}, under correct model specification and \textit{consistency}, and evaluated at the true SNMM parameter $\bm{\beta}^{\ast}_k$, we have the following conditional mean equivalence between the blipped-down outcome and the counterfactual outcome under the regime that sets adherence to zero from time $j$ onward:
\begin{equation}
\mathbb{E}[Y^{(j)}(\bm{\beta}^\ast_k)\mid \bm{H}_j, A_j]
=
\mathbb{E}[Y(\bar {\bm{A}}_{j-1},0,\bm{0}_{k-j})\mid \bm{H}_j, A_j] \text{.}
\label{eq:cond-mean-bridge}
\end{equation}
Furthermore, if \textit{sequential exchangeability} holds (together with \textit{positivity} and \textit{consistency}), it follows from (\ref{eq:cond-mean-bridge}) and standard SNMM results that the blipped-down outcome is mean independent of the current adherence conditional on the observed history \citep{robins1997causal}:
\begin{equation}
\mathbb{E}[Y^{(j)}(\bm{\beta}^\ast_k)\mid \bm{H}_j, A_j]
=
\mathbb{E}[Y^{(j)}(\bm{\beta}^\ast_k)\mid \bm{H}_j].
\label{eq:mean-ind}
\end{equation}
For details and proofs of Equations (\ref{eq:cond-mean-bridge}) and (\ref{eq:mean-ind}), see Section~8.3 of \citet{robins1997causal} or Sections~5.1 and 6.1 of \citet{vansteelandt2014structural}.\\
\\
Based on the mean-independence restriction in Equation~(12), G-estimation can be formulated in terms of unconditional moment restrictions. In particular, Equation~(\ref{eq:mean-ind}) implies that the blipped-down outcome is orthogonal to the residualized adherence, yielding the moment condition:
\begin{equation}\label{eq:moment-condition}
    \mathbb{E}\Big[ \{A_j - \mathbb{E}(A_j \mid \bm{H}_j)\}\, Y^{(j)}(\bm{\beta}^\ast_k) \Big] = 0,
\qquad j = 1,\ldots,k.
\end{equation}
Building on \eqref{eq:moment-condition}, statistical efficiency can be improved by additionally residualizing the blipped-down outcome with respect to $\bm{H}_j$, yielding the alternative moment condition
\begin{equation}
\mathbb{E}\left[\left\{A_j-\mathbb{E}(A_j\mid \bm{H}_j)\right\}
\left\{Y^{(j)}(\bm{\beta}_k^\ast)-\mathbb{E}\left(Y^{(j)}(\bm{\beta}_k^\ast)\mid \bm{H}_j\right)\right\}\right]=0,
\qquad j=1,\ldots,k.
\label{eq:moment-condition-eff}
\end{equation}
Estimation based on \eqref{eq:moment-condition-eff} attains local semiparametric efficiency under homoscedasticity \citep{vansteelandt2014structural}. A brief derivation of \eqref{eq:moment-condition} and justification of the efficiency property of \eqref{eq:moment-condition-eff} are provided in Appendix A.2.
The moment-condition formulation highlights the close connection between G-estimation and GMM framework, and allows estimation and inference to be carried out using standard GMM machinery including sandwich variance estimators \citep{hansen1982large,newey1994large}. 
In the traditional recursive implementation of G-estimation, estimation proceeds backwards in time: at each time point $j$, one updates the blipped-down outcome and solves the corresponding estimating equation for the current blip parameter (see Figure~\ref{fig:all-methods-subfig}, middle panel (b)). Alternatively, the moment conditions in Equation~(\ref{eq:moment-condition}) and (\ref{eq:moment-condition-eff}) can be stacked over $j = 1, ..., k$ and solved jointly within a GMM framework, yielding a single estimator for $\bm{\beta}_k$.
Section~\ref{sec:implementation} provides further implementation details.\\
\\
In summary, both IPTW and G-estimation rely on the identification assumptions of sequential exchangeability, consistency and positivity. In addition, IPTW depends on the correct specification of the MSM, whereas G-estimation requires correct specification of the SNMM blip function. 
The two approaches differ in how the adherence process enters estimation: IPTW requires modelling the conditional density of adherence given the observed history through the weights, while G-estimation only requires a model for the conditional mean of adherence given the history in order to construct the residualized adherence term \citep{vansteelandt2016revisiting}.
When these models are correctly specified, both approaches target the same causal estimand and yield consistent estimation, but they can differ in statistical efficiency and in finite-sample performance, as we will demonstrate in Section~\ref{sec:main_simu}.

\subsection{G-estimation via Instrumental Variables}\label{sec:method-iv}
When the NUC assumption is doubtful, an alternative is to augment the SNMM with an instrumental variable for adherence and use an IV version of G-estimation to identify the same hypothetical estimand. In clinical trials, randomized assignment to a treatment arm naturally plays this role, and has been used to relax NUC when estimating hypothetical estimands both for single-time outcomes \citep{bowden2021connecting} and, more recently, for longitudinal outcomes \citep{Bowden2025}. We will use the IV framework proposed in \citet{Bowden2025} as a basis for the longitudinal analysis in the observational data setting. In previous sections we suppressed the outcome time index $k$ and wrote $Y$ for a generic chosen horizon, since the standard SNMM and G-estimation arguments apply identically for any fixed $k$. In the IV extension, however, identification relies on moment conditions indexed by the outcome time point. We therefore reintroduce the index $k$ and write $Y_k$ for $k=1,\ldots,K$ throughout this section. Correspondingly, we index the SNMM parameters by the outcome time point and write $\beta_k(j)$ for $j\le k$ as the causal parameter of $A_j$ on $Y_k$.
\\
\\
Without randomized assignment in observational studies, we must identify an IV for adherence that satisfies the three key IV assumptions \citep{angrist1996identification,HernanRobins2020, Bowden2025}. It should:
\begin{itemize}
    \item Be sufficiently correlated with $\bar{\bm{A}}_k$.
    \item Be independent of unmeasured confounders of the $A_j$--$Y_k$ relationship for any $j \leq k$, conditional on baseline covariates.
    % \item Not be influenced by any variables that confound the relationship between $A_j$ and $Y_k$ for any $j \leq k$.
    \item Only influence $Y_k$ through $\bar{\bm{A}}_k$.
\end{itemize}
To mimic the applied analysis setting of this paper, let $G$ be a germ line genetic variant or score that is a valid IV for adherence, which naturally is time-fixed and determined at baseline. It does not predict treatment initiation, but does influence the amount of drug taken, for example by affecting its bio-availability \citep{Bowden2021TWIST}. Using the notation of \citet{tompsett2025instrumental}, the latter two IV assumptions can be summarized as 
\begin{equation*}\label{eq:IV-assumption}
    Y_k (\bm{\bar{a}}_k) \perp\!\!\!\perp G \mid \bm{\tilde{L}}_{0},S=1 \quad \text{for all } \bm{\bar{a}}_k,
\end{equation*}
where $\bm{\tilde{L}}_{0}$ is a sufficient subset of the baseline covariates to ensure the validity of $G$. We will assume implicit conditioning on $\bm{\tilde{L}}_{0}$ throughout and therefore omit it from the conditioning set in what follows. 
The condition $S = 1$ ensures that $G$ is a valid instrument for the treated-only population and it will also be implicitly assumed, with further discussion provided in Section~\ref{sec:application-iv}.
Following the practice in, e.g., Matsouaka and Tchetgen Tchetgen (2017), Shi et al. (2021), and Tompsett et al. (2025), the SNMM can be extended to incorporate the instrument $G$ by defining the blip at time $j$ as
\begin{equation*}
    \mathbb{E}[Y_k(\bar {\bm{a}}_{j-1}, a_j,\bm{0}_{k-j})-Y_k(\bar {\bm{a}}_{j-1}, 0, \bm{0}_{k-j})\mid \bm{H}_j, A_j=a_j, G] = \psi_k(j)(a_j,\bm{H}_j, G).
\end{equation*}
Additionally, we assume no effect modification of the blip by the instrument, so that the blip function retains the form $\psi_k(j)(a_j,\bm{H}_j, G)=\beta_k(j) a_j$. Under this specification, all of the arguments in Section~\ref{sec:MSM-and-SNMM} continue to hold and we still obtain
\begin{equation}\label{eq:snmm-iv}
    \mathbb{E}[Y_k(\bar {\bm{a}}_k) - Y_k(\bm{0}_k)] = \sum_{j=1}^k \beta_k(j) a_j,
\end{equation}
with $\bm{\beta}_k=(\beta_k(1),\ldots,\beta_k(k))$ having the same causal interpretation as before.\\
\\
Without \textit{sequential exchangeability}, but assuming a valid IV $G$, we now show that $\bm{\beta}_k$ can be identified as follows. Maintaining the \textit{consistency} assumption, it follows from Equation~(\ref{eq:snmm-iv}) that for the observed adherence path  $\bar {\bm{A}}_k$:
\begin{equation*}
    \mathbb{E}[Y_k - Y_k(\bm{0}_k) \mid \bar {\bm{A}}_k] = \sum_{j=1}^k \beta_k(j) A_j.
\end{equation*}
We can therefore write
\begin{equation*}
    Y_k - Y_k(\bm{0}_k) = \sum_{j=1}^k \beta_k(j) A_j + e_k
\end{equation*}
where $e_k$ is a random error and $\mathbb{E}[e_k \mid \bar {\bm{A}}_k] = 0$ by definition. By the IV validity assumption that $Y_k(\bar {\bm{a}}_k) \perp\!\!\!\perp G$ for all $\bar {\bm{a}}_k$, in particular $Y_k(\bm{0}_k) \perp\!\!\!\perp G$, it follows that the centred instrument $G - \mathbb{E}(G)$ satisfies the moment condition
\begin{equation}\label{eq:iv-moment}
    \mathbb{E}\Big[ (G - \mathbb{E}(G)) \Big\{ Y_{k} - \sum_{j=1}^k \beta_k(j) A_j \Big\} \Big] = 0 \quad \text{for} \quad k=1,...,K.
\end{equation}
This provides the basis for the G-estimation formula based on the following score equation $S(\bm{\theta})$ proposed in \citet{Bowden2025}, where $\bm{\theta}$ collects all parameters $\beta_k(j)$ $(j\le k,\; k=1,\ldots,K)$:
\begin{eqnarray}
S(\bm{\theta}) &=& \sum^{n}_{i=1}(G_{i}-\bar{G})\boldsymbol{\Sigma^{-1}}\begin{pmatrix}
Y_{i,1}&-&\beta_{1}(1)A_{i,1}\\
Y_{i,2}&-&\left\{\beta_{2}(1)A_{i,1}+\beta_{2}(2)A_{i,2}\right\}\\
&&.\\
&&.\\
&&.\\
Y_{i,K}&-&\sum^{K}_{j=1}\beta_{K}(j)A_{i,j} 
\end{pmatrix},
\label{eq:G-estimation}
\end{eqnarray}
where $K$ is the total number of time points under study; $\bar{G}$ is the mean value of $G_i$ across all $n$ individuals and $\bm{\Sigma}$ is an arbitrary $K \times K$ positive definite matrix (e.g., the $K$-dimensional identity matrix). The parameters $\beta_k(j)$ ($j \leq k$) can then be estimated via minimization of $S^{T}(\bm{\theta})S(\bm{\theta})$. Here, G-estimation is again applied backward in time but is always in a single joint step (see Figure~\ref{fig:all-methods-subfig}, bottom panel (c)). \\
\\
A key distinction between NUC-based G-estimation in (\ref{eq:moment-condition}) and IV-based G-estimation in (\ref{eq:iv-moment}) is that (\ref{eq:iv-moment}) is under-identified in general for $K>1$, since the number of parameters $K(K+1)/2$ exceeds the number of moment conditions $K$. Intuitively, in the standard SNMM setting (\ref{eq:moment-condition}) the residualized adherence terms $\{A_j-\mathbb{E}(A_j\mid \bm{H}_j)\}$ provide a distinct set of instruments at each time point, whereas in the IV setting a single time-fixed instrument $G$ yields only one moment restriction per outcome time. As a result, IV-based G-estimation requires stacking moment conditions across $k=1,\ldots,K$, but still provides fewer moment restrictions than unknown parameters when $K>1$. Therefore, we need to impose further parametric assumptions on the $\beta_k(j)$ structure to achieve identification. For example, the two-parameter model proposed in \citet{Bowden2025}, which assumes: $\beta_k(j) = \beta \alpha^{t_k - t_j}$, where $t_k - t_j$ measures the lag between time point $k$ and $j$. 
Hypothetical estimands can therefore be quantified after estimation of $\bm{\theta} = \left(\beta, \alpha \right)$. 
In the time-saturated SNMM, each $\beta_k(j)$ is a distinct parameter, so identification requires sufficient variation in adherence at every time point across histories appearing in the model (i.e. positivity). By contrast, the decay model $\beta_k(j)=\beta\alpha^{t_k-t_j}$ specifies a smooth trajectory of effects over time using only two parameters. This non-saturated parameterization allows information to be “shared” across time points, so that lack of variation in adherence at a particular time is less problematic. We show in Section 2.5 that this same smoothing trick can also be applied to IPTW and NUC-based G-estimation, if needed.

\subsection{Implementation of the Methods}\label{sec:implementation}
For IPTW, several methods have been proposed in the literature for estimating the inverse probability weights defined in (\ref{eq:weight}). Here we focus on covariate balancing propensity score (CBPS) methods, which are available for continuous exposures \citep{fong2018covariate}. 
In the parametric CBPS for a continuous treatment, the conditional distribution of $A_j$ given covariates is modelled as approximately normal with mean specified by a linear predictor in the covariates,
%(in practice including transformations such as squared terms and, if needed, a Box–Cox transformation of $A_j$), 
and the parameters are chosen so that sample covariate–treatment associations are minimized subject to the implied moment conditions for covariate balance \citep{fong2018covariate}.  
Nonparametric CBPS extensions are also available for continuous exposures, but we focus on the parametric formulation in this paper.\\
\\
In the longitudinal setting, CBPS can be applied at each time point to estimate the conditional densities $f(A_{s}\mid \bm{H}_{s})$ and hence the stabilized weights $w_{s}$, which are then multiplied to obtain the cumulative weights $W_{j}$ in (\ref{eq:weight}).  
Following the standard continuous-treatment CBPS formulation, we take $I_s=\emptyset$ so that the numerator is the marginal density $f(A_s)$.
Alternative methods estimate the weights jointly across all time points, for example the longitudinal CBPS for marginal structural models of \citet{imai2015robust} or the Stable Estimand Adaptive IPTW of \citet{avagyan2021stable}. However, these developments are primarily formulated for binary or categorical treatments and are therefore less directly applicable to our continuous adherence measure.  
As noted by \citet{imai2015robust}, fitting exposure models separately at each time point may entail some loss of efficiency, but still yields accurate estimators when the exposure models are correctly specified.\\
\\
\citet{austin2019assessing} suggests to assess covariate balance by examining the absolute weighted correlation between the exposure and each covariate, taking values below 0.1 to indicate adequate balance. Any covariates with an absolute weighted correlation exceeding 0.1 are then additionally included as adjustment variables in the second-stage weighted regression. 
To assess the stability of the IPTW weights and potential practical violations of the positivity assumption, we examined several summary diagnostics of the cumulative weights in our simulation studies and real-world application. 
First, we computed an effective sample size (ESS) measure, which provides an intuitive summary of how much information remains after weighting. 
Second, we calculated the proportion of total weight carried by the largest 1\% of observations, which reflects whether a small subset of individuals dominates the weighted pseudo-population. 
Finally, we examined the upper tail of the weight distribution using a high percentile (e.g., the 99.9th percentile) to assess the extent of extreme weights \citep{shook2022power, cole2008constructing, lee2011weight, austin2015moving}.
Full definitions and additional details are provided in Appendix B.2.\\
\\
As we demonstrate in simulation studies, the standard errors obtained from the second-stage weighted regression tend to underestimate the true variability of the IPTW estimates, as they do not account for the uncertainty introduced by the weighting process. Following the recommendation of \citet{daniel2013methods} and \citet{blackwell2018make}, one option to estimate the standard errors is using a block bootstrap procedure, in which individuals are resampled as blocks, preserving all of their associated time points. A more computationally efficient alternative is to compute sandwich standard errors from the weighted regression, see Appendix A.5 for details. \\
\\
G-estimation can be implemented by the classic sequential recursive procedure which solves the moment conditions in (\ref{eq:moment-condition}) backwards over time (see the summary of the procedure in Appendix A.3). 
Alternatively, we can estimate all components of $\bm{\beta}_k = (\beta_1,\ldots,\beta_k)$ jointly. For each $j$ we form the sample analogue of the moment condition in (\ref{eq:moment-condition}) as:
\begin{equation*}
    \hat m_j(\bm{\beta}_k)
    = \frac{1}{n}\sum_{i=1}^n
      \{A_{i,j} - \widehat{\mathbb{E}(A_{i,j}\mid \mathbf H_{i,j})}\}
      \,Y^{(j)}_i(\bm{\beta}_k),
\end{equation*}
where $Y^{(j)}_i(\bm{\beta}_k) = Y_i - \sum_{s=j}^k \beta_s A_{i,s}$, and $\widehat{\mathbb{E}(A_{i,j}\mid \mathbf H_{i,j})}$ is obtained by fitting a regression model for $A_{i,j}$ given $\bm{H}_{i,j}$ (e.g., a linear model).
Stacking these moment conditions across $j = 1, ..., k$ yields the $k\times 1$ vector
\begin{equation*}
    \hat{\bm{m}}_k(\bm{\beta})
=
\big(
\hat m_1(\bm{\beta}_k),
\ldots,
\hat m_k(\bm{\beta}_k)
\big)^\top .
\end{equation*}
The basic joint GMM estimator chooses
$\bm{\beta}_k$ to make all these sample moments close to zero by minimizing 
\begin{equation}\label{eq:GMM-mini}
     \hat{\bm{\beta}}^{\text{GMM}}_k
    = \arg\min_{\bm{\beta}_k}
\;
\hat{\bm{m}}(\bm{\beta}_k)^\top
\bm{\Sigma}_g
\hat{\bm{m}}(\bm{\beta}_k)
\end{equation}
where $\bm{\Sigma}_g$ is a $k \times k$ positive definite weighting matrix. In practice, we can set $\bm{\Sigma}_g = \bm{I}_k$ (the identity matrix) and compute $\hat{\bm{\beta}}_k^{\text{GMM}}$ using numerical optimization, with user-specified starting values. 
This joint approach treats all time points symmetrically. In Appendix A.4 we describe a more efficient two-step GMM estimator based on the double-residualized moment condition in \eqref{eq:moment-condition-eff}.
Appendix A.2 provides a brief justification that the algorithm in Appendix A.4 is a direct implementation of (\ref{eq:moment-condition-eff}).
A practical advantage of the joint GMM formulation over the purely recursive approach is that it naturally yields a joint variance–covariance matrix for all components of $\hat{\bm{\beta}}_k$ via the usual sandwich estimator for GMM (see Appendix A.5 for details).  This allows us to construct standard errors and confidence intervals for the hypothetical estimand, as described in Section~\ref{sec:application}.\\
\\
So far we have treated the MSM and SNMM as saturated in time with each $\eta_k(j)$ and $\beta_k(j)$ a distinct parameter.
As in the IV setting, we can also impose a lower–dimensional parametric structure on these effects, for example the two-parameter decay model $\beta_k(j) = \beta \alpha^{t_k - t_j}$ introduced in Section~\ref{sec:method-iv}, with parameter $\bm{\theta} = (\beta,\alpha)$, if we believe the decay form is a reasonable approximation. This simply amounts to replacing the saturated coefficients in the minimization problem in (\ref{eq:IPTW_mini}) and (\ref{eq:GMM-mini}) by the chosen decay model, and minimizing over the low-dimensional parameter $\bm{\theta}$ instead. In this setup, identification no longer requires full positivity at every time point, therefore extending its applicability in practice.\\
\\
In the score equation (\ref{eq:G-estimation}) for the IV method, $\bm{\Sigma}$ can be any arbitrary $K \times K$ positive definite matrix, with the identity matrix being a convenient choice. However, to improve the efficiency of the parameter estimates, a more appropriate choice for $\bm{\Sigma}$ should account for the correlation structure \citep{wang2014generalized}. Specifically, it can be the variance-covariance matrix of the $K$ residuals, $Y_{i,k} - \sum_{j=1}^{k} \beta_{k}(j)A_{i,j}$, where $k = 1, ..., K$ \textcolor{black}{\citep{Bowden2025}}. 
To enhance efficiency, \citet{Bowden2025} proposed the ``partial-out'' estimation strategy: regressing each $A_{k}$ and $Y_{k}$ ($k = 1, ..., K$) on the baseline covariates $\bm{L}_{0}$, and then replacing $A_{k}$ and $Y_{k}$ with the residuals from these regressions. See Section 2.6.2 in \citet{Bowden2025} for further details. The variance–covariance matrix of $\widehat{\bm{\theta}} = (\widehat{\beta}, \widehat{\alpha})$ is obtained using the sandwich formula of \citet{Bowden2025}; details are provided in Appendix A.5.

%% file: 03-Simulations.tex
\section{Monte Carlo Simulations}\label{sec:main_simu}
\subsection{Data Generation Process and Simulation Scenarios}\label{sec:simu_setup}
We perform Monte Carlo simulations to evaluate the performance of IPTW, G-estimation and IV in different settings by generating data for $K = 3$ time points across $n = 5,000$ individuals. For simplicity, we make the time intervals equally spaced, so that $t_{k} = k$. Based on the longitudinal structure shown in Figure~\ref{fig:causal-DAG}, assume both the adherence and outcome are continuous, and that:
\begin{equation}\label{eq:simu_a_1}
    A_{i, k} = \gamma_G G_{i} + \eta_A A_{i, k-1} + \eta_Y Y_{i, k-1} + \eta_{1} F1_{i} + \eta_{2} F2_{i} + \eta_V V_{i, k} + \xi_{i, k}.
\end{equation}
\begin{equation}\label{eq:simu_y_1}
    Y_{i, k} = \sum_{j = 1}^{k} \pi_{k}(j) A_{i, j} + \beta_Y Y_{i, k-1} + \beta_{F1} F1_{i} + \beta_{F2} F2_{i} + \beta_V V_{i, k} + \epsilon_{i, k}.
\end{equation}
for $k = 1, 2, 3$. In the adherence equation (\ref{eq:simu_a_1}), we assume that the initial adherence for $k =1$ is determined as follows:
\begin{itemize}
    \item $A_{i, 0} = Y_{i, 0} = 0$ and  $\eta_A = 0.2$, $\eta_Y = 0.1$.
    \item The instrumental variable $G_i$ is generated from a Binomial(2, $p$) distribution with $p$=0.2 and set $\gamma_G = 0.5$.
    \item $F1_{i}$ and $F2_{i}$ are time fixed baseline covariates following $N(0.2, 1)$ and $N(-0.1, 1)$ distributions respectively. Set $\eta_1 = 0.2$ and $\eta_2 = 0.3$.
    \item The baseline value for the time-varying covariate, denoted by $BV_{i, k}$, is generated from an AR(1) process as $BV_{i, k} = 0.98 \times BV_{i, k-1} + e_t$ where $e_t$ follows a $N(0, 0.2)$ distribution and $V_{i, 0} = 0$. Since the time-varying covariate can be affected by the past exposure, we set $V_{i, k} = BV_{i, k} + \tau A_{i, k-1}$ with varying $\tau$. Set $\eta_V = 0.5$. 
    \item The random errors $\epsilon_{i, k}$ and $\xi_{i, k}$ follow a $N(0, 1)$ distribution.
\end{itemize}
In the outcome equation (\ref{eq:simu_y_1}), we set $\beta_Y= 0.5$, $\beta_{F1} = 0.4$, $\beta_{F2} = -0.1$ and $\beta_V = -0.6$. For simplicity, under model (\ref{eq:simu_a_1}) and (\ref{eq:simu_y_1}), we assume that only the immediate previous adherence and outcome exert direct effects on the current $A$ and $Y$, but this can be easily extend to the full past history. 
We aim to estimate the causal effect of $A_{j}$ on $Y_{k}$, denoted by $\beta_k(j)$, with $j \leq k$. As discussed in Section \ref{sec:method}, the effect $\beta_k(j)$ incorporates both the direct effect $\pi_{k}(j)$ and indirect effects. 
To make the IV model identifiable and the estimation results comparable across methods, we maintain the assumption that the overall effect $\beta_k(j)$ follows a causal decay relation as $\beta_k(j) = \beta \alpha^{k - j}$ with $\beta = -1.1$ and $\alpha = 0.95$. We set the values of $\pi_{k}(j)$ accordingly so the total effect $\beta_k(j)$ fits into the decay structure. See the Appendix B.1 for further details. Under this specification, IPTW and G-estimation estimate a six-parameter time-saturated model with each $\beta_k(j)$ while IV targets the estimation of only $\beta$ and $\alpha$.\\
\\
We evaluated the performance of the methods under three simulation scenarios:
\begin{enumerate}
    \item No unobserved confounding with
    \begin{enumerate}
        \item strong feedback between exposure $A$ and time-varying covariate $V$ ($\tau = 0.8$);
        \item no exposure-covariate feedback ($\tau = 0$).
    \end{enumerate}
    \item Unobserved confounding present where we treat $F1$ as the unobserved confounder and omit it from the estimation for all methods. For this design, we additionally examine the IV method with two variations:
    \begin{enumerate}
        \item Make $G$ a weak instrument by setting $\gamma_G = 0.05$. Under this setting, the partial first-stage $F$-statistics for the association between $G$ and $A$, conditional on covariates, are well below the conventional rule-of-thumb threshold of 10, indicating a weak instrument \citep{StaigerStock1997, StockYogo2005}.
        \item In addition to regressing $A$ and $Y$ on baseline covariates in the ``partial-out'' process, we incorrectly adjust for the time-varying covariates as well.
    \end{enumerate}
     % \item Unobserved confounding present with a null causal effect, i.e.\ $\beta = 0$.
    \item No unobserved confounding, but with model misspecification: $\beta_1(1) = -1.1$, $\beta_2(1) = -0.6$, $\beta_2(2) = -1.05$, $\beta_3(1) = -0.2$, $\beta_3(2) = -0.55$ and $\beta_3(3) = -0.1$ do not follow the decay model $\beta_k(j) = \beta \alpha^{k-j}$.
\end{enumerate}

\subsection{Simulation Results}\label{sec:simu_results}
For each method, we summarize their performance using statistics averaged over 1{,}000 Monte Carlo replications. For IPTW and G-estimation, we report the Monte Carlo mean of the point estimates (\textit{Estimate}), the mean absolute error (\textit{MAE}), and the root mean squared error (\textit{RMSE}) relative to the true causal effects. To assess inferential performance, we report the empirical standard deviation of the estimates across replications (\textit{EmpSD}), the average sandwich standard errors (\textit{Sand SE}), and the coverage rates of the 95\% confidence intervals constructed using these sandwich standard errors (\textit{Cov(sand)}). For IPTW, we additionally report the standard errors from the weighted regression (\textit{Reg SE}) and the corresponding 95\% confidence interval coverage (\textit{Cov(reg)}). For G-estimation, results are presented for both the basic and two-step efficient GMM implementations, which yield identical point estimates but differ in their estimated variances.
For the IV approach implemented under a two-parameter decay model, we report analogous summaries for $\beta$ and $\alpha$, including the Monte Carlo mean of the point estimates, MAE, RMSE, empirical standard deviation, sandwich standard errors, and 95\% confidence interval coverage. \\
\\
Table \ref{tab:simulation_summary_swapped} provides a high level summary of the simulation study main findings, with full details provided in Appendix B.2, Table S1 -- Table S14. In Table S6, we report three summary diagnostics of the IPTW weights: 
the effective sample size (ESS) of the cumulative inverse probability weights, 
the share of total weight carried by the largest 1\% of observations (Top 1\% weight share), 
and the upper tail of the weight distribution ($q_{0.999}$). See Appendix B.2 for detailed definitions of the statistics and further discussion of the results. 

\begin{table}[!htbp]
\centering
\small
\setlength{\tabcolsep}{6pt}
\renewcommand{\arraystretch}{1.2}

\caption{{Summary of main simulation findings across designs and approaches. Each cell summarizes estimation accuracy, finite-sample performance, and inferential behavior based on the detailed results reported in Table S1 -- Table S14 in Appendix B.2.}}
\label{tab:simulation_summary_swapped}

\begin{tabular}{>{\raggedright\arraybackslash}p{3.0cm}
                >{\raggedright\arraybackslash}p{4.2cm}
                >{\raggedright\arraybackslash}p{4.2cm}
                >{\raggedright\arraybackslash}p{4.2cm}}
\toprule
\textbf{Method} & \textbf{Design 1} & \textbf{Design 2} & \textbf{Design 3} \\
\midrule
\textbf{IPTW} &
Finite-sample performance degrades under strong exposure--confounder feedback due to near-positivity violations. Instability and bias are most pronounced for lagged effects. Without exposure--confounder feedback, IPTW estimates are close to the truth but remain less accurate than G-estimation, with slightly larger bias. Regression-based SE severely undercovers, while sandwich SE is conservative. &
Inconsistent due to violation of NUC. Regression SE undercovers; sandwich SE is more conservative but cannot correct inconsistency arising from identification failure. &
Consistent for each parameter in the time-saturated six-parameter model, provided NUC holds. \\
\midrule
\textbf{G-estimation (basic GMM)} &
Stable and essentially unbiased across scenarios; sandwich standard errors are conservative and systematically exceed empirical SD, leading to near-100\% coverage. &
Inconsistent under NUC violation, reflecting failure of identification rather than a variance-estimation issue. &
Consistent for each parameter; sandwich SE tends to be conservative. \\
\midrule
\textbf{G-estimation (efficient GMM)} &
Unbiased across scenarios. Sandwich standard errors closely match empirical variability, yielding near-nominal coverage. &
Also inconsistent; efficiency gains do not mitigate bias arising from identification failure, and coverage can remain very poor. &
Consistent for each parameter with improved efficiency; sandwich SE is well calibrated to empirical variability. \\
\midrule
\textbf{IV G-estimation} &
Consistent and robust to practical positivity problems, but less efficient than basic G-estimation when NUC holds. &
Remains consistent under valid IV assumptions in the baseline setting. Performance deteriorates sharply under weak instruments and when incorrectly adjusting for time-varying covariates. &
Misspecified under decay violation. Implied cumulative effects provide interpretable summaries of total effects. \\
\bottomrule
\end{tabular}
\end{table}

%% file: 04-Application.tex
\section{Applied Example}\label{sec:application}
\subsection{Data Description}
\begin{figure}[htbp]
  \centering
  
  \begin{subfigure}[t]{0.59\textwidth}
    \centering
    \includegraphics[width=\linewidth]{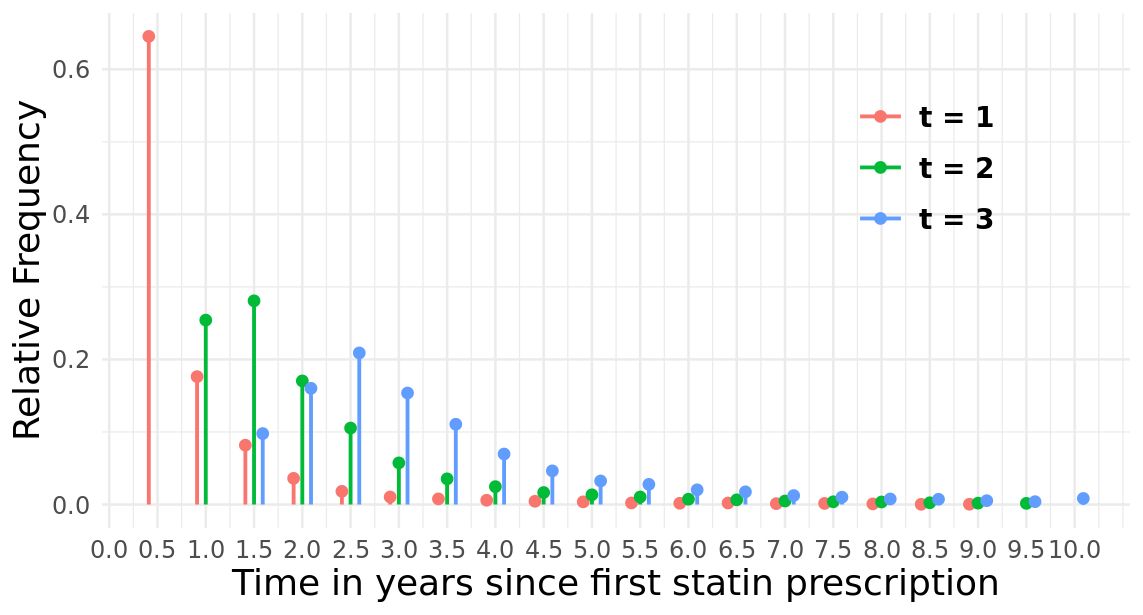}
    \caption{{\small Frequency distribution of the first three post-statin LDL-c measurement times. 
    The x-axis shows time since statin initiation (years); the y-axis shows relative frequency.}}
    \label{fig:followup-time}
  \end{subfigure}
  \hfill
  \begin{subfigure}[t]{0.4\textwidth}
    \centering
    \includegraphics[width=\linewidth]{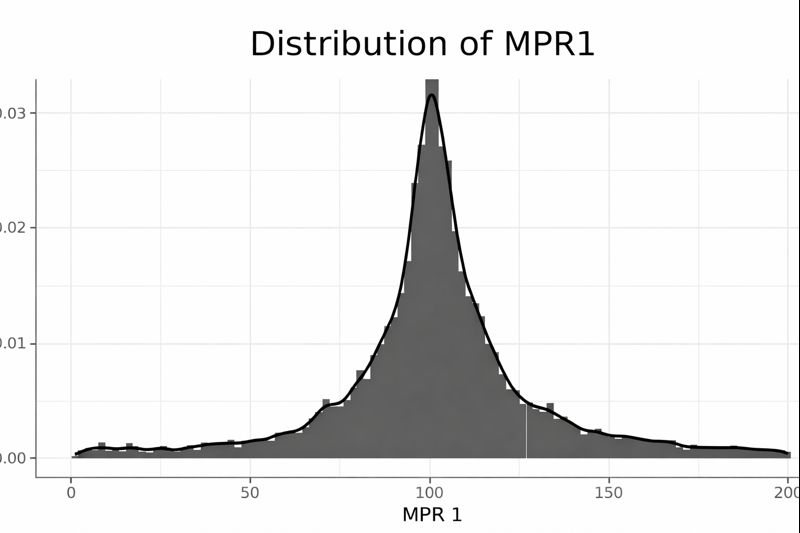}
    \caption{{\small Distribution of MPR from baseline to the first follow-up interval. 
    The x-axis shows individual MPR values; the y-axis shows the estimated density.}}
    \label{fig:MPR1-dist}
  \end{subfigure}
  
  \caption{{\small Distributions of (a) LDL-c follow-up times and (b) MPR from baseline to the first LDL-c follow-up.}}
  \label{fig:followup-time-MPR1}
\end{figure}

 We illustrate the application of the three approaches by applying them to a real-world data example to estimate the effect of sustained adherence to statin medication on LDL-c control. This builds on our own recent work detailed in \citet{turkmen2025understanding} that used a cohort of $\approx$ 69,000 UK Biobank participants with detailed linked electronic primary care health records who were prescribed statin medication. We specifically focused on individuals prescribed simvastatin or atorvastatin (the most commonly prescribed statins), who in addition had at least one pre-treatment and three post-treatment low density lipoprotein (LDL-c) measurements; accordingly, the longitudinal dataset consists of three post-treatment time points ($k = 1, 2, 3$). This reduced the sample size to approximately 13,000 patients.\\
 \\
 Following \citet{turkmen2025understanding}, adherence is proxied using patients' prescription records and the time they are treated.
 In this paper, we refer to this measure as the medication possession ratio (MPR), defined as the proportion of days within a given period for which medication is available \citep{andrade2006methods, hess2006measurement}. In practice, and following common implementations in the literature (e.g., \citet{shields2020identifying, cordioli2023socio}), days’ supply is inferred from prescribed drug quantity under standard dosing assumptions, such that MPR is calculated as
\begin{equation*}
\mathrm{MPR}
=
\frac{\text{Number of tablets prescribed in study period}}
     {\text{Length of study period (days)}}
\times 100.
\end{equation*}
In our previous work \citep{turkmen2025understanding}, we referred to this quantity as the proportion of days covered (PDC); here, we adopt the term MPR for greater precision.
Values of MPR can exceed 100\% either because patients are prescribed more than one tablet per day during a given period, or because of early refills or stockpiling. In the latter case, excess supply in one period may be offset by lower MPR values in subsequent periods, such that overall adherence is still appropriately reflected.\\
\\
The outcome variable at time point $k$ is defined as the percentage change in LDL-c relative to the pre-treatment baseline, calculated as:
\begin{equation*}
  Y_{k} =  \frac{LDL_k - LDL_0}{LDL_k},
\end{equation*}
where $LDL_k$ is the post-statin LDL-c level for time point $k$ and $LDL_0$ is the pre-treatment baseline level (measured <180 days prior to statin initiation).
In addition to adherence and LDL-c measurements, our analysis dataset also included the following baseline covariates $\bm{L}_0$: age (in years) at first statin prescription, biological sex (male or female), education level (measured as self-reported highest qualification) and assessment center. The timing of $Y_{k}$, $t_{k}$ (in years) was also recorded and used as a time-varying covariate. For each follow-up time, adherence was calculated as the average MPR over the interval between the current and previous follow-up time points. Following \citet{turkmen2025understanding}, we excluded individuals with extreme adherence values (here any MPR > 200).
We additionally restricted the analysis to individuals with valid polygenic score (PGS) values, which serve as the instrumental variable for statin adherence (see Section~\ref{sec:application-iv} for details), yielding a final longitudinal sample of n=13,285 participants. For these individuals, we constructed a longitudinal dataset with the variables ($Y_{1},Y_{2},Y_{3},MPR_{1},MPR_{2},MPR_{3},t_{1},t_{2},t_{3},\bm{L}_0$). 
For each time point $k$, we can then define an observed history set $\bm{H}_k$, for example, $\bm{H}_3 = (Y_{1},Y_{2},MPR_{1},MPR_{2},t_{1},t_{2},t_{3},\bm{L}_0)$.\\
\\
Figure~\ref{fig:followup-time-MPR1} Panel (a) shows the distribution of all follow-up times: $t_{1}$ was highly concentrated within one year with an approximate mean of $0.5$ and the average value of $t_{3}$ was approximately 3 years (with a maximum of 10). 
Figure~\ref{fig:followup-time-MPR1} Panel (b) shows the distribution of MPR$_{1}$, which is centered close to $100\%$ but shows that a noticeable subset of patients already have substantially lower MPR values in this early period.

\subsection{IPTW and G-estimation Analysis}
We apply IPTW and G-estimation of the SNMM to estimate a six-parameter saturated model: the causal effects of $MPR_1$ on $Y_1$; $MPR_1$ and $MPR_2$ on $Y_2$; $MPR_1$, $MPR_2$ and $MPR_3$ on $Y_3$. For IPTW, in the first stage, we estimate inverse probability weights using the CBPS method \citep{fong2018covariate}, adjusting for the observed history $\bm{H}_k$.
We consider two CBPS options: the method = ``exact'' specification chooses weights so that the selected covariates are exactly balanced in the weighted sample, whereas method = ``over'' imposes additional (over-identifying) moment conditions that encourage balance while allowing some slack to potentially improve efficiency \citep{CBPSpackage}.
In the second stage, for each time point we estimate the causal effects by performing weighted regression of $Y$ on the current and lagged MPR simultaneously, using the CBPS-derived weights. For G-estimation, we use the joint GMM procedure described in Section~\ref{sec:implementation}. Specifically, we treat $Y_1$, $Y_2$ and $Y_3$ in turn as the final outcome and stack the corresponding moment conditions into a single GMM problem, as defined in (\ref{eq:GMM-mini}), so that all causal parameters are estimated jointly. We report results from both the basic and the two-step efficient versions of the joint GMM estimation.\\
\\
Tables S15 and S16 in Appendix C.2 report the estimated effects of MPR at each time point on LDL-c change, obtained using IPTW and G-estimation respectively, together with their sandwich standard errors. 
Tables S18, S19, and S20 report diagnostic statistics for the CBPS weights and exposure–confounder correlations after weighting, with nearly all correlations falling below the 0.1 threshold.
Almost all coefficients are negative and highly statistically significant, indicating that higher adherence at any of the three windows is associated with larger LDL-c reduction. Our primary estimand of interest is the overall LDL-c reduction difference between two hypothetical adherence regimes: full adherence at any time (all $MPR = 100\%$) and non-adherence (all $MPR = 0$). 
As each coefficient in Tables S15--S16 represents the effect per one percentage increase of MPR, the implied contrast at time $k$ is obtained by summing the relevant coefficients for $Y_k$ and multiplying by $100$.
More generally, the estimated coefficients can define contrasts between any two adherence regimes (e.g., $\text{MPR}=50\%$ versus $\text{MPR}=100\%$); here, we focus on the contrast between full adherence and non-adherence for interpretability.
For example, if the effects of $MPR_1$, $MPR_2$ and $MPR_3$ on $Y_3$ are denoted $\beta_{31}$, $\beta_{32}$ and $\beta_{33}$, then the contrast of full versus non-adherence for the third LDL-c reduction would be $\hat{\Delta}_3 = (100\times\sum^{3}_{k=1}\hat{\beta}_{3k})$. \\
\\
To quantify uncertainty for this estimand, we use a parametric bootstrap to obtain the $95\%$ confidence intervals. For each method and each time point $k$, we treat the vector of coefficients $\bm{\hat{\beta}}_k$ as approximately multivariate normal with mean equal to the point estimates and variance-covariance matrix given by the sandwich estimator. We then draw $1000$ samples of $\bm{\hat{\beta}}_k$ from this distribution, compute the corresponding $\Delta_k$ for each draw, and summarize the simulated values by their mean and standard deviation. The $95\%$ confidence intervals for $\Delta_k$ are then obtained from these simulated values, see Figure~\ref{fig:estimand-all} and Table S17 in Appendix C.2.\\

\begin{figure}[htbp]
    \centering

    %------------ (a) CIs over time ------------
    \begin{subfigure}{\textwidth}
        \centering
        \includegraphics[width=\textwidth]{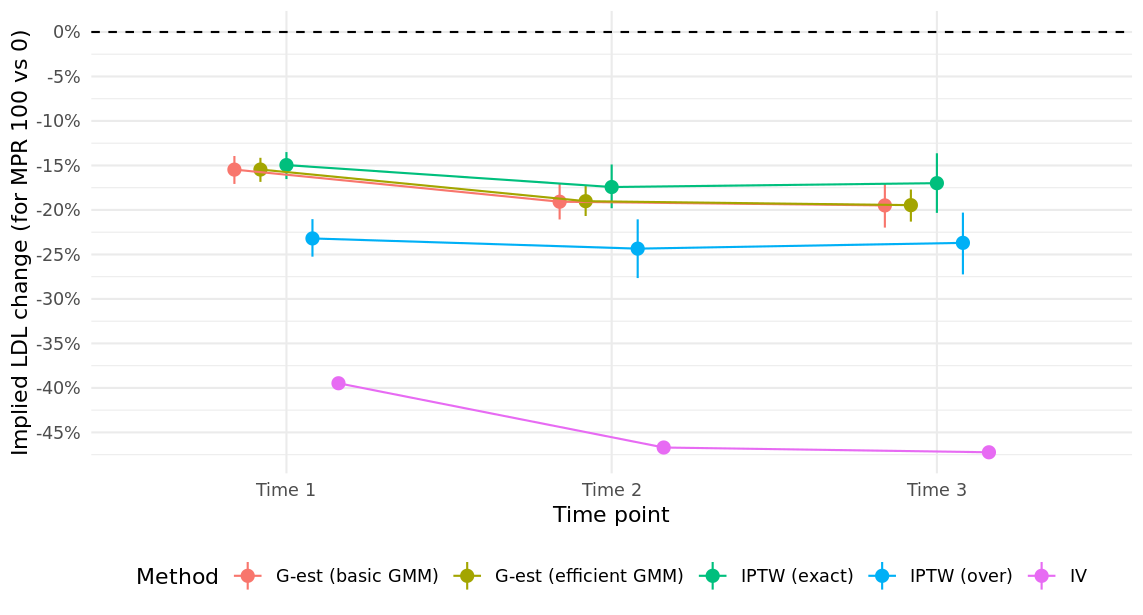}
        \subcaption{%
        {\small The 95\% confidence intervals of LDL-c reduction difference between
        MPR = 100 versus MPR = 0 implied by estimates from IPTW (with
        ``exact'' and ``over'' weights) and G-estimation (with basic and
        efficient GMM). We also show the point estimates of the LDL-c reduction difference from the IV approach.}}
        \label{fig:estimand-all}
    \end{subfigure}

    \vspace{0.7cm}

    %------------ (b) Slopes for each MPR_t (IPTW, G-est, IV) ------------
    \begin{subfigure}{\textwidth}
        \centering

        \begin{minipage}{0.48\textwidth}
            \centering
            \includegraphics[width=\linewidth]{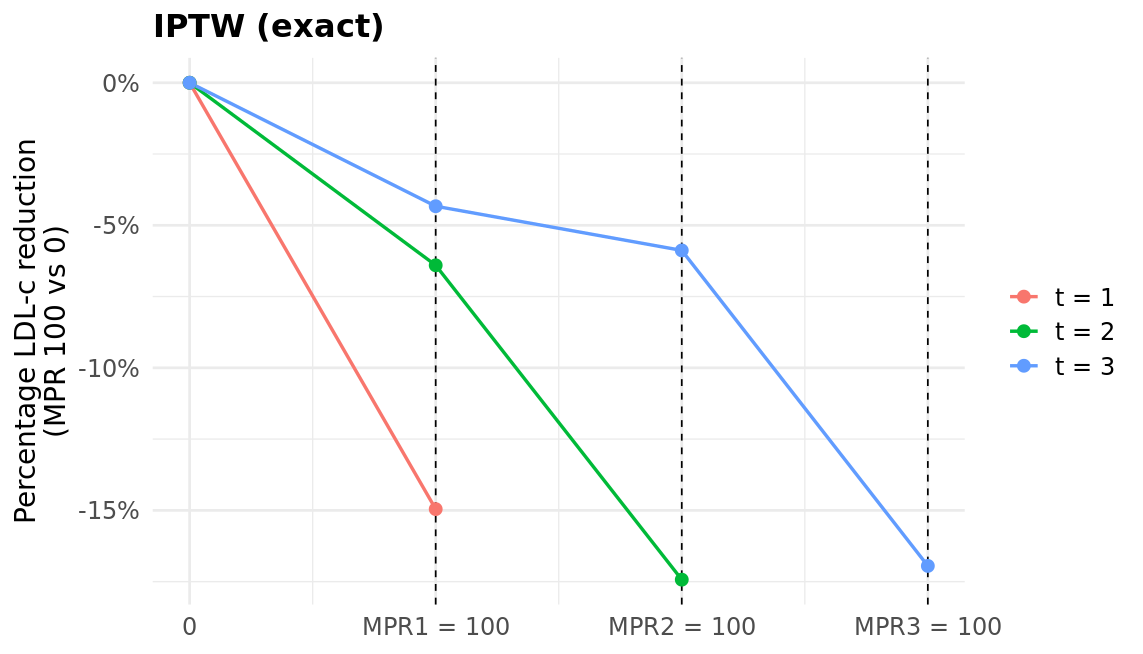}
        \end{minipage}
        \hfill
        \begin{minipage}{0.48\textwidth}
            \centering
            \includegraphics[width=\linewidth]{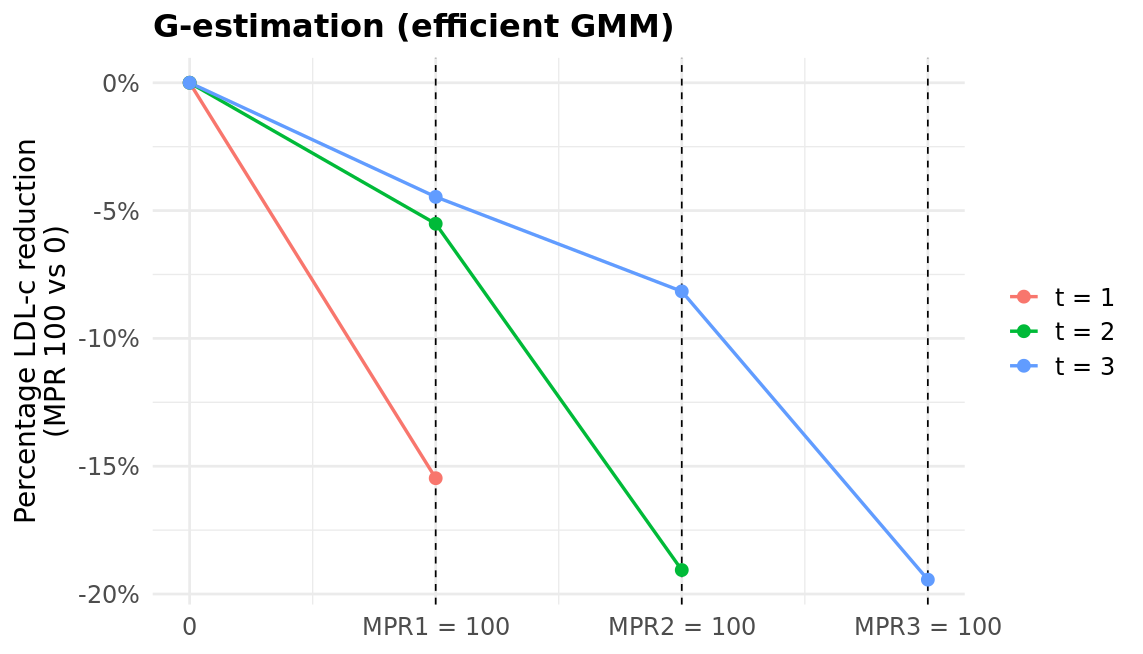}
        \end{minipage}
        \hfill
        \begin{minipage}{0.5\textwidth}
            \centering
            \includegraphics[width=\linewidth]{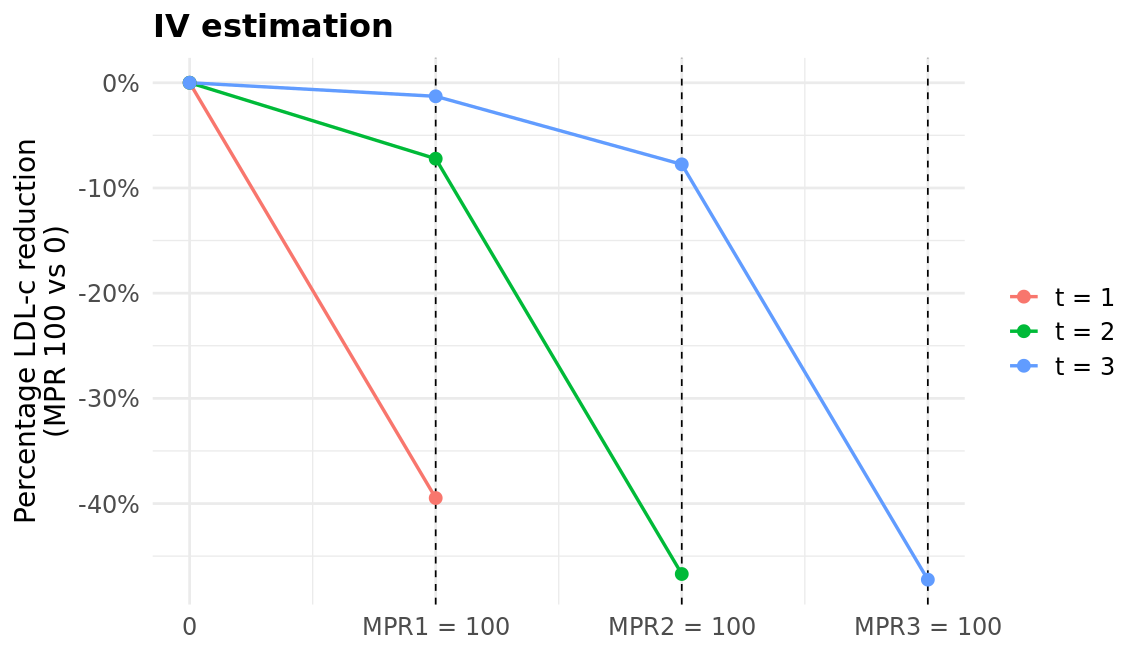}
        \end{minipage}

        \subcaption{%
        {\small Estimated trajectories of effects of statin adherence on LDL-c change. The red line shows the effect of MPR$_1$ on $Y_1$; the two
        segments of the green line correspond to MPR$_1$ and MPR$_2$ on $Y_2$;
        and the three segments of the blue line correspond to MPR$_1$, MPR$_2$
        and MPR$_3$ on $Y_3$. The top left, top right and bottom panels show IPTW with
        ``exact'' weights, efficient G-estimation, and the IV approach,
        respectively.}}
        \label{fig:LDL-reduction-slopes}
    \end{subfigure}

    \caption{Estimation results of IPTW, G-estimation and IV.}
    \label{fig:iptw-g-iv-results}
\end{figure}

\noindent Across methods, full adherence versus no adherence is consistently associated with substantially larger LDL-c reduction. IPTW with the ``exact'' weighting option yields estimated differences of roughly $15\%$ to $17\%$, whereas IPTW with ``over'' weighting gives larger effects of about $23\%$ to $24\%$ LDL-c reduction. G-estimation produces estimates in between these two IPTW specifications, with efficient GMM and basic GMM giving identical point estimates (around 15\% larger LDL-c reduction at time point 1 and 19–20\% at the later time points). As expected, the efficient GMM yields slightly narrower confidence intervals. Across all methods, the total effect (expressed as a summation across the time points) remained remarkably similar.\\
\\
The slope plots (from IPTW with ``exact'' weights and efficient GMM) in Figure \ref{fig:LDL-reduction-slopes} further decompose the overall contrast into contributions from each adherence window. They show a clear timing pattern: adherence closest to the LDL-c measurement ($MPR_3$) has the largest impact on $Y_3$, followed by initial adherence ($MPR_1$), while mid-period adherence ($MPR_2$) contributes less. This pattern is broadly consistent across both IPTW and G-estimation, highlighting the importance of early follow-up monitoring \citep{turkmen2025understanding}.\\
\\
The comparison of methods highlights that the IPTW results can be sensitive to the choice of CBPS weighting option: ``over'' weighting accentuates the estimated adherence effect and yields wider intervals, whereas the ``exact'' option produces more moderate estimates. The magnitudes from G-estimation and IPTW with “exact’’ weights are similar: for each time point the implied LDL-c reduction difference from full versus non-adherence is all in the $15-20\%$ range and the 95\% confidence intervals overlap substantially. Therefore, while the precise effect size to some extent depends on the estimation approach, the qualitative conclusions that substantial overall benefit of adherence and greater impact of recent adherence, are consistent across methods.

\subsection{IV Analysis}\label{sec:application-iv}
 To construct an IV for statin adherence, we attempted to use the PGS for on-statin LDL-c response developed by \citet{mayerhofer2022genetically}. This score is based on genetic associations reported by the Genomic Investigation of Statin Therapy Consortium, who conducted a pharmacogenetic meta-analysis of genome-wide association studies (GWAS) of on-statin LDL-c response in $40,914$ statin-treated individuals of European ancestry \citep{postmus2014pharmacogenetic}. A total of 35 independent single nucleotide polymorphisms (SNPs) were selected for inclusion in the PGS, with pairwise linkage disequilibrium $r^2<0.001$. All selected SNPs were associated with on-statin LDL-c response at genome-wide significance ($p < 5 \times 10^{-8}$) \citep{mayerhofer2022genetically}. 

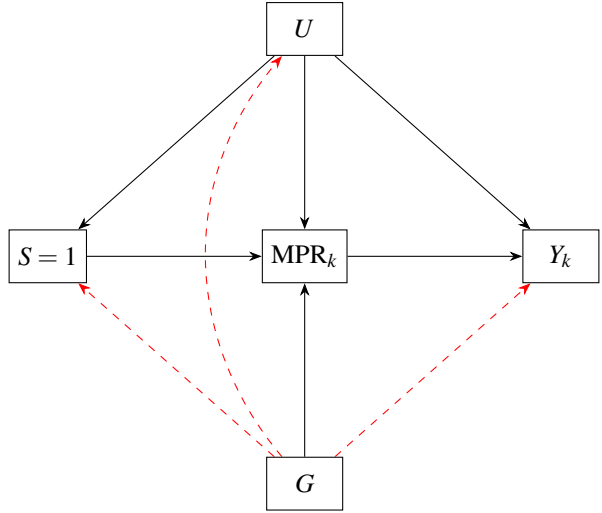
\begin{figure}[H]
    \centering
   \begin{tikzpicture}[
    node distance = 2.3cm,
    >=Stealth,
    every node/.style = {rectangle, draw, minimum width=10mm, minimum height=7mm}
]
\node (S)  {$S = 1$};
\node (P)  [right=of S] {$\mathrm{MPR}_k$};
\node (Y)  [right=of P] {$Y_k$};
\node (U)  [above=of P] {$U$};
\node (G)  [below=of P] {$G$};
\draw[->] (S) -- (P);
\draw[->] (P) -- (Y);
\draw[->] (U) -- (S);
\draw[->] (U) -- (P);
\draw[->] (U) -- (Y);
\draw[->] (G) -- (P);
\draw[->, red, dashed] (G) -- (S);
\draw[->, red, dashed] (G) to (Y);
\draw[->, red, dashed, bend left=40] (G) to (U);

\end{tikzpicture}
    \caption{Conceptual illustration of conditioning on statin initiation. $S$ indicates statin initiation. Solid black arrows represent the assumed causal structure. Conditioning on $S = 1$ (restricting to treated individuals) can induce additional associations between the instrument $G$, unobserved confounder $U$ and outcome $Y_k$ (shown as red dashed arrows), potentially violating the IV independence and exclusion assumptions within the selected sample.} 
    \label{fig:treated-only}
\end{figure}
\noindent To enhance the validity of the PGS as an instrument for adherence, we follow the practice in \citet{mayerhofer2022genetically} and exclude 4 SNPs that are also associated with off-statin LDL-c levels, thereby minimizing the risk of direct effects on the outcome induced by conditioning on statin initiation (see Figure~\ref{fig:treated-only} and discussion in Appendix C.1). Our PGS was constructed using the {\verb `ukbrapR`}  package v0.3.9 (\url{https://github.com/lcpilling}) for the final 31 variant PGS from UK Biobank-provided DRAGEN variant calls from the whole genome sequencing data on 490,000 participants (field:24309). Crucially, this sample does not overlap with the GWAS meta-analysis sample used to derive the associations. See \citet{mayerhofer2022genetically} and \citet{postmus2014pharmacogenetic} for details; a list of the 35 SNPs and the 4 excluded SNPs can be found in Supplementary Table S5 in \citet{mayerhofer2022genetically}. \\
\\
As explained in Section~\ref{sec:method-iv}, the proposed instrument is designed to influence the LDL-c at time point $k$, $Y_{k}$, via $MPR_{k}$ but not $Y_{k}$ directly, and to be independent of unmeasured confounders of statin initiation, $MPR_{k}$ and $Y_{k}$ (Figure~\ref{fig:treated-only}). In practice, however, only a very small association between the instrument and adherence was present (partial F-statistics = $0.422$, $1.041$ and $0.856$ for MPR1, MPR2 and MPR3 respectively), indicating that the instrument is empirically extremely weak in this dataset. As a consequence, the IV coefficient estimates are accompanied by very large standard errors and conventional weak-instrument concerns apply.\\
\\
For illustration purposes only, we present the estimation results from the two-parameter decay specification as described in Section~\ref{sec:method-iv}, where we estimate the `main effect' $\beta$ and the decaying factor $\alpha$. 
Figure \ref{fig:iptw-g-iv-results}b displays the same “MPR = 100 vs MPR = 0” contrasts used earlier, translated from the IV point estimates ($\hat{\beta} = -0.0039$ and $\hat{\alpha} = 0.1293$). The trajectories are directionally consistent with the IPTW and G-estimation results in the sense that higher adherence is associated with larger LDL-c reduction, and the estimated effect increases over time, with the largest LDL-c reduction attributed to adherence closest to the LDL-c measurement. Numerically, the LDL-c reduction difference between full adherence versus non-adherence implied by the IV point estimates is roughly 40–50\% (see Figure~\ref{fig:iptw-g-iv-results}a), which is qualitatively closer to the 30-50\% LDL-c reduction often cited in treatment guidelines for high-risk patients shortly after statin initiation \citep{NICE2023NG238, Virani2023CCDGuideline, Arnett2019PrimaryPrevention}
than the 15-20\% reductions estimated by IPTW and G-estimation. However, because the instrument is so weak, these point estimates are extremely imprecise, with very wide confidence intervals (not shown). 

%% file: 05-Discussion.tex
\section{Discussion}
In this paper, we compared three longitudinal causal methods for estimating time-varying causal effects in settings with a continuous exposure, motivated by drug adherence and long-term health outcomes.
Under a unified longitudinal modelling framework and a common target hypothetical estimand, we clarified the identification assumptions required by each approach and described how point estimates and confidence intervals can be obtained in practice. We provided implementation details and illustrated the methods in an applied analysis of UK Biobank participants with linked primary care data on statin adherence and LDL-c control. 
We conducted a simulation study to evaluate their finite-sample performance under various scenarios. 
Under the “no unmeasured confounding” assumption, our simulation results showed that G-estimation of the SNMM was generally more statistically efficient than IPTW estimation of the MSM. In addition, we found that IPTW could be fragile in settings with a continuous exposure, particularly when there is strong exposure–covariate feedback, where practical violations of positivity and unstable weights can arise. These findings are consistent with existing theoretical and empirical work highlighting challenges for weighting-based approaches with continuous time-varying exposures, and motivate the use of G-estimation as a natural alternative in such settings \citep{vansteelandt2014structural, vansteelandt2016revisiting}.\\
\\
In practice, IPTW benefits from mature and widely available software for implementation, whereas G-estimation is less commonly implemented in standard applied workflows. A contribution of our work is therefore to provide accessible \textsf{R} code implementing both the basic and efficient GMM formulations of longitudinal G-estimation, alongside the simulated dataset used in our study as a resource for future methodological comparisons. From a modelling perspective, SNMMs parameterize causal effects through conditional blip functions given the observed history, which in principle allows interactions between adherence and time-varying covariates to be incorporated more directly than in MSMs, which target marginal effects \citep{robins2000marginal,daniel2013methods}. Exploring such effect modification represents a natural direction for future work. Finally, when the “no unmeasured confounding” assumption is doubtful, our results illustrate how an IV-based extension of G-estimation can be used to relax this assumption under appropriate instrument conditions. Although motivated by drug adherence, the modelling framework and estimation strategies considered here apply more broadly to longitudinal settings with continuous time-varying exposures, such as blood pressure trajectories.\\
\\
In the real-world example where we applied the methods to estimate the effect of sustained statin adherence on LDL-c reduction with a UK Biobank cohort, the absolute size of the estimated LDL-c reduction from IPTW and G-estimation (15–20\%) was smaller than the 30–50\% LDL-c reduction guideline targets \citep{NICE2023NG238, Virani2023CCDGuideline, Arnett2019PrimaryPrevention}.
However, this difference might be due to the fact that the MPR measure is only a proxy for true adherence based on prescription records. It crucially does not capture whether tablets are dispensed or actually taken. Also, the guidelines are based largely on randomized trials of high-intensity statin therapy in carefully selected patients with tightly controlled follow-up, whereas our estimates come from routinely collected data on a much more heterogeneous patient cohort. For example, our data almost certainly includes patients whose cholesterol levels are relatively low on their own, but are prescribed a statin because of a perceived high risk of cardiovascular disease, as measured by their QRISK. The true effect of full adherence to treatment on their LDL-c levels may be lower than in other patient groups. In this more noisy observational data setting, although IPTW and G-estimation adjust for a set of observed covariates, residual confounding and measurement error in both the adherence and outcome measures are still potentially present, which could lead to dilution in the estimated adherence effect.\\
\\
Our IPTW estimates are also smaller than those from our previous work \citet{turkmen2025understanding}, where the IPTW analysis suggested an LDL-c reduction of around $35\%$. In that earlier analysis we fitted the outcome model without an intercept, implicitly assuming that LDL-c reduction would be zero when statin adherence (MPR) is zero. Our subsequent exploration of untreated individuals (those not on statins) showed that LDL-c levels still reduce on average after GP visits even when statins are not prescribed and adherence is by definition zero. This could be due to non-statin health interventions such as diet and lifestyle change resulting from patients being made aware of their current disease risk. In the present analysis we therefore reintroduced the intercept, allowing for non–statin-related changes in LDL-c at $MPR = 0$.\\
\\
It should be noted that our analysis is conducted in a selected sample. Unlike in randomized controlled trials, where individuals are assigned to treatment and control groups, meaningful measures of statin adherence are only defined among individuals who initiate treatment. Consequently, our causal estimands pertain to treated individuals only. In addition, to construct a longitudinal dataset with multiple time points, we further restrict attention to individuals with at least three post-treatment LDL-c measurements, which imposes an additional selection criterion. While in principle, individuals with fewer follow-up measurements could be incorporated via imputation, this would require jointly imputing missing LDL-c values, adherence measures, and irregular follow-up times, introducing substantial additional uncertainty and modelling complexity. For these reasons, we focus on a well-defined longitudinal subset for which adherence, outcomes, and follow-up timing were all directly observed.\\
\\
In our real-world application, the genetic instrument for statin adherence was weak, with partial F statistics well below conventional thresholds. Using synthetic data calibrated to our setting, we estimated that a sample approximately nine times larger would be required to achieve partial F statistics above the standard threshold of $10$ at each time point, giving a rough indication of the sample size needed for a reliable IV analysis of this type. As a result, our IV estimates should be viewed as exploratory, illustrating the potential scale of adherence effects rather than providing precise effect sizes. Future work using larger and richer longitudinal resources such as Our Future Health \citep{cook2025cohort} will soon dramatically increase the size of genotyped individuals with linked medication records to over 5 million individuals. There is also the potential to increase power further by synthesising data from other US and European cohorts such as All of Us and FinnGen \citep{allofus2019, kurki2023finngen}. Our hope is that researchers will be able to revisit the analyses described here to obtain much more robust IV-based estimates of longitudinal adherence effects once such data become available. 

\section*{Data Availability}
The empirical data used in the applied analysis were obtained from the UK Biobank, approved by the North West Multi-Centre Research Ethics Committee (Research Ethics Committee reference 11/NW/0382). Due to the data access agreement and strict privacy regulations, the authors are not permitted to share the raw patient-level data publicly. Qualified researchers can apply to access the UK Biobank resource directly via their official portal (www.ukbiobank.ac.uk). The R code used to implement the methods and generate the simulated datasets is freely available on GitHub at: https://github.com/xiaoran-liang/longitudinal-causal

%% file: 06-Appendix.tex
\doublespacing
\vspace{5em}
\thispagestyle{empty}
\begin{center}
    \Large{Appendix to ``Living forwards or understanding backwards? A comparison of Inverse Probability of Treatment Weighting and G-estimation methods for targeting hypothetical full adherence estimands in longitudinal cohort studies ''}\\
\vspace{1em}\small{Xiaoran Liang$^{a}$,
Deniz Türkmen$^{a}$, Jane A H Masoli$^{a, b}$, Luke C Pilling$^{a}$, Jack Bowden$^{a, c}$}\vspace{1em}\\
%EndAName
$^{a}${\small Department of Clinical and Biomedical Sciences, University of Exeter, Exeter, UK}\\
$^{b}${\small Royal Devon University Healthcare NHS Foundation Trust, Barrack Road, Exeter, UK}\\
$^{c}${\small Novo Nordisk Research Centre (NNRCO), Oxford, U.K}\\\vspace{1em}
%\today
\date{\vspace{-0.9em}6 March, 2026}\endgraf\bigskip\bigskip
\end{center}

\pagebreak

\appendix

% to reset figure and table counters in Supplementary section
\newcommand{\beginsupplement}{%
        \setcounter{table}{0}
        \renewcommand{\thetable}{S\arabic{table}}%
        \setcounter{figure}{0}
        \renewcommand{\thefigure}{S\arabic{figure}}%
        \def\theequation{S\arabic{equation}}
        \setcounter{equation}{0}
     }
     
\beginsupplement
\setcounter{page}{1}

\section{Modelling and Method Details}\label{app:model}
\subsection{Linking SNMM Parameters to the Causal Estimand}\label{app:model-causal-estimand}
Following the argument in \citet{tompsett2025instrumental}, we show that under the linear blip specification in (\ref{eq:blip}) and the NCTI assumption, the estimand in (\ref{eq:estimand}) can be written as a function of the SNMM parameters $\bm{\beta} = (\beta_1,\ldots,\beta_k)$.
To fix ideas, take $k=3$ and consider the contrast between the regimes $\bar {\bm{a}}=(1,1,1)$ and $\bar{\bm{a}}^{\prime}=(1,1,0)$ at the final time point. From the definition of the blip at time $3$ in (\ref{eq:SNMM}), we have
\begin{equation*}
    \mathbb{E}[Y(1, 1, 1)-Y(1, 1, 0)\mid \bm{H}_3, A_3=1] = \beta_3
\end{equation*}
where $\bm{H}_3 = \left(A_1, A_2, L_1, L_2, L_3\right)$ denotes the observed history up to time $3$. Under NCTI, the expected difference in potential outcomes is independent of observed adherence $A_3$ so that
\begin{equation*}
    \mathbb{E}[Y(1, 1, 1)-Y(1, 1, 0)\mid \bm{H}_3, A_3=1] = \mathbb{E}[Y(1, 1, 1)-Y(1, 1, 0)\mid \bm{H}_3] = \beta_3.
\end{equation*}
By the law of iterated expectations it follows
\begin{align*}
    &\mathbb{E}[Y(1, 1, 1)-Y(1, 1, 0)\mid A_1, L_1, L_2, L_3] \\
    &= \mathbb{E}_{A_2} \{\mathbb{E}[Y(1, 1, 1)-Y(1, 1, 0)\mid \bm{H}_3 = \left(A_1, A_2, L_1, L_2, L_3\right)] \}\\
    &= \mathbb{E}_{A_2}(\beta_3) = \beta_3.
\end{align*}
Repeating the same procedure for each variable in $\bm{H}_3$ and eventually we achieve
\begin{equation*}
    \mathbb{E}[Y(1, 1, 1)-Y(1, 1, 0)] = \beta_3.
\end{equation*}
From the MSM in (2), the same contrast $\mathbb{E}[Y(1,1,1) - Y(1,1,0)]$ equals $\eta_3$, so under the specifications above we have $\beta_3 = \eta_3$.\\
\\
Repeating the same argument for times $j=2$ and $j=1$ yields
\begin{equation*}
    \mathbb{E}[Y(1, 1, 0)-Y(1, 0, 0)] = \beta_2
\end{equation*}
and
\begin{equation*}
    \mathbb{E}[Y(1, 0, 0)-Y(0, 0, 0)] = \beta_1.
\end{equation*}
Summing these three contrasts gives
\begin{equation*}
    \mathbb{E}[Y(1, 1, 1)-Y(0, 0, 0)] = \beta_1 + \beta_2 + \beta_3.
\end{equation*}
More generally, for $k$ time points we obtain
\begin{equation*}
    \mathbb{E}[Y(\bm{1}_k) - Y(\bm{0}_k)]
= \sum_{j=1}^k \beta_j,
\end{equation*}
which coincides with the causal estimand defined in (\ref{eq:estimand}). In fact, the same argument gives, for any adherence history $\bar {\bm{a}}_k$,
\begin{equation*}
    \mathbb{E}[Y(\bar {\bm{a}}_k) - Y(\bm{0}_k)]
= \sum_{j=1}^k \beta_j a_j.
\end{equation*}

\subsection{Moment Conditions for G-estimation}\label{app:moment-condition}
We show that the conditional mean independence assumption in (\ref{eq:mean-ind}):
\begin{equation*}
\mathbb{E}[Y^{(j)}(\bm{\beta}_k^\ast) \mid \bm{H}_j, A_j] = \mathbb{E}[Y^{(j)}(\bm{\beta}_k^\ast) \mid \bm{H}_j] 
\end{equation*}
implies the moment condition in (\ref{eq:moment-condition}):
\begin{equation*}
\mathbb{E}\left[\{A_j - \mathbb{E}(A_j \mid \bm{H}_j)\} Y^{(j)}(\bm{\beta}_k^\ast)\right] = 0. 
\end{equation*}
As $Y^{(j)}(\bm{\beta}_k^\ast)$ and $A_j$ are mean independent given $\bm{H}_j$, we have
\begin{equation*}
\mathbb{E}\!\left[ A_j\,Y^{(j)}\!\left(\bm{\beta}_k^\ast\right) \mid \bm{H}_j\right]
=
\mathbb{E}\!\left( A_j \mid \bm{H}_j\right)\;
\mathbb{E}\!\left[ Y^{(j)}\!\left(\bm{\beta}_k^\ast\right) \mid \bm{H}_j\right].
\end{equation*}
Then we have
\begin{equation*}
\mathbb{E}\!\left[ A_j\,Y^{(j)}\!\left(\bm{\beta}_k^\ast\right) \mid \bm{H}_j\right]
-
\mathbb{E}\!\left( A_j \mid \bm{H}_j\right)\;
\mathbb{E}\!\left[ Y^{(j)}\!\left(\bm{\beta}_k^\ast\right) \mid \bm{H}_j\right]
=0
\end{equation*}
and
\begin{equation*}
\mathbb{E}\!\left[ A_j\,Y^{(j)}\!\left(\bm{\beta}_k^\ast\right) \mid \bm{H}_j\right]
-
\mathbb{E}\!\left[ \mathbb{E}\!\left( A_j \mid \bm{H}_j\right)\; Y^{(j)}\!\left(\bm{\beta}_k^\ast\right) \mid \bm{H}_j\right]
=0.
\end{equation*}
It follows that
\begin{equation*}
\mathbb{E}\!\left[
\{A_j-\mathbb{E}\!\left(A_j\mid \bm{H}_j\right)\}
Y^{(j)}\!\left(\bm{\beta}_k^\ast\right)
\,\middle|\, \bm{H}_j
\right]
=0.
\end{equation*}
By the Law of Iterated Expectations, we have the unconditional moment condition
\begin{equation*}
\mathbb{E}\!\left[
\{A_j-\mathbb{E}\!\left(A_j\mid \bm{H}_j\right)\}
Y^{(j)}\!\left(\bm{\beta}_k^\ast\right)
\right]
=0.
\end{equation*}

\noindent \citet[Equation~(33)]{vansteelandt2014structural} considers a general class of estimating equations for SNMM parameters of the form
\begin{equation*}
\mathbb{E}\!\left(
\left[d_j(\bm{H}_j,A_j)-\mathbb{E}\!\left\{d_j(\bm{H}_j,A_j)\mid \bm{H}_j\right\}\right]
\left[Y^{(j)}(\bm{\beta}_k)-\mathbb{E}\!\left\{Y^{(j)}(\bm{\beta}_k)\mid \bm{H}_j\right\}\right]
\right)=0,
\end{equation*}
where $d_j(\bm{H}_j,A_j)$ is an arbitrary function of the appropriate dimension. Under homoscedasticity, local semiparametric efficiency is attained by choosing
\begin{equation*}
d_j(\bm{H}_j,A_j)
=
\mathbb{E}\!\left[
\left.
\frac{\partial Y^{(j)}(\bm{\beta}_k^\ast)}{\partial \beta_j}
\right|\, \bm{H}_j,A_j
\right].
\end{equation*}
\noindent Under the linear SNMM blip function used in this paper, $\psi_j(a_j,\bm{H}_j)=\beta_j a_j$, the blipped-down outcome is linear in $\beta_j$, and thus
\begin{equation*}
\frac{\partial Y^{(j)}(\bm{\beta}_k^\ast)}{\partial \beta_j} = -A_j.
\end{equation*}
Therefore, the efficient choice of $d_j(\bm{H}_j,A_j)$ is proportional to $A_j$ (up to sign), which yields the double-residualized moment condition in \eqref{eq:moment-condition-eff}.\\
\\
The efficient moment condition in \eqref{eq:moment-condition-eff} can be written as
\begin{equation}
\mathbb{E}\!\left[
\left\{A_j-\mathbb{E}(A_j\mid \bm{H}_j)\right\}
\left\{Y^{(j)}(\bm{\beta}_k^\ast)-\mathbb{E}\!\left(Y^{(j)}(\bm{\beta}_k^\ast)\mid \bm{H}_j\right)\right\}
\right]=0.
\label{eq:mc_eff_repeat}
\end{equation}
Appendix~\ref{app:model-effG} implements \eqref{eq:mc_eff_repeat} by constructing a residualized blipped-down outcome.
Specifically, for each $j=1,\ldots,k$, define the residualized outcome and exposures as
\begin{equation*}
\widetilde{Y}_j = Y-\mathbb{E}(Y\mid \bm{H}_j),
\qquad
\widetilde{A}_{j,s} = A_s-\mathbb{E}(A_s\mid \bm{H}_j),
\quad s=j,\ldots,k.
\end{equation*}
The residualized blipped-down outcome in Appendix~\ref{app:model-effG} is then defined as
\begin{equation*}
\widetilde{Y}^{(j)}(\bm{\beta}_k)
=
\widetilde{Y}_j-\sum_{s=j}^k \beta_s\,\widetilde{A}_{j,s}.
\end{equation*}
Substituting the definitions of $\widetilde{Y}_j$ and $\widetilde{A}_{j,s}$ gives
\begin{align*}
\widetilde{Y}^{(j)}(\bm{\beta}_k)
&=
\left\{Y-\mathbb{E}(Y\mid \bm{H}_j)\right\}
-
\sum_{s=j}^k \beta_s
\left\{A_s-\mathbb{E}(A_s\mid \bm{H}_j)\right\} \\
&=
\left\{Y-\sum_{s=j}^k \beta_s A_s\right\}
-
\left\{\mathbb{E}(Y\mid \bm{H}_j)-\sum_{s=j}^k \beta_s \mathbb{E}(A_s\mid \bm{H}_j)\right\}.
\end{align*}
By linearity of conditional expectation,
\begin{equation*}
\mathbb{E}(Y\mid \bm{H}_j)-\sum_{s=j}^k \beta_s \mathbb{E}(A_s\mid \bm{H}_j)
=
\mathbb{E}\!\left(Y-\sum_{s=j}^k \beta_s A_s \,\middle|\, \bm{H}_j\right)
\end{equation*}
Therefore,
\begin{equation*}
\widetilde{Y}^{(j)}(\bm{\beta}_k)
=
Y^{(j)}(\bm{\beta}_k)-\mathbb{E}\!\left\{Y^{(j)}(\bm{\beta}_k)\mid \bm{H}_j\right\},
\end{equation*}
where $Y^{(j)}(\bm{\beta}_k)=Y-\sum_{s=j}^k \beta_s A_s$ denotes the blipped-down outcome under the linear SNMM.
Consequently, the moment condition implemented in Appendix~\ref{app:model-effG} as 
\begin{equation*}
\mathbb{E}\!\left[\widetilde{A}_{j,j}\,\widetilde{Y}^{(j)}(\bm{\beta}_k^\ast)\right]=0,
\end{equation*}
is exactly equivalent to \eqref{eq:mc_eff_repeat}.

\subsection{The Sequential G-estimation Procedure}\label{app:model-G}
For the implementation of the G-estimation, here we summarize the classic sequential G-estimation procedure, which solves the moment conditions in (\ref{eq:moment-condition}) backwards over time. Let $Y$ denote the chosen final outcome at time $k$ and let $\mathbf H_j$ be the recorded history up to time $j$. The basic procedure can be summarized as follows \citep{blackwell2018make}:
\begin{enumerate}
  \item \textbf{Final time point.}  
  Regress the observed outcome $Y$ on the final adherence $A_k$,
  adjusting for $\mathbf H_k$, and the coefficient of $A_k$ from this regression is taken as the estimate
  $\hat\beta_k$ of the contemporaneous adherence effect.

  \item \textbf{Blip down the effect of $A_k$.}  
  Construct a “blipped-down” outcome that removes the estimated effect
  of adherence at time $k$:
  \[
    Y^{(k-1)} = Y - \hat\beta_k A_k.
  \]

  \item \textbf{Previous time point.}  
  Regress the blipped-down outcome $Y^{(k-1)}$ on adherence at the
  previous time point $A_{k-1}$, again adjusting for the corresponding
  history $\mathbf H_{k-1}$ and take the coefficient of $A_{k-1}$ as $\hat\beta_{k-1}$.

  \item \textbf{Iterate backwards.}  
  Update the blipped outcome by subtracting the newly estimated effect
  of $A_{k-1}$,
  \[
    Y^{(k-2)} = Y^{(k-1)} - \hat\beta_{k-1} A_{k-1},
  \]
  and repeat step 3 for $j = k-2,k-3,\ldots,1$, each time regressing the
  current blipped outcome $Y^{(j)}$ on $A_j$ and $\mathbf H_j$ and taking
  the coefficient of $A_j$ as $\hat\beta_j$.
\end{enumerate}
At the end of this backward sequence we obtain $\hat{\bm{\beta}}^{\text{rec}}_k = (\hat\beta_1,\ldots,\hat\beta_k)$, which are the recursive G-estimates of the SNMM parameters.

\subsection{Efficient GMM estimation}\label{app:model-effG}
Fix the final outcome at time point $k$ and let $Y \equiv Y_k$.
\begin{enumerate}
\item \textbf{Residualize the outcome.}
For each $j = 1,\dots,k$, regress $Y$ on the corresponding observed history $\bm{H}_j$ and obtain the fitted value $\widehat{E}(Y \mid \bm{H}_j)$. Define the residualized outcome as
\[
\widetilde Y_j = Y - \widehat{E}(Y \mid \bm{H}_j).
\]

\item \textbf{Residualize the exposures.}
For each $j = 1,\dots,k$, and for each $s = j,\dots,k$, regress $A_s$ on $\bm{H}_j$ and obtain the fitted value $\widehat{E}(A_s \mid \bm{H}_j)$. Define the residualized exposure as
\[
\widetilde A_{j,s} = A_s - \widehat{E}(A_s \mid \bm{H}_j).
\]

\item \textbf{Construct residualized blipped-down outcomes.}
For each $j = 1,\dots,k$, construct the residualized blipped-down outcome
\[
\widetilde Y^{(j)}(\bm{\beta})
=
\widetilde Y_j
-
\sum_{s=j}^k \beta_s \widetilde A_{j,s}.
\]

\item \textbf{Efficient moment conditions.}
For each $j = 1,\dots,k$, use the residualized exposure $\widetilde A_{j,j}$ and the residulized blipp-down outcome $\widetilde Y^{(j)}(\bm{\beta})$, define the moment condition
\[
E\!\left[
\widetilde A_{j,j}\,
\widetilde Y^{(j)}(\bm{\beta}^\ast)
\right]
= 0.
\]

The corresponding sample moment is
\[
\hat m^{\mathrm{eff}}_j(\bm{\beta})
=
\frac{1}{n}\sum_{i=1}^n
\widetilde A_{i,j,j}\,
\widetilde Y^{(j)}_i(\bm{\beta}),
\qquad j = 1,\dots,k.
\]

\item \textbf{Joint GMM estimation.}
Stack the moment conditions as
\[
\hat{\bm m}^{\mathrm{eff}}(\bm{\beta})
=
\big(
\hat m^{\mathrm{eff}}_1(\bm{\beta}),
\ldots,
\hat m^{\mathrm{eff}}_k(\bm{\beta})
\big)^\top,
\]
and estimate $\bm{\beta}$ by solving the joint GMM minimization problem
\[
\hat{\bm{\beta}}^{\mathrm{eff}}
=
\arg\min_{\bm{\beta}}
\;
\hat{\bm m}^{\mathrm{eff}}(\bm{\beta})^\top
\bm{W}_{eff}
\hat{\bm m}^{\mathrm{eff}}(\bm{\beta}),
\]
where $\bm{W}_{eff}$ is a $k \times k$ positive definite weighting matrix which can be set to the identity matrix.\\
\\
The efficient GMM estimation can be implemented using a two-step procedure. In the first step, the parameter vector $\bm{\beta}$ is estimated by minimizing the GMM objective with the identity matrix as the weighting matrix, yielding a preliminary estimate $\hat{\bm{\beta}}^{(1)}$. 
Using $\hat{\bm{\beta}}^{(1)}$, define the $k$-dimensional vector
\[
\hat{\bm g}_i(\bm{\beta})
=
\begin{pmatrix}
\widetilde A_{i,1,1}\,\widetilde Y^{(1)}_i(\bm{\beta})\\
\vdots\\
\widetilde A_{i,k,k}\,\widetilde Y^{(k)}_i(\bm{\beta})
\end{pmatrix}.
\]
Note that the stacked sample moments can be written as
\[
\hat{\bm m}^{\mathrm{eff}}(\bm{\beta})
=
\frac{1}{n}\sum_{i=1}^n \hat{\bm g}_i(\bm{\beta}).
\]
Estimate the covariance matrix of the moment vector by
\[
\hat\Omega
=
\frac{1}{n}\sum_{i=1}^n
\Big(\hat{\bm g}_i(\hat{\bm{\beta}}^{(1)})-\bar{\bm g}\Big)
\Big(\hat{\bm g}_i(\hat{\bm{\beta}}^{(1)})-\bar{\bm g}\Big)^\top,
\]
where $\bar{\bm g}
=
\frac{1}{n}\sum_{i=1}^n \hat{\bm g}_i(\hat{\bm{\beta}}^{(1)})$. In words, $\hat\Omega$ is the sample variance-covariance matrix of
$\hat{\bm g}_i(\hat{\bm{\beta}}^{(1)})$. We then update the weighting matrix to $\bm{W}_{eff} = \hat{\bm{\Omega}}^{-1}$ and use $\hat{\bm{\beta}}^{(1)}$ as the initial value to solve the minimization and obtain the two-step efficient GMM estimator $\hat{\bm{\beta}}^{(2)}$.

\end{enumerate}

\subsection{Sandwich Variance Estimators}\label{app:model-inference}
All IPTW and NUC-based G-estimation estimators considered in this paper are just-identified and can be written as solutions to a system of estimating equations or equivalently, as generalized method of moments (GMM) estimators. In this setting, all estimators admit a heteroskedasticity robust (HC1) sandwich variance estimator which can be generally defined as \citep{hansen1982large,newey1994large}:
\begin{equation}\label{eq:general-sandwich}
   \widehat{\mathrm{Var}}_{\mathrm{HC1}}(\hat{\bm{\varphi}})
=
\frac{1}{n}
\hat{\bm J}^{-1}
\hat{\bm\Omega}_{\mathrm{HC1}}
\hat{\bm J}^{-\top}, 
\end{equation}
where $\hat{\bm J}$ is the Jacobian matrix of the stacked sample moment conditions evaluated at the estimated parameter vector $\hat{\bm{\varphi}}$, and $\hat{\bm\Omega}_{\mathrm{HC1}}$ is the HC1-corrected variance-covariance matrix of the moment conditions.
Specifically, let
\[
\hat{\bm m}(\bm{\varphi})
=
\frac{1}{n}\sum_{i=1}^n \bm g_i(\bm{\varphi})
\]
denote the vector of stacked sample moments.
The Jacobian matrix is
\[
\hat{\bm J}
=
\left.
\frac{\partial \hat{\bm m}(\bm{\varphi})}
{\partial \bm{\varphi}^\top}
\right|_{\bm{\varphi}=\hat{\bm{\varphi}}},
\]
and the moment variance-covariance matrix is estimated as
\[
\hat{\bm\Omega}_{\mathrm{HC0}}
=
\frac{1}{n}\sum_{i=1}^n
\big(\bm g_i(\hat{\bm{\varphi}})-\bar{\bm g}\big)
\big(\bm g_i(\hat{\bm{\varphi}})-\bar{\bm g}\big)^\top,
\qquad
\bar{\bm g}=\frac{1}{n}\sum_{i=1}^n \bm g_i(\hat{\bm{\varphi}}),
\]
with the HC1 correction
\[
\hat{\bm\Omega}_{\mathrm{HC1}}
=
\frac{n}{n-q}\,\hat{\bm\Omega}_{\mathrm{HC0}},
\]
where $q$ is the number of moment conditions.\\
\\
For the IPTW, parameters are estimated by solving the estimating equations
\[
\frac{1}{n}\sum_{i=1}^{n}
W_{i,k}\,\Tilde{A}_{i,k}
\left(
Y_{i,k}-\Tilde{A}_{i,k}^{\top}\bm{\eta}_k
\right)
= \bm{0}.
\]
In the notation of Equation~\ref{eq:general-sandwich}, the moment function is therefore
\[
\bm{g}_i(\bm{\eta}_k)
=
W_{i,k}\,\Tilde{A}_{i,k}
\left(
Y_{i,k}-\Tilde{A}_{i,k}^{\top}\bm{\eta}_k
\right),
\]
and the sandwich variance estimator is obtained by evaluating the general HC1 formula in Equation~\ref{eq:general-sandwich} using these estimating equations. In practice, the HC1 sandwich variance for the IPTW can be implemented via the \texttt{vcovHC} function in \texttt{R} with the \texttt{type = "HC1"}.
\\
\\
For the basic G-estimation procedure, define the blipped-down outcomes
\[
U_{i,j}(\bm{\beta})
=
Y_{i,k}
-
\sum_{s=j}^{k}\beta_s A_{i,s},
\qquad j=1,\dots,k.
\]
Let $R_{i,j}=A_{i,j}-E(A_{i,j}\mid \bm{H}_{i,j})$ denote the residualized exposure at time $j$. The estimating equations are then given by
\[
\frac{1}{n}\sum_{i=1}^{n}
R_{i,j}\,U_{i,j}(\bm{\beta})
=0,
\qquad j=1,\dots,k.
\]
In the notation of Equation~\ref{eq:general-sandwich}, the moment function is therefore
\[
\bm{g}_i(\bm{\beta})
=
\big(
R_{i,1}U_{i,1}(\bm{\beta}),
\;\dots,\;
R_{i,k}U_{i,k}(\bm{\beta})
\big)^{\top}.
\]
The sandwich variance estimator for $\hat{\bm{\beta}}$ is obtained by evaluating the general HC1 formula in Equation~S1 using these moment conditions. \\
\\
For efficient G-estimation, the moment conditions are constructed using residualized outcomes and residualized exposures. Let
\[
\widetilde Y_{i,j}
=
Y_{i,k}-E(Y_{i,k}\mid \bm{H}_{i,j}),
\qquad
\widetilde A_{i,j,s}
=
A_{i,s}-E(A_{i,s}\mid \bm{H}_{i,j}),
\]
and define the residualized blipped-down outcomes
\[
\widetilde U_{i,j}(\bm{\beta})
=
\widetilde Y_{i,j}
-
\sum_{s=j}^{k}\beta_s\,\widetilde A_{i,j,s},
\qquad j=1,\dots,k.
\]
The efficient estimating equations are
\[
\frac{1}{n}\sum_{i=1}^{n}
\widetilde A_{i,j,j}\,
\widetilde U_{i,j}(\bm{\beta})
=0,
\qquad j=1,\dots,k.
\]
In the notation of Equation~\ref{eq:general-sandwich}, the corresponding moment function is
\[
\bm{g}_i^{\mathrm{eff}}(\bm{\beta})
=
\big(
\widetilde A_{i,1,1}\widetilde U_{i,1}(\bm{\beta}),
\;\dots,\;
\widetilde A_{i,k,k}\widetilde U_{i,k}(\bm{\beta})
\big)^{\top}.
\]
The sandwich variance estimator is obtained by evaluating the general HC1 formula
in Equation~S1 using these efficient moment conditions.\\
\\
For IV estimation, in the two-parameter decay model with $\beta_k(j) = \beta \alpha^{t_k - t_j}$, the variance-covariance matrix of of the causal estimates $\widehat{\theta} = \left(\widehat{\beta}, \widehat{\alpha} \right)$ can be estimated with the following sandwich formula proposed in \citet{Bowden2025}:
\begin{equation*}\label{eq:sandwich}
    \text{Var}(\widehat{\theta}) = \frac{1}{n} G_{\text{Gen}}^{-1}(\widehat{\theta}) \, \text{Var}[S_i(\widehat{\theta})] \, G_{\text{Gen}}^{-1}(\widehat{\theta})'\text{,}
\end{equation*}
where
\begin{itemize}
    \item \( \text{Var}[S_i(\widehat{\theta})] \) is the sample variance of the individual scores $S_i(\widehat{\theta})$ evaluated at $\widehat{\theta}$.
    \item \( G_{\text{Gen}}(\widehat{\theta}) \) is the sample average of the gradient matrix, where each row corresponds to the derivative of a score component with respect to each element of \( \theta \).
    \item The outer terms represent the Moore-Penrose generalized inverse of \( G_{\text{Gen}}(\widehat{\theta}) \) evaluated at \( \hat{\theta} \).
\end{itemize}

\section{Simulation Studies}\label{app:simulation}
\subsection{Parameter Specification}\label{app:simulation-parameter}
We derive the direct causal effect $\pi_{k}(j)$ given the total effect $\beta_{k}(j)$, based on the following data generation process:
\begin{equation}\label{eq:app-simu_a_1}
    A_{i, k} = \gamma_G G_{i} + \eta_A A_{i, k-1} + \eta_Y Y_{i, k-1} + \eta_{1} F1_{i} + \eta_{2} F2_{i} + \eta_V V_{i, k} + \xi_{i, k}.
\end{equation}
\begin{equation}\label{eq:app-simu_y_1}
    Y_{i, k} = \sum_{j = 1}^{k} \pi_{k}(j) A_{i, j} + \beta_Y Y_{i, k-1} + \beta_{F1} F1_{i} + \beta_{F2} F2_{i} + \beta_V V_{i, k} + \epsilon_{i, k}.
\end{equation}
It follows directly that for contemporaneous effects, we have $\pi_1(1) = \beta_1(1)$, $\pi_2(2) = \beta_2(2)$ and $\pi_3(3) = \beta_3(3)$ as there are no indirect effects in this case.\\
\\
For time point $k = 2$, we have
\begin{align*}
    Y_{i, 2} &= \beta_2(2) A_{i, 2} + \pi_2(1) A_{i, 1} + \beta_Y Y_{i, 1} + \beta_{F1} F1_{i} + \beta_{F2} F2_{i} + \beta_V V_{i, 2} + \epsilon_{i, 2}\\
    & = \beta_2(2) A_{i, 2} + \pi_2(1) A_{i, 1} + \beta_Y \left(\beta_1(1) A_{i, 1} + \beta_{F1} F1_{i} + \beta_{F2} F2_{i} + \beta_V V_{i, 1} + \epsilon_{i, 1}\right)\\
    &+ \beta_{F1} F1_{i} + \beta_{F2} F2_{i} + \beta_V V_{i, 2} + \epsilon_{i, 2}\\
     & = \beta_2(2) A_{i, 2} + \pi_2(1) A_{i, 1} + \beta_Y \left(\beta_1(1) A_{i, 1} + \beta_{F1} F1_{i} + \beta_{F2} F2_{i} + \beta_V V_{i, 1} + \epsilon_{i, 1}\right)\\
    &+ \beta_{F1} F1_{i} + \beta_{F2} F2_{i} + \beta_V (BV_{i, 2} + \tau A_{i, 1}) + \epsilon_{i, 2}\\
    &= \beta_2(2) A_{i, 2} + \left(\pi_2(1) + \beta_Y\beta_1(1) + \beta_V\tau\right)A_{i, 1} \\
    & + \beta_{F1}\left(\beta_Y + 1\right)F1_{i} + \beta_{F2}\left(\beta_Y + 1\right)F2_{i} + \beta_Y\beta_{V}V_{i, 1} + \beta_V BV_{i,2} + e_{i,2}
\end{align*}
where the random error term $e_{i,2}$ is implicitly defined with $\epsilon_{i, 1}$, $\epsilon_{i, 2}$. Given the total effect $\beta_2(1)$, for example, $\beta_2(1) = \beta \alpha$, we impose
\begin{equation*}
    \pi_2(1) = \beta_2(1) - \beta_Y \beta_1(1) - \beta_V\tau.
\end{equation*}
Similarly, for time point $t = 3$, we have
\begin{equation*}
    \pi_3(2) = \beta_3(2) - \beta_Y \beta_2(2) - \beta_V\tau
\end{equation*}
and
\begin{equation*}
    \pi_3(1) = \beta_3(1) - \beta_Y \pi_2(1) - \beta_Y^2 \beta_1(1) - \beta_V \beta_Y \tau.
\end{equation*}

\subsection{Simulation Results}\label{app:simulation-results}
Simulation results are presented in Table~\ref{tab:d1_tau08_ar098_acc} to Table~\ref{tab:iv_d4_tau0_ar098}. In terms of estimation accuracy, we report the mean absolute error (MAE) and the root mean squared error (RMSE), defined as follows: Suppose $R$ Monte Carlo replications are performed, indexed by $r=1,\dots,R$.
Let $\hat{\beta}^{(r)}$ denote the estimate obtained from replication $r$, and let $\beta$ denote the true parameter value. The MAE is defined as
\[
\mathrm{MAE}
=
\frac{1}{R}\sum_{r=1}^{R}
\left|
\hat{\beta}^{(r)} - \beta
\right|.
\]
The RMSE is defined as
\[
\mathrm{RMSE}
=
\left(
\frac{1}{R}\sum_{r=1}^{R}
\left(\hat{\beta}^{(r)} - \beta\right)^2
\right)^{1/2}.
\]
To understand the finite-sample behavior of IPTW under strong exposure-confounder feedback, we examined diagnostics for \emph{practical} violations of the positivity assumption. In particular, we summarized the effective sample size (ESS) of the cumulative inverse probability weights, the concentration of total weight mass among the largest weights, and the upper tail of the weight distribution. The ESS, defined as
\[
\left(\sum_i w_i\right)^2 \Big/ \sum_i w_i^2,
\]
provides a measure of the amount of information retained after weighting and is widely used to quantify efficiency loss and instability induced by highly variable weights in marginal structural models \citep{kish1965survey, shook2022power}. 
Weight concentration was quantified as the proportion of total weight carried by the largest 1\% of observations ordered by weights, which offers an intuitive assessment of whether a small subset of individuals dominates the weighted pseudo-population \citep{lee2011weight,cole2008constructing}.
Finally, the 99.9th percentile of the weight distribution was used to characterize the heaviness of the upper tail, which is known to drive finite-sample instability and sensitivity of IPTW estimation to a small number of influential observations \citep{austin2015moving, cole2008constructing}.\\
\\
In our simulation design, the identification assumptions of no unmeasured confounding and consistency are satisfied, and the positivity assumption holds in theory. However, under strong exposure-confounder feedback ($\tau_v = 0.8$) and high covariate persistence (AR(1) $= 0.98$), the conditional adherence density becomes extremely small for certain histories, leading to practical positivity problems in finite samples. This is reflected empirically by a substantial reduction in ESS for the time point 3 cumulative weights, pronounced concentration of total weight mass among a small number of observations, and heavy-tailed weight distributions. As either the feedback strength or covariate persistence was reduced, or when a less restrictive covariate-balancing procedure was used, the ESS increased and weight concentration decreased correspondingly. See Table~\ref{tab:d1_iptw_sensitivity_weights}.\\
\\
These patterns closely mirrored the observed finite-sample bias and increased RMSE of IPTW for lagged treatment effects (see Table~\ref{tab:d1_iptw_sensitivity_acc}), indicating that the  degraded performance of IPTW in these settings is driven by unstable weights arising from practical positivity problems rather than by misspecification of the causal estimand. In contrast, G-estimation remained stable across all configurations considered, reflecting its reduced sensitivity to near-violation of positivity. Together, these results highlight that even when identification assumptions are satisfied, IPTW can exhibit substantial finite-sample bias in longitudinal settings with strong exposure--confounder feedback, whereas G-estimation may offer improved robustness in such scenarios.

\begin{table}[!htbp]
\centering
\caption{Design 1 ($\tau_v=0.8$, AR(1)=0.98): estimation accuracy for IPTW and G-estimation (Monte Carlo mean over 1{,}000 replications).}
\label{tab:d1_tau08_ar098_acc}
\begin{tabular}{lrrrrrrr}
\toprule
 & \multicolumn{1}{c}{Truth} & \multicolumn{3}{c}{IPTW (CBPS, exact)} & \multicolumn{3}{c}{G-estimation (GMM)} \\
\cmidrule(lr){2-2}\cmidrule(lr){3-5}\cmidrule(lr){6-8}
Estimand & Truth & Estimate & MAE & RMSE & Estimate & MAE & RMSE \\
\midrule
$\beta_1(1)$ & -1.100 & -1.100 & 0.018 & 0.023 & -1.100 & 0.011 & 0.014 \\
$\beta_2(1)$ & -1.045 & -1.073 & 0.058 & 0.075 & -1.046 & 0.013 & 0.017 \\
$\beta_2(2)$ & -1.100 & -1.096 & 0.041 & 0.055 & -1.099 & 0.011 & 0.014 \\
$\beta_3(1)$ & -0.993 & -1.032 & 0.092 & 0.118 & -0.994 & 0.015 & 0.018 \\
$\beta_3(2)$ & -1.045 & -1.061 & 0.089 & 0.119 & -1.044 & 0.014 & 0.018 \\
$\beta_3(3)$ & -1.100 & -1.101 & 0.073 & 0.099 & -1.100 & 0.011 & 0.014 \\
\bottomrule
\end{tabular}
\end{table}

\begin{table}[!htbp]
\centering
\caption{Design 1 ($\tau_v=0.8$, AR(1)=0.98): inferential performance. EmpSD is Monte Carlo SD of estimates; SE columns are averages across replications; coverage is empirical coverage of 95\% CIs.}
\label{tab:d1_tau08_ar098_inf_compact}

\begin{tabular}{lrrrrrrr}
\toprule
 & \multicolumn{3}{c}{IPTW (CBPS, exact)} & \multicolumn{2}{c}{GMM (basic)} & \multicolumn{2}{c}{GMM (efficient)} \\
\cmidrule(lr){2-4}\cmidrule(lr){5-6}\cmidrule(lr){7-8}
Estimand & EmpSD & Reg SE & Sand SE & EmpSD & Sand SE & EmpSD & Sand SE \\
\midrule
$\beta_1(1)$ & 0.023 & 0.015 & 0.028 & 0.014 & 0.017 & 0.014 & 0.014 \\
$\beta_2(1)$ & 0.069 & 0.018 & 0.063 & 0.017 & 0.027 & 0.017 & 0.017 \\
$\beta_2(2)$ & 0.055 & 0.017 & 0.058 & 0.014 & 0.030 & 0.014 & 0.014 \\
$\beta_3(1)$ & 0.111 & 0.020 & 0.086 & 0.018 & 0.030 & 0.018 & 0.018 \\
$\beta_3(2)$ & 0.118 & 0.018 & 0.086 & 0.018 & 0.040 & 0.018 & 0.017 \\
$\beta_3(3)$ & 0.099 & 0.018 & 0.077 & 0.014 & 0.046 & 0.014 & 0.014 \\
\bottomrule
\end{tabular}

\vspace{0.6em}

\begin{tabular}{lrrrrr}
\toprule
 & \multicolumn{2}{c}{IPTW (CBPS, exact)} & \multicolumn{1}{c}{GMM (basic)} & \multicolumn{1}{c}{GMM (efficient)} \\
\cmidrule(lr){2-3}\cmidrule(lr){4-4}\cmidrule(lr){5-5}
Estimand & Cov(Reg) & Cov(Sand) & Cov & Cov \\
\midrule
$\beta_1(1)$ & 0.797 & 0.973 & 0.989 & 0.955 \\
$\beta_2(1)$ & 0.378 & 0.908 & 1.000 & 0.956 \\
$\beta_2(2)$ & 0.520 & 0.962 & 1.000 & 0.937 \\
$\beta_3(1)$ & 0.278 & 0.847 & 0.999 & 0.947 \\
$\beta_3(2)$ & 0.282 & 0.883 & 1.000 & 0.945 \\
$\beta_3(3)$ & 0.326 & 0.900 & 1.000 & 0.937 \\
\bottomrule
\end{tabular}

\end{table}

\begin{table}[!htbp]
\centering
\caption{Design 1 ($\tau_v=0$, AR(1)=0.98): estimation accuracy for IPTW and G-estimation (Monte Carlo mean over 1{,}000 replications).}
\label{tab:d1_tau0_ar098_acc}
\begin{tabular}{lrrrrrrr}
\toprule
 & \multicolumn{1}{c}{Truth} & \multicolumn{3}{c}{IPTW (CBPS, exact)} & \multicolumn{3}{c}{G-estimation (GMM)} \\
\cmidrule(lr){2-2}\cmidrule(lr){3-5}\cmidrule(lr){6-8}
Estimand & Truth & Estimate & MAE & RMSE & Estimate & MAE & RMSE \\
\midrule
$\beta_1(1)$ & -1.100 & -1.100 & 0.018 & 0.023 & -1.100 & 0.011 & 0.014 \\
$\beta_2(1)$ & -1.045 & -1.049 & 0.029 & 0.039 & -1.045 & 0.012 & 0.015 \\
$\beta_2(2)$ & -1.100 & -1.100 & 0.029 & 0.039 & -1.099 & 0.011 & 0.014 \\
$\beta_3(1)$ & -0.993 & -0.996 & 0.043 & 0.058 & -0.994 & 0.013 & 0.016 \\
$\beta_3(2)$ & -1.045 & -1.048 & 0.042 & 0.056 & -1.044 & 0.013 & 0.016 \\
$\beta_3(3)$ & -1.100 & -1.098 & 0.041 & 0.060 & -1.100 & 0.011 & 0.014 \\
\bottomrule
\end{tabular}
\end{table}

\begin{table}[!htbp]
\centering
\caption{Design 1 ($\tau_v=0$, AR(1)=0.98): inferential performance. EmpSD is Monte Carlo SD of estimates; SE columns are averages of estimated standard errors; coverage is the empirical coverage of 95\% CIs.}
\label{tab:d1_tau0_ar098_inf_compact}

\begin{tabular}{lrrrrrrr}
\toprule
 & \multicolumn{3}{c}{IPTW (CBPS, exact)} & \multicolumn{2}{c}{GMM (basic)} & \multicolumn{2}{c}{GMM (efficient)} \\
\cmidrule(lr){2-4}\cmidrule(lr){5-6}\cmidrule(lr){7-8}
Estimand & EmpSD & Reg SE & Sand SE & EmpSD & Sand SE & EmpSD & Sand SE \\
\midrule
$\beta_1(1)$ & 0.023 & 0.015 & 0.028 & 0.014 & 0.017 & 0.014 & 0.014 \\
$\beta_2(1)$ & 0.039 & 0.018 & 0.042 & 0.015 & 0.022 & 0.015 & 0.015 \\
$\beta_2(2)$ & 0.039 & 0.018 & 0.042 & 0.014 & 0.030 & 0.014 & 0.014 \\
$\beta_3(1)$ & 0.058 & 0.020 & 0.053 & 0.016 & 0.024 & 0.016 & 0.016 \\
$\beta_3(2)$ & 0.055 & 0.020 & 0.052 & 0.016 & 0.032 & 0.016 & 0.015 \\
$\beta_3(3)$ & 0.060 & 0.021 & 0.052 & 0.014 & 0.041 & 0.014 & 0.014 \\
\bottomrule
\end{tabular}

\vspace{0.6em}

\begin{tabular}{lrrrr}
\toprule
 & \multicolumn{2}{c}{IPTW (CBPS, exact)} & \multicolumn{1}{c}{GMM (basic)} & \multicolumn{1}{c}{GMM (efficient)} \\
\cmidrule(lr){2-3}\cmidrule(lr){4-4}\cmidrule(lr){5-5}
Estimand & Cov(Reg) & Cov(Sand) & Cov & Cov \\
\midrule
$\beta_1(1)$ & 0.797 & 0.973 & 0.989 & 0.955 \\
$\beta_2(1)$ & 0.697 & 0.982 & 0.997 & 0.962 \\
$\beta_2(2)$ & 0.693 & 0.971 & 1.000 & 0.937 \\
$\beta_3(1)$ & 0.563 & 0.959 & 0.997 & 0.947 \\
$\beta_3(2)$ & 0.563 & 0.955 & 0.999 & 0.940 \\
$\beta_3(3)$ & 0.640 & 0.949 & 1.000 & 0.937 \\
\bottomrule
\end{tabular}

\end{table}

\begin{table}[!htbp]
\centering
\caption{Design 1 (IPTW sensitivity analysis): estimation accuracy across alternative settings. Panel A reports Monte Carlo mean estimates. Panel B reports MAE/RMSE, where each entry is \textit{MAE}/\textit{RMSE}. All summaries are averaged over 1{,}000 Monte Carlo replications.}
\label{tab:d1_iptw_sensitivity_acc}

\begin{tabular}{lrrrrrr}
\toprule
 & \multicolumn{1}{c}{Truth} 
 & \multicolumn{1}{c}{$\tau_v=0$} 
 & \multicolumn{1}{c}{$\tau_v=0.2$} 
 & \multicolumn{1}{c}{$\tau_v=0.8$} 
 & \multicolumn{1}{c}{$\tau_v=0.8$} 
 & \multicolumn{1}{c}{$\tau_v=0.8$} \\
 &  
 & \multicolumn{1}{c}{AR(1)=0.98} 
 & \multicolumn{1}{c}{AR(1)=0.98} 
 & \multicolumn{1}{c}{AR(1)=0.98} 
 & \multicolumn{1}{c}{AR(1)=0.1} 
 & \multicolumn{1}{c}{AR(1)=0.98 (over)} \\
\midrule
\multicolumn{7}{l}{\textbf{Panel A: Monte Carlo mean of point estimates}}\\
\midrule
$\beta_1(1)$ & -1.100 & -1.100 & -1.100 & -1.100 & -1.100 & -1.107 \\
$\beta_2(1)$ & -1.045 & -1.049 & -1.051 & -1.073 & -1.041 & -1.053 \\
$\beta_2(2)$ & -1.100 & -1.100 & -1.098 & -1.096 & -1.101 & -1.109 \\
$\beta_3(1)$ & -0.993 & -0.996 & -0.999 & -1.032 & -0.985 & -1.005 \\
$\beta_3(2)$ & -1.045 & -1.048 & -1.048 & -1.061 & -1.031 & -1.052 \\
$\beta_3(3)$ & -1.100 & -1.098 & -1.098 & -1.101 & -1.100 & -1.110 \\
\midrule
\multicolumn{7}{l}{\textbf{Panel B: MAE/RMSE}}\\
\midrule
$\beta_1(1)$ &  & 0.018/0.023 & 0.018/0.023 & 0.018/0.023 & 0.013/0.016 & 0.021/0.026 \\
$\beta_2(1)$ &  & 0.029/0.039 & 0.031/0.041 & 0.058/0.075 & 0.032/0.042 & 0.052/0.071 \\
$\beta_2(2)$ &  & 0.029/0.039 & 0.031/0.042 & 0.041/0.055 & 0.024/0.031 & 0.050/0.065 \\
$\beta_3(1)$ &  & 0.043/0.058 & 0.048/0.063 & 0.092/0.118 & 0.050/0.069 & 0.076/0.101 \\
$\beta_3(2)$ &  & 0.042/0.056 & 0.048/0.062 & 0.089/0.119 & 0.053/0.071 & 0.086/0.115 \\
$\beta_3(3)$ &  & 0.041/0.060 & 0.045/0.065 & 0.073/0.099 & 0.039/0.051 & 0.069/0.092 \\
\bottomrule
\end{tabular}

\end{table}

\begin{table}[!htbp]
\centering
\caption{Design 1 (IPTW sensitivity analysis): diagnostics of cumulative inverse probability weights at time 3. Entries are median [Q1, Q3] across 1{,}000 Monte Carlo replications.}
\label{tab:d1_iptw_sensitivity_weights}

\begin{tabular}{lrrr}
\toprule
Scenario & ESS (median [IQR]) & Top 1\% share (median [IQR]) & $q_{0.999}$ (median [IQR]) \\
\midrule
$\tau_v=0$, AR(1)=0.98, exact       
& 1126.3 [811.3, 1384.3] 
& 0.133 [0.122, 0.147] 
& 20.893 [18.461, 24.003] \\
$\tau_v=0.2$, AR(1)=0.98, exact     
& 917.2 [638.6, 1129.7] 
& 0.152 [0.140, 0.168] 
& 24.353 [21.485, 27.809] \\
$\tau_v=0.8$, AR(1)=0.98, exact     
& 246.3 [138.2, 372.9] 
& 0.282 [0.254, 0.320] 
& 47.510 [41.054, 55.544] \\
$\tau_v=0.8$, AR(1)=0.1, exact      
& 580.6 [369.7, 791.8] 
& 0.188 [0.170, 0.210] 
& 30.396 [26.347, 35.272] \\
$\tau_v=0.8$, AR(1)=0.98, over      
& 376.9 [206.2, 544.7] 
& 0.229 [0.206, 0.262] 
& 38.360 [32.766, 45.069] \\
\bottomrule
\end{tabular}

\end{table}

\begin{table}[!htbp]
\centering
\caption{Design 2 ($\tau_v=0.8$, AR(1)=0.98): estimation accuracy for IPTW and G-estimation (Monte Carlo mean over 1{,}000 replications).}
\label{tab:d2_tau08_ar098_acc}
\begin{tabular}{lrrrrrrr}
\toprule
 & \multicolumn{1}{c}{Truth} & \multicolumn{3}{c}{IPTW (CBPS, exact)} & \multicolumn{3}{c}{G-estimation (GMM)} \\
\cmidrule(lr){2-2}\cmidrule(lr){3-5}\cmidrule(lr){6-8}
Estimand & Truth & Estimate & MAE & RMSE & Estimate & MAE & RMSE \\
\midrule
$\beta_1(1)$ & -1.100 & -1.028 & 0.072 & 0.075 & -1.028 & 0.072 & 0.073 \\
$\beta_2(1)$ & -1.045 & -1.008 & 0.063 & 0.084 & -0.972 & 0.073 & 0.075 \\
$\beta_2(2)$ & -1.100 & -1.039 & 0.068 & 0.082 & -1.043 & 0.057 & 0.059 \\
$\beta_3(1)$ & -0.993 & -0.973 & 0.088 & 0.116 & -0.918 & 0.075 & 0.078 \\
$\beta_3(2)$ & -1.045 & -1.008 & 0.098 & 0.129 & -0.987 & 0.058 & 0.061 \\
$\beta_3(3)$ & -1.100 & -1.056 & 0.082 & 0.108 & -1.055 & 0.045 & 0.048 \\
\bottomrule
\end{tabular}
\end{table}

\begin{table}[!htbp]
\centering
\caption{Design 2 ($\tau_v=0.8$, AR(1)=0.98): inferential performance. EmpSD is Monte Carlo SD of estimates; SE columns are averages of estimated standard errors; coverage is the empirical coverage of 95\% CIs.}
\label{tab:d2_tau08_ar098_inf_compact}

\begin{tabular}{lrrrrrrr}
\toprule
 & \multicolumn{3}{c}{IPTW (CBPS, exact)} & \multicolumn{2}{c}{GMM (basic)} & \multicolumn{2}{c}{GMM (efficient)} \\
\cmidrule(lr){2-4}\cmidrule(lr){5-6}\cmidrule(lr){7-8}
Estimand & EmpSD & Reg SE & Sand SE & EmpSD & Sand SE & EmpSD & Sand SE \\
\midrule
$\beta_1(1)$ & 0.023 & 0.015 & 0.028 & 0.014 & 0.017 & 0.014 & 0.014 \\
$\beta_2(1)$ & 0.075 & 0.018 & 0.066 & 0.019 & 0.028 & 0.019 & 0.019 \\
$\beta_2(2)$ & 0.055 & 0.017 & 0.058 & 0.015 & 0.030 & 0.015 & 0.014 \\
$\beta_3(1)$ & 0.115 & 0.020 & 0.088 & 0.021 & 0.026 & 0.021 & 0.021 \\
$\beta_3(2)$ & 0.124 & 0.018 & 0.088 & 0.019 & 0.033 & 0.019 & 0.019 \\
$\beta_3(3)$ & 0.099 & 0.019 & 0.078 & 0.015 & 0.033 & 0.021 & 0.020 \\
\bottomrule
\end{tabular}

\vspace{0.6em}

\begin{tabular}{lrrrr}
\toprule
 & \multicolumn{2}{c}{IPTW (CBPS, exact)} & \multicolumn{1}{c}{GMM (basic)} & \multicolumn{1}{c}{GMM (efficient)} \\
\cmidrule(lr){2-3}\cmidrule(lr){4-4}\cmidrule(lr){5-5}
Estimand & Cov(Reg) & Cov(Sand) & Cov & Cov \\
\midrule
$\beta_1(1)$ & 0.030 & 0.201 & 0.003 & 0.002 \\
$\beta_2(1)$ & 0.367 & 0.902 & 0.185 & 0.032 \\
$\beta_2(2)$ & 0.251 & 0.797 & 0.568 & 0.028 \\
$\beta_3(1)$ & 0.307 & 0.895 & 0.227 & 0.054 \\
$\beta_3(2)$ & 0.262 & 0.845 & 0.899 & 0.129 \\
$\beta_3(3)$ & 0.302 & 0.866 & 0.999 & 0.124 \\
\bottomrule
\end{tabular}

\end{table}

\begin{table}[!htbp]
\centering
\caption{Design 2 ($\tau_v=0$, AR(1)=0.98): estimation accuracy for IPTW and G-estimation (Monte Carlo mean over 1{,}000 replications).}
\label{tab:d2_tau0_ar098_acc}
\begin{tabular}{lrrrrrrr}
\toprule
 & \multicolumn{1}{c}{Truth} & \multicolumn{3}{c}{IPTW (CBPS, exact)} & \multicolumn{3}{c}{G-estimation (GMM)} \\
\cmidrule(lr){2-2}\cmidrule(lr){3-5}\cmidrule(lr){6-8}
Estimand & Truth & Estimate & MAE & RMSE & Estimate & MAE & RMSE \\
\midrule
$\beta_1(1)$ & -1.100 & -1.028 & 0.072 & 0.075 & -1.028 & 0.072 & 0.073 \\
$\beta_2(1)$ & -1.045 & -0.957 & 0.086 & 0.093 & -0.950 & 0.095 & 0.097 \\
$\beta_2(2)$ & -1.100 & -1.046 & 0.057 & 0.065 & -1.043 & 0.057 & 0.059 \\
$\beta_3(1)$ & -0.993 & -0.893 & 0.101 & 0.113 & -0.886 & 0.107 & 0.109 \\
$\beta_3(2)$ & -1.045 & -0.980 & 0.071 & 0.084 & -0.969 & 0.076 & 0.078 \\
$\beta_3(3)$ & -1.100 & -1.052 & 0.059 & 0.072 & -1.055 & 0.045 & 0.048 \\
\bottomrule
\end{tabular}
\end{table}

\begin{table}[!htbp]
\centering
\caption{Design 2 ($\tau_v=0$, AR(1)=0.98): inferential performance. EmpSD is Monte Carlo SD of estimates; SE columns are averages of estimated standard errors; coverage is the empirical coverage of 95\% CIs.}
\label{tab:d2_tau0_ar098_inf_compact}

\begin{tabular}{lrrrrrrr}
\toprule
 & \multicolumn{3}{c}{IPTW (CBPS, exact)} & \multicolumn{2}{c}{GMM (basic)} & \multicolumn{2}{c}{GMM (efficient)} \\
\cmidrule(lr){2-4}\cmidrule(lr){5-6}\cmidrule(lr){7-8}
Estimand & EmpSD & Reg SE & Sand SE & EmpSD & Sand SE & EmpSD & Sand SE \\
\midrule
$\beta_1(1)$ & 0.023 & 0.015 & 0.028 & 0.014 & 0.017 & 0.014 & 0.014 \\
$\beta_2(1)$ & 0.036 & 0.019 & 0.040 & 0.017 & 0.022 & 0.017 & 0.017 \\
$\beta_2(2)$ & 0.036 & 0.019 & 0.041 & 0.015 & 0.030 & 0.015 & 0.014 \\
$\beta_3(1)$ & 0.053 & 0.021 & 0.053 & 0.019 & 0.025 & 0.019 & 0.018 \\
$\beta_3(2)$ & 0.054 & 0.021 & 0.052 & 0.017 & 0.032 & 0.017 & 0.017 \\
$\beta_3(3)$ & 0.052 & 0.021 & 0.050 & 0.015 & 0.041 & 0.015 & 0.014 \\
\bottomrule
\end{tabular}

\vspace{0.6em}

\begin{tabular}{lrrrr}
\toprule
 & \multicolumn{2}{c}{IPTW (CBPS, exact)} & \multicolumn{1}{c}{GMM (basic)} & \multicolumn{1}{c}{GMM (efficient)} \\
\cmidrule(lr){2-3}\cmidrule(lr){4-4}\cmidrule(lr){5-5}
Estimand & Cov(Reg) & Cov(Sand) & Cov & Cov \\
\midrule
$\beta_1(1)$ & 0.029 & 0.205 & 0.003 & 0.002 \\
$\beta_2(1)$ & 0.061 & 0.361 & 0.000 & 0.000 \\
$\beta_2(2)$ & 0.332 & 0.768 & 0.554 & 0.028 \\
$\beta_3(1)$ & 0.096 & 0.502 & 0.000 & 0.000 \\
$\beta_3(2)$ & 0.259 & 0.738 & 0.231 & 0.007 \\
$\beta_3(3)$ & 0.407 & 0.835 & 0.990 & 0.124 \\
\bottomrule
\end{tabular}

\end{table}

\begin{table}[!htbp]
\centering
\caption{Design 3 ($\tau_v=0$, AR(1)=0.98): estimation accuracy for IPTW and G-estimation (Monte Carlo mean over 1{,}000 replications).}
\label{tab:d4_tau0_ar098_acc}
\begin{tabular}{lrrrrrrr}
\toprule
 & \multicolumn{1}{c}{Truth} & \multicolumn{3}{c}{IPTW (CBPS, exact)} & \multicolumn{3}{c}{G-estimation (GMM)} \\
\cmidrule(lr){2-2}\cmidrule(lr){3-5}\cmidrule(lr){6-8}
Estimand & Truth & Estimate & MAE & RMSE & Estimate & MAE & RMSE \\
\midrule
$\beta_1(1)$ & -1.100 & -1.100 & 0.018 & 0.023 & -1.100 & 0.011 & 0.014 \\
$\beta_2(1)$ & -0.600 & -0.604 & 0.029 & 0.039 & -0.600 & 0.012 & 0.015 \\
$\beta_2(2)$ & -1.050 & -1.050 & 0.029 & 0.039 & -1.049 & 0.011 & 0.014 \\
$\beta_3(1)$ & -0.200 & -0.203 & 0.042 & 0.056 & -0.201 & 0.013 & 0.016 \\
$\beta_3(2)$ & -0.550 & -0.554 & 0.041 & 0.055 & -0.549 & 0.013 & 0.016 \\
$\beta_3(3)$ & -1.000 & -0.998 & 0.041 & 0.060 & -1.000 & 0.011 & 0.014 \\
\bottomrule
\end{tabular}
\end{table}

\begin{table}[!htbp]
\centering
\caption{Design 3 ($\tau_v=0$, AR(1)=0.98): inferential performance. EmpSD is Monte Carlo SD of estimates; SE columns are averages of estimated standard errors; coverage is the empirical coverage of nominal 95\% CIs.}
\label{tab:d4_tau0_ar098_inf_compact}

\begin{tabular}{lrrrrrrr}
\toprule
 & \multicolumn{3}{c}{IPTW (CBPS, exact)} & \multicolumn{2}{c}{GMM (basic)} & \multicolumn{2}{c}{GMM (efficient)} \\
\cmidrule(lr){2-4}\cmidrule(lr){5-6}\cmidrule(lr){7-8}
Estimand & EmpSD & Reg SE & Sand SE & EmpSD & Sand SE & EmpSD & Sand SE \\
\midrule
$\beta_1(1)$ & 0.023 & 0.015 & 0.028 & 0.014 & 0.017 & 0.014 & 0.014 \\
$\beta_2(1)$ & 0.039 & 0.018 & 0.042 & 0.015 & 0.022 & 0.015 & 0.015 \\
$\beta_2(2)$ & 0.039 & 0.018 & 0.042 & 0.014 & 0.025 & 0.014 & 0.014 \\
$\beta_3(1)$ & 0.056 & 0.020 & 0.053 & 0.016 & 0.024 & 0.016 & 0.016 \\
$\beta_3(2)$ & 0.055 & 0.020 & 0.052 & 0.016 & 0.025 & 0.016 & 0.016 \\
$\beta_3(3)$ & 0.060 & 0.020 & 0.052 & 0.014 & 0.028 & 0.014 & 0.014 \\
\bottomrule
\end{tabular}

\vspace{0.6em}

\begin{tabular}{lrrrr}
\toprule
 & \multicolumn{2}{c}{IPTW (CBPS, exact)} & \multicolumn{1}{c}{GMM (basic)} & \multicolumn{1}{c}{GMM (efficient)} \\
\cmidrule(lr){2-3}\cmidrule(lr){4-4}\cmidrule(lr){5-5}
Estimand & Cov(Reg) & Cov(Sand) & Cov & Cov \\
\midrule
$\beta_1(1)$ & 0.797 & 0.973 & 0.989 & 0.955 \\
$\beta_2(1)$ & 0.697 & 0.982 & 0.997 & 0.962 \\
$\beta_2(2)$ & 0.693 & 0.971 & 1.000 & 0.937 \\
$\beta_3(1)$ & 0.579 & 0.965 & 0.997 & 0.945 \\
$\beta_3(2)$ & 0.559 & 0.956 & 0.999 & 0.940 \\
$\beta_3(3)$ & 0.624 & 0.948 & 1.000 & 0.937 \\
\bottomrule
\end{tabular}

\end{table}

\begin{table}[!htbp]
\centering
\caption{IV results under $\tau_v=0.8$ and AR(1)=0.98. Columns correspond to Design~1 and Design~2 under three specifications: baseline, weak instrument ($\gamma_G=0.05$), and an incorrectly adjusted model that additionally controls for time-varying covariates. Results are summarized over 1{,}000 replications.}
\label{tab:iv_d1_d2_tau08_ar098}

% -------------------- Panel A: Accuracy --------------------
\begin{tabular}{lrrrrrrrr}
\toprule
 & \multicolumn{2}{c}{Design 1} & \multicolumn{2}{c}{Design 2 (baseline)} & \multicolumn{2}{c}{Design 2 (weak IV)} & \multicolumn{2}{c}{Design 2 (wrong adj.)} \\
\cmidrule(lr){2-3}\cmidrule(lr){4-5}\cmidrule(lr){6-7}\cmidrule(lr){8-9}
 & $\beta$ & $\alpha$ & $\beta$ & $\alpha$ & $\beta$ & $\alpha$ & $\beta$ & $\alpha$ \\
\midrule
Truth & -1.100 & 0.950 & -1.100 & 0.950 & -1.100 & 0.950 & -1.100 & 0.950 \\
\midrule
\multicolumn{9}{l}{\textbf{Estimation accuracy}} \\
Estimate & -1.099 & 0.953 & -1.098 & 0.953 & -1.146 & 2.709 & -0.976 & 0.520 \\
MAE      &  0.037 & 0.033 &  0.039 & 0.034 &  0.464 & 2.497 &  0.125 & 0.430 \\
RMSE     &  0.046 & 0.043 &  0.049 & 0.043 &  0.828 & 15.313 & 0.137 & 0.434 \\
\midrule
\multicolumn{9}{l}{\textbf{Inferential performance}} \\
Empirical SD & 0.046 & 0.043 & 0.049 & 0.043 & 0.827 & 15.219 & 0.057 & 0.059 \\
Sandwich SE  & 0.045 & 0.041 & 0.048 & 0.042 & 1.154 & 143.323 & 0.052 & 0.053 \\
CI Coverage  & 0.948 & 0.951 & 0.944 & 0.944 & 0.967 & 0.986 & 0.352 & 0.000 \\
\bottomrule
\end{tabular}

\end{table}

\begin{table}[!htbp]
\centering
\caption{IV results for Design 3 ($\tau_v=0$, AR(1)=0.98). The decay restriction is violated in the DGP, so $(\beta,\alpha)$ are reported as \emph{working-model} parameters (no truth-based metrics). We additionally report implied cumulative effects $C_2=\beta(1+\alpha)$ and $C_3=\beta(1+\alpha+\alpha^2)$, for which truth is well-defined from the six total effects. Summaries are over 1{,}000 replications.}
\label{tab:iv_d4_tau0_ar098}

% -------------------- Panel A: Working-model parameters --------------------
\begin{tabular}{lrr}
\toprule
 & $\beta$ & $\alpha$ \\
\midrule
% \multicolumn{3}{l}{\textbf{One-step estimation}} \\
Estimate       & -1.130 & 0.408 \\
Empirical SD   & 0.058  & 0.041 \\
Sandwich SE    & 0.057  & 0.042 \\
\bottomrule
\end{tabular}

\vspace{0.9em}

% -------------------- Panel B: Implied cumulative effects --------------------
\begin{tabular}{lrrrrr}
\toprule
 & \multicolumn{1}{c}{Truth} & \multicolumn{1}{c}{Estimate} & \multicolumn{1}{c}{MAE} & \multicolumn{1}{c}{RMSE} & \multicolumn{1}{c}{EmpSD} \\
\midrule
\multicolumn{6}{l}{\textbf{Implied cumulative effects}} \\
$C_2=\beta(1+\alpha)$                 & -1.650 & -1.590 & 0.070 & 0.086 & 0.061 \\
$C_3=\beta(1+\alpha+\alpha^2)$        & -1.750 & -1.779 & 0.062 & 0.077 & 0.071 \\
\bottomrule
\end{tabular}

\end{table}

\section{Further Details of the Applied Example}\label{app:application}

\subsection{Validity of the Genetic Instrument for Treat-only Cohorts}\label{app:application-IV}
One potential concern is that our IV analysis conditions on statin initiation (i.e., restricting to treated individuals only), so the instrument $G$ is effectively used to predict adherence among initiators. As illustrated in Figure~\ref{fig:treated-only}, if there exist unmeasured common causes $U$ of statin initiation, adherence and the outcome, and if $G$ is also related to the statin initiation decision $S$, then conditioning on $S = 1$ (i.e., treated individuals), can open spurious pathways between $G$, $U$ and $Y_k$ through the collider $S$ \citep{bowden2021triangulation}. This would violate the IV independence and potentially exclusion assumptions in the selected sample, even if they held in the full population. To explore this, we constructed an expanded dataset of UK Biobank individuals, including both statin initiators and non-initiators, and examined the association between $G$ and statin initiation. Empirically this association appeared very weak, suggesting that any collider bias induced by conditioning on $S = 1$ is unlikely to be large. However, because the instrument is already extremely weak for adherence in our data, these diagnostics are themselves noisy, and the IV estimates should be interpreted with caution as estimates of adherence effects specifically among statin initiators. 

\subsection{Application Results}\label{app:results}

\begin{table}[htbp]
\centering
\caption{Estimated effects of adherence (MPR) on LDL-c reduction from IPTW. Entries are point estimates with sandwich standard errors in parentheses. 
All effects are per 1\% increase in MPR.}
\label{tab:adherence-effects-iptw}
\begin{tabular}{llcc}
\hline
Outcome & Adherence 
& IPTW (exact) 
& IPTW (over) \\
\hline
$Y_1$ & $MPR_1$ 
& $-1.50\times 10^{-3}$ ($7.82\times 10^{-5}$)
& $-2.31\times 10^{-3}$ ($1.06\times 10^{-4}$) \\[0.4em]

$Y_2$ & $MPR_1$ 
& $-6.41\times 10^{-4}$ ($1.94\times 10^{-4}$)
& $-6.35\times 10^{-4}$ ($2.15\times 10^{-4}$) \\[0.4em]

$Y_2$ & $MPR_2$ 
& $-1.10\times 10^{-3}$ ($1.59\times 10^{-4}$)
& $-1.81\times 10^{-3}$ ($1.71\times 10^{-4}$) \\[0.4em]

$Y_3$ & $MPR_1$ 
& $-4.33\times 10^{-4}$ ($1.24\times 10^{-4}$)
& $-6.83\times 10^{-4}$ ($1.26\times 10^{-4}$) \\[0.4em]

$Y_3$ & $MPR_2$ 
& $-1.55\times 10^{-4}$\hspace{0.3em} ($1.16\times 10^{-4}$)
& $-7.30\times 10^{-4}$ ($1.37\times 10^{-4}$) \\[0.4em]

$Y_3$ & $MPR_3$ 
& $-1.11\times 10^{-3}$ ($1.22\times 10^{-4}$)
& $-9.63\times 10^{-4}$ ($1.10\times 10^{-4}$) \\
\hline
\end{tabular}
\end{table}

\begin{table}[htbp]
\centering
\caption{Estimated effects of adherence (MPR) on LDL-c reduction from G-estimation. Entries are point estimates with sandwich standard errors in parentheses.
All effects are per $1\%$ increase in MPR.}
\label{tab:adherence-effects-g}
\begin{tabular}{llcc}
\hline
Outcome & Adherence 
& G-est (basic GMM) 
& G-est (efficient GMM) \\
\hline
$Y_1$ & $MPR_1$ 
& $-1.55\times 10^{-3}$ ($7.94\times 10^{-5}$)
& $-1.55\times 10^{-3}$ ($6.93\times 10^{-5}$) \\[0.4em]

$Y_2$ & $MPR_1$ 
& $-5.52\times 10^{-4}$ ($8.71\times 10^{-5}$)
& $-5.52\times 10^{-4}$ ($6.66\times 10^{-5}$) \\[0.4em]

$Y_2$ & $MPR_2$ 
& $-1.35\times 10^{-3}$ ($9.12\times 10^{-5}$)
& $-1.35\times 10^{-3}$ ($6.90\times 10^{-5}$) \\[0.4em]

$Y_3$ & $MPR_1$ 
& $-4.46\times 10^{-4}$ ($8.58\times 10^{-5}$)
& $-4.46\times 10^{-4}$ ($6.28\times 10^{-5}$) \\[0.4em]

$Y_3$ & $MPR_2$ 
& $-3.70\times 10^{-4}$ ($9.88\times 10^{-5}$)
& $-3.70\times 10^{-4}$ ($6.32\times 10^{-5}$) \\[0.4em]

$Y_3$ & $MPR_3$ 
& $-1.13\times 10^{-3}$ ($9.86\times 10^{-5}$)
& $-1.13\times 10^{-3}$ ($6.59\times 10^{-5}$) \\
\hline
\end{tabular}
\end{table}

\begin{table}[h]
    \centering
    \caption{Estimated LDL-c reduction (MPR = 100 vs MPR = 0) at each outcome time point:
    point estimates and 95\% confidence intervals from IPTW and G-estimation.}
    \label{tab:ci-estimand}
    \begin{tabular}{llccc}
        \toprule
        Method & Time point & Estimate & 95\% CI (lower) & 95\% CI (upper) \\
        \midrule
        IPTW (over)          & 1 & $-0.23$ & $-0.25$ & $-0.21$ \\
        IPTW (over)          & 2 & $-0.24$ & $-0.28$ & $-0.21$ \\
        IPTW (over)          & 3 & $-0.24$ & $-0.27$ & $-0.20$ \\
        IPTW (exact)         & 1 & $-0.15$ & $-0.17$ & $-0.13$ \\
        IPTW (exact)         & 2 & $-0.17$ & $-0.20$ & $-0.150$ \\
        IPTW (exact)         & 3 & $-0.17$ & $-0.20$ & $-0.14$ \\
        G-est (basic GMM)    & 1 & $-0.15$ & $-0.17$ & $-0.14$ \\
        G-est (basic GMM)    & 2 & $-0.19$ & $-0.21$ & $-0.17$ \\
        G-est (basic GMM)    & 3 & $-0.19$ & $-0.22$ & $-0.17$ \\
        G-est (efficient GMM)& 1 & $-0.16$ & $-0.17$ & $-0.14$ \\
        G-est (efficient GMM)& 2 & $-0.19$ & $-0.21$ & $-0.18$ \\
        G-est (efficient GMM)& 3 & $-0.19$ & $-0.21$ & $-0.18$ \\
        \bottomrule
    \end{tabular}
\end{table}

\begin{table}[!htbp]
\centering
\caption{IPTW weight diagnostic statistics in the statin adherence and LDL-c reduction analysis. We report effective sample size (ESS), top 1\% weight share, and the 99.9th percentile of the cumulative weight distribution at each time point, comparing CBPS \textit{exact} versus \textit{over}-identified weighting.}
\label{tab:realworld_iptw_diagnostics}
\begin{tabular}{lrrr}
\toprule
Time point & ESS & share\_top1 & $q_{0.999}$ \\
\midrule
\multicolumn{4}{l}{\textbf{Panel A: CBPS (exact)}}\\
\midrule
$W_1$ & 12640.21 & 0.02 & 3.26 \\
$W_2$ &  7843.19 & 0.04 & 5.35 \\
$W_3$ &  7493.43 & 0.05 & 6.75 \\
\midrule
\multicolumn{4}{l}{\textbf{Panel B: CBPS (over)}}\\
\midrule
$W_1$ & 11674.73 & 0.02 & 2.07 \\
$W_2$ & 10241.26 & 0.02 & 2.33 \\
$W_3$ & 10059.88 & 0.02 & 2.52 \\
\bottomrule
\end{tabular}
\end{table}

\begin{table}[!htbp]
\centering
\caption{Absolute weighted correlations between adherence measures (MPR1--MPR3) and covariates under CBPS weighting (method = exact). Each cell reports the absolute weighted Pearson and Spearman correlations (Pearson / Spearman). Cells marked \texttt{NA} indicate that the correlation is not applicable (e.g., correlations with future variables or self-correlations). Values exceeding the 0.1 threshold are highlighted in bold.}
\label{tab:cbps_exact_corr}
%\footnotesize
\begin{tabular}{lccc}
\toprule
 & MPR1 & MPR2 & MPR3 \\
\midrule
\multicolumn{4}{l}{\textit{Baseline covariates}} \\
Age & 0.000/0.020 & 0.005/0.011 & 0.004/0.014 \\
Sex & 0.000/0.013 & 0.013/0.016 & 0.005/0.004 \\
Education & 0.000/0.005 & 0.016/0.018 & 0.009/0.000 \\
Assessment centre & 0.000/0.016 & 0.006/0.040 & 0.002/0.003 \\
Baseline LDL-c & 0.000/0.017 & 0.016/0.012 & 0.008/0.010 \\
\addlinespace
\multicolumn{4}{l}{\textit{Follow-up time}} \\
Follow-up time 1 & 0.001/\textbf{0.173} & 0.072/0.010 & 0.080/0.010 \\
Follow-up time 2 & NA & 0.048/0.034 & 0.047/0.010 \\
Follow-up time 3 & NA & NA & 0.040/0.007 \\
\addlinespace
\multicolumn{4}{l}{\textit{Past adherence (MPR)}} \\
MPR1 & NA & 0.049/0.060 & 0.050/0.073 \\
MPR2 & NA & NA & 0.058/0.047 \\
\addlinespace
\multicolumn{4}{l}{\textit{Past LDL-c percentage change}} \\
LDL-c percentage change 1 & NA & 0.000/0.037 & 0.002/0.010 \\
LDL-c percentage change 2 & NA & NA & 0.001/0.030 \\
\bottomrule
\end{tabular}
\end{table}

\begin{table}[!htbp]
\centering
\caption{Absolute weighted correlations between adherence measures (MPR1--MPR3) and covariates under CBPS weighting (method = over). Each cell reports the absolute weighted Pearson and Spearman correlations (Pearson / Spearman). Cells marked \texttt{NA} indicate that the correlation is not applicable (e.g., correlations with future variables or self-correlations). Values exceeding the 0.1 threshold are highlighted in bold.}
\label{tab:cbps_over_corr}
\begin{tabular}{lccc}
\toprule
 & MPR1 & MPR2 & MPR3 \\
\midrule
\multicolumn{4}{l}{\textit{Baseline covariates}} \\
Age & 0.033/0.040 & 0.047/0.049 & 0.008/0.027 \\
Sex & 0.024/0.036 & 0.018/0.024 & 0.000/0.006 \\
Education & 0.000/0.006 & 0.010/0.019 & 0.009/0.006 \\
Assessment centre & 0.002/0.021 & 0.012/0.024 & 0.014/0.013 \\
Baseline LDL-c & 0.021/0.034 & 0.010/0.015 & 0.003/0.006 \\
\addlinespace
\multicolumn{4}{l}{\textit{Follow-up time}} \\
Follow-up time 1 & 0.001/\textbf{0.144} & 0.010/0.002 & 0.012/0.007 \\
Follow-up time 2 & NA & 0.026/0.026 & 0.020/0.005 \\
Follow-up time 3 & NA & NA & 0.004/0.002 \\
\addlinespace
\multicolumn{4}{l}{\textit{Past adherence (MPR)}} \\
MPR1 & NA & \textbf{0.172}/\textbf{0.185} & 0.023/0.086 \\
MPR2 & NA & NA & 0.070/\textbf{0.126} \\
\addlinespace
\multicolumn{4}{l}{\textit{Past LDL-c percentage change}} \\
LDL-c percentage change 1 & NA & 0.060/0.072 & 0.012/0.009 \\
LDL-c percentage change 2 & NA & NA & 0.007/0.029 \\
\bottomrule
\end{tabular}
\end{table}